\documentclass[11pt,a4paper]{article}
\pdfoutput=1
\usepackage[english]{babel}
\usepackage{amsfonts,amsbsy,bm,euscript,mathrsfs,bbm}
\usepackage{amssymb,stmaryrd,faktor,slashed}
\usepackage[x11names]{xcolor}
\usepackage[tbtags]{amsmath}
\usepackage[bookmarks=true,colorlinks=true,linkcolor=black,citecolor=black,urlcolor=black,bookmarksnumbered]{hyperref}
\usepackage[nosort]{cite}
\usepackage{tensor,braket}
\usepackage{tablefootnote}
\usepackage{graphicx}

\numberwithin{equation}{section}

\makeatletter
\renewcommand\section{\@startsection {section}{1}{\z@}
{-3.5ex \@plus -1ex \@minus -.2ex}
{2.3ex \@plus.2ex}
{\normalfont\Large\bfseries}}
\renewcommand\subsection{\@startsection{subsection}{2}{\z@}
{-3.25ex\@plus -1ex \@minus -.2ex}
{1.5ex \@plus.2ex}
{\normalfont\large\bfseries}}
\makeatother

\def\ads{{\rm AdS}_5\times {\rm S}^5}

\newcommand{\com}[2]{\left[#1,#2\right]}

\newcommand{\tr}{\mathrm{tr}}

\newcommand{\Tr}[1]{\tr\left(#1\right)}

\newcommand{\form}[1]{\mathrm{d}#1}
\newcommand{\scom}[2]{\left[#1\stackrel{\star}{,}#2\right]}

\begin{document}

\setcounter{equation}{0}
\setcounter{footnote}{0}
\setcounter{section}{0}

\thispagestyle{empty}

\begin{flushright}
\texttt{
HU-EP-23/11-RTG
}
\end{flushright}

\begin{center}

\vspace{1.5truecm}

{\LARGE \bf Gauge theory on twist-noncommutative spaces}

\vspace{1.5truecm}

{Tim Meier and Stijn J. van Tongeren}

\vspace{1.0truecm}

{\em Institut f\"ur Mathematik und Institut f\"ur Physik, Humboldt-Universit\"at zu Berlin, \\ IRIS Geb\"aude, Zum Grossen Windkanal 6, 12489 Berlin, Germany}

\vspace{1.0truecm}

{{\tt tmeier@physik.hu-berlin.de // svantongeren@physik.hu-berlin.de}}

\vspace{1.0truecm}
\end{center}

\begin{abstract}
We construct actions for four dimensional noncommutative Yang-Mills theory with star-gauge symmetry, with non-constant noncommutativity, to all orders in the noncommutativity. Our construction covers all noncommutative spaces corresponding to Drinfel'd twists based on the Poincar\'e algebra, including nonabelian ones, whose $r$ matrices are unimodular. This includes particular Lie-algebraic and quadratic noncommutative structures. We prove a planar equivalence theorem for all such noncommutative field theories, and discuss how our actions realize twisted Poincar\'e symmetry, as well as twisted conformal and twisted supersymmetry, when applicable. Finally, we consider noncommutative versions of maximally supersymmetric Yang-Mills theory, conjectured to be AdS/CFT dual to certain integrable deformations of the AdS$_5\times$S$^5$ superstring.
\end{abstract}

\newpage

\setcounter{equation}{0}
\setcounter{footnote}{0}
\setcounter{section}{0}

\tableofcontents

\section{Introduction}

Noncommutativity between space-time coordinates is a likely feature of quantum gravity \cite{Doplicher:1994tu}, where our picture of spacetime as a differentiable manifold would break down at the Planck scale, and is actively studied in this regard \cite{Arzano:2021scz,Addazi:2021xuf}. In string theory in particular, noncommutative gauge theory appears in the low energy dynamics of open strings stretching between branes \cite{Seiberg:1999vs,Douglas:2001ba,Szabo:2001kg}, and thereby in the context of the AdS/CFT correspondence \cite{Maldacena:1997re,Maldacena:1999mh,Hashimoto:1999ut}. It is however not known how to construct star-gauge theory actions on arbitrary noncommutative spaces, to all orders in the noncommutativity. In this paper we answer this question in a restricted setting, by constructing all-order actions for four dimensional Yang-Mills theory that are invariant under star-gauge symmetry, for all noncommutative spaces described via Drinfel'd twists based on the Poincar\'e algebra. We also include matter, and discuss the twisted symmetry, the structure of planar Feynman diagrams, and potential AdS/CFT applications, of these theories.

Noncommutative field theory has a history dating back to early work by Snyder \cite{Snyder:1946qz}, who first suggested replacing the Cartesian coordinate fields of Minkowski space by operators with constant nonzero commutators. In the spirit of Weyl quantization, such noncommuting field operators can be traded for a noncommutative product -- the star product -- between regular commutative fields, see e.g. \cite{Madore:2000en,Szabo:2001kg}. In this picture the noncommutative version of a manifold is modeled by equipping the algebra of functions on this manifold by an associative, but noncommutative, (star) product. The original quantized spacetime proposed by Snyder corresponds to the Groenewold-Moyal star product, but in general a wide variety of star products is possible \cite{Kontsevich:1997vb}. In this paper we focus on noncommutative spaces described by star products that arise from Drinfel'd twists \cite{drinfeld_YBESolutions_1983}. Our reason for this is twofold. First, this twist-based approach has direct appeal in noncommutative field theory, for instance offering all order star products with clear algebraic properties, a natural twisted differential calculus \cite{Aschieri:2009zz}, and the notion of twisted (Poincar\'e) symmetry \cite{Chaichian:2004za,Wess:2003da,Chaichian:2004yh}. Second, in the context of AdS/CFT, there are integrable (Yang-Baxter) deformations of the AdS$_5\times$S$^5$ superstring with Drinfel'd twisted symmetry \cite{Vicedo:2015pna,vanTongeren:2015uha,Borsato:2021fuy}, and their field theory duals are conjectured to be maximally-supersymmetric Yang-Mills theory on correspondingly twisted noncommutative spacetimes \cite{vanTongeren:2015uha,vanTongeren:2016eeb}.

Star products arising from Drinfel'd twists come in various forms, including the basic Groenewold-Moyal one. The latter is special in the sense that the associated noncommutativity is constant, in Cartesian coordinates, and relatedly that the underlying algebraic structure is abelian. Other Drinfel'd twists have non-constant Cartesian noncommutativity, and can be non-abelian. In the Groenewold-Moyal case it is well known how to construct an action for Yang-Mills theory invariant under star-gauge transformations, as reviewed in \cite{Szabo:2001kg}. Beyond this case however, it is generally not clear how to define a suitable dual field strength tensor, and construct an action for noncommutative Yang-Mills theory. This question was previously investigated for Yang-Mills theory on $\kappa$-Minkowski space in \cite{Dimitrijevic:2011jg,Dimitrijevic:2014dxa} for example, where the various approaches were necessarily perturbative, and solved to leading order in the noncommutativity.\footnote{See \cite{Hersent:2022gry} for a recent review on noncommutative gauge theories, including how $\kappa$-Minkowski gauge theory can be formulated in five dimensions \cite{Mathieu:2020ccc}.} The only non-constant case known to all orders, to our knowledge, is the $U(1)$ Yang-Mills theory studied in \cite{Ciric:2017rnf} for a particular angular twist with linear noncommutativity, where standard Hodge duality suffices. In this paper we reconsider this question for general Drinfel'd twists, and give a construction that yields all-order star-gauge invariant Yang-Mills actions for all twists based on the Poincar\'e algebra, whose $r$ matrices are unimodular. Our present construction does not cover $\kappa$-Minowski space, but it does include related linearly noncommutative spaces. It also covers cases with quadratic noncommutativity, such as the Lorentz deformation we previously considered in \cite{Meier:2023kzt}.

Our construction relies on a twisted notion of Hodge duality that naturally incorporates the noncommutative structure, building on the natural twist deformation of the Levi-Civita symbol. Requiring this notion of Hodge duality to be left and right star linear -- essential in constructing a noncommutative Yang-Mills action -- restricts us to twists based on the Poincar\'e algebra. Further requiring the similarly crucial cyclicity of products under an integral, means that the $r$ matrix associated to our twist has to be unimodular. Beyond defining Yang-Mills theory, we discuss how to couple it to fundamental and adjoint matter, introducing suitable index-free notation for fermions. As a first step towards investigating these theories at the quantum level, we consider Filk's theorem \cite{Filk:1996dm}, which for the Groenewold-Moyal deformation shows that planar Feynman diagrams are given by their undeformed versions up to star products between the external fields. By recasting its proof in position space, we are able to generalize this theorem to a planar equivalence theorem for any of our deformed theories. We then also discuss the twisted symmetry of our models, including twisted conformal and twisted supersymmetry. Finally, with an eye towards AdS/CFT, we define noncommutative versions of maximally supersymmetric Yang-Mills theory (SYM), having in mind its planar integrable structure \cite{Beisert:2010jr}, which we are aiming to extend to our noncommutative cases with our planar equivalence theorem \cite{NCintegrabilitypaper}.  As a byproduct of our investigation, we also find what we believe to be a new explicit type of Drinfel'd twist which is fundamentally rank four, in the process completing the construction of twists for the Poincar\'e algebra of \cite{Tolstoy:2008zz}.

\section{Twist-noncommutative spaces}

We will be working with noncommutative spaces in terms of star products, and limiting ourselves to those defined by Drinfel'd twists. We start from the Lie algebra of vector fields on Minkowski space, and twist its natural Hopf algebra structure, leading to noncommutative versions of Minkowski space. We give a higher-than-standard level of detail in our review of Drinfel'd twists and the associated twisted differential calculus, both to clearly establish our general approach and notation, and with the aim to provide an accessible discussion for readers with a background in e.g. AdS/CFT and integrable deformations of string sigma models.

Let us start by reviewing Drinfel'd twisted Hopf algebras as relevant in our context. For further details we refer to \cite{Chari:1994pz}.

\subsection{Drinfel'd twists}

\paragraph{Hopf algebra of vector fields} The space of vector fields $\mathfrak{g}$ on a smooth manifold $M$, together with the commutator bracket
\begin{align*}
\left[.,.\right]:~TM\times TM\rightarrow TM;~\left[a,b\right]=ab-ba,
\end{align*}
forms a Lie algebra. This algebra can be extended to a Hopf algebra to act on fields and their products in an ordinary field theory by introducing the standard coproduct, counit and antipode on the universal enveloping algebra $\mathcal{U}(\mathfrak{g})$ as
\begin{align*}
\Delta(X)&=1\otimes X+X\otimes1&\Delta(1)&=1\otimes1\\
S(X)&=-X&S(1)&=1\\
\epsilon(X)&=0&\epsilon(1)&=1,
\end{align*}
for all $X\in \mathfrak{g}$, extended to $\mathcal{U}(\mathfrak{g})$. The coproduct extending the action of $X$ to an $n$-fold tensor product is $\Delta^{(n)} = (1\otimes \ldots \otimes 1 \otimes \Delta) \Delta^{(n-1)}$, with $\Delta^{(2)} = \Delta$.

We will consider vector fields on four dimensional Minkowski space. While we initially consider arbitrary vector fields, a central role will be played by the subalgebra of isometries, the Poincar\'e algebra $\mathcal{P}$. We take this to be generated by the translation generators $p_\mu$ and Lorentz generators $M_{\mu\nu}$
\begin{equation}
p_\mu = - i \partial_\mu, \qquad M_{\mu\nu} = i (x_\mu \partial_\nu - x_\nu \partial_\mu).
\end{equation}
At times we exchange (Lie derivatives along) vector fields for abstract generators, we hope this will be clear from context in cases where we fail to indicate this explicitly.

\paragraph{Drinfel'd twist} A Drinfel'd twist $\mathcal{F}$ is an invertible element of  $\mathcal{U}(\mathfrak{g}) \otimes \mathcal{U}(\mathfrak{g})$ fulfilling the cocycle and normalization condition \cite{drinfeld_YBESolutions_1983}
\begin{align}
\label{eq:cocycle}
\left(\mathcal{F}\otimes1\right)\left(\Delta\otimes1\right)\mathcal{F}&=\left(1\otimes\mathcal{F}\right)\left(1\otimes\Delta\right)\mathcal{F}\\
\left(1\otimes\epsilon\right)\mathcal{F}=(\epsilon&\otimes1)\mathcal{F}=1\otimes1\notag
\end{align}
that is expandable in a deformation parameter $\lambda$
\begin{align*}
\mathcal{F}=1\otimes1+\lambda \mathcal{F}^{(1)}+\mathcal{O}\left(\lambda^2\right).
\end{align*}
The antisymmetrization of the leading order term gives the classical $r$ matrix $r \in \Lambda^2{\mathfrak{g}}$
\begin{align*}
r_{12}=\mathcal{F}^{(1)}_{21}-\mathcal{F}^{(1)} _{12},
\end{align*}
which satisfies the classical Yang Baxter equation (CYBE)
\begin{align*}
[r_{12},r_{13}]+[r_{13},r_{23}]+[r_{12},r_{23}]=0.
\end{align*}
A Drinfel'd twist exists for every such $r$ matrix, although an explicit general formula is not known. By a suitable algebra homomorphism any twist can be made $r$ symmetric, meaning $\mathcal{F}^{(1)}_{12} = -\tfrac{r_{12}}{2}$ \cite{drinfeld_YBESolutions_1983}, see also \cite{Giaquinto:1994jx,Tolstoy:2008zz}.

\paragraph{Twisted Hopf algebra} To define an associated twisted Hopf algebra, we write $\mathcal{F}$ as a sum of tensor products between distinct elements $f^\alpha,~\bar{f}^\alpha$ of $\mathcal{U}(\mathfrak{g})$:
\begin{align*}
\mathcal{F}&=f^\alpha\otimes f_\alpha&\mathcal{F}^{-1}&=\bar{f}^\alpha\otimes \bar{f}_\alpha.
\end{align*}
We can then define a new coproduct and antipode
\begin{align*}
\Delta_\mathcal{F}(X)&=\mathcal{F}\Delta(X)\mathcal{F}^{-1}\\
S_\mathcal{F}(X)&=f^\alpha S(f_\alpha)S(X)S(\bar{f}^\beta)\bar{f}_\beta
\end{align*}
\paragraph{Twisted module and product} The algebra of fields with the standard associative and commutative product between fields
\begin{align*}
\mu(f,g)(x):=f(x)g(x)
\end{align*}
respects the Hopf algebra on  $\mathcal{U}(\mathfrak{g})$, i.e.:
\begin{equation}
\label{eq:muIsEquiv}
X\mu(f,g)=\mu\left(\Delta(X)(f,g)\right) \qquad \forall X\in  \mathcal{U}(\mathfrak{g}).
\end{equation}
To be compatible with the twisted Hopf algebra on  $\mathcal{U}(\mathfrak{g})$, the product between functions need to be deformed to
\begin{equation}
\mu_\mathcal{F}(f,g)(x)=\mu\left(\mathcal{F}^{-1}(f,g)\right)(x) \equiv f(x)\star g(x) = \bar{f}^\alpha(f) \bar{f}_\alpha(g),
\end{equation}
so that
\begin{equation}
\label{eq:twistedprodrule}
X\mu_\mathcal{F}(f,g)=\mu_\mathcal{F}\left(\Delta_\mathcal{F}(X)(f,g)\right)\qquad \forall X\in \mathcal{U}(\mathfrak{g}).
\end{equation}
This product is conventionally called the star product. It is not commutative, but it is associative, since
\begin{equation}
\begin{aligned}
f \star \left(g \star h\right)&=\mu_{\mathcal{F}}\left(f,\mu_{\mathcal{F}}\left(g,h\right)\right)\\
&=\mu(\mathcal{F}^{-1}\left(1\otimes \mu\right)\left(1\otimes \mathcal{F}^{-1}\right)\left(f,g,h\right))\\
&=\mu(\left(1\otimes\mu\right)\left(1 \otimes \Delta\right)\left(\mathcal{F}^{-1}\right)\left(1\otimes\mathcal{F}^{-1}\right)\left(f,g,h\right))\\
&=\mu(\left(\mu \otimes 1\right)\left(\Delta \otimes 1\right)\left(\mathcal{F}^{-1}\right)\left(\mathcal{F}^{-1} \otimes 1 \right)\left(f,g,h\right))\\
&=\mu(\mathcal{F}^{-1} \left(\mu \otimes 1 \right) \left(\mathcal{F}^{-1} \otimes 1\right)\left(f,g,h\right))\\
&=\mu_\mathcal{F}\left(\mu_\mathcal{F}\left(f,g\right),h\right)\\
&=\left(f \star g\right) \star h,
\end{aligned}
\end{equation}
due to property \eqref{eq:muIsEquiv}, the cocycle condition \eqref{eq:cocycle} and the associativity of the ordinary point wise product $\mu$. The noncommutativity is captured by the R Matrix
\begin{align}
\label{eq:Rmatrixdef}
\mathcal{R} & = \mathcal{F}_{op}\mathcal{F}^{-1}=R^\alpha \otimes R_\alpha\\
f \star g & = \mu\left(\mathcal{F}^{-1}(f,g)\right)=\mu\left(\mathcal{F}_{op}^{-1}\mathcal{R}(f,g)\right) = R_\alpha g \star R^\alpha f,\notag
\end{align}
where $\mathcal{R}$ satisfies the (quantum) Yang-Baxter equation thanks to the cocycle condition on the twist. Our twisted Hopf algebra is triangular since
\begin{equation}
\mathcal{R}_{op} = \mathcal{R}^{-1}.
\end{equation}
We can then similarly write
\begin{align*}
f \star g  = \mu\left(\mathcal{F}^{-1}\mathcal{R}^{-1} (g,f)\right) = \bar{R}^\alpha g \star \bar{R}_\alpha f.
\end{align*}
In line with this, for notational convenience, from here on out we use bars to denote inverses on the twist and $\mathcal{R}$ matrix, i.e.
\begin{equation*}
\bar{\mathcal{F}} \equiv \mathcal{F}^{-1}, \qquad \bar{\mathcal{R}} \equiv \mathcal{R}^{-1},
\end{equation*}
in line with the bars on $f^\alpha$ and $R^\alpha$.

To leading order, $\mathcal{R} = 1+ \lambda r$, and the noncommutativity is captured by the classical $r$ matrix through
\begin{equation}
[x^\mu\stackrel{\star}{,}x^\nu] \equiv  x^\mu \star x^\nu - x^\nu \star x^\mu = \lambda \mu(r(x^\mu \otimes x^\nu)) + \mathcal{O}(\lambda^2) = \lambda r^{\mu\nu} + \mathcal{O}(\lambda^2).
\end{equation}
Here $r^{\mu\nu}$ are the components of the $r$ matrix in the vector field representation
\begin{equation}
r = \frac{1}{2} r^{\mu\nu}\partial_\mu \wedge \partial_\nu,
\end{equation}
where we use the antisymmetric wedge product
\begin{equation}
a \wedge b = a \otimes b - b \otimes a.
\end{equation}
The components $\lambda r^{\mu\nu}$ are also frequently denoted $\theta^{\mu\nu}$.\footnote{Thanks to the classical Yang-Baxter equation, $r^{\mu\nu}$ defines a Poisson structure, a necessary condition for any star product \cite{Kontsevich:1997vb}.}

\subsection{Examples of twisted structures}
\label{subsec:twistexamples}

Here we discuss some examples of noncommutative structures, that our construction below will cover.

\paragraph{Groenewold-Moyal noncommutative space} The best understood example of a noncommutative product arising from a Drinfel'd twist is the so-called Moyal product, obtained from the twist
\begin{align*}
\mathcal{F}&=\exp\left(-\frac{i\theta^{\mu\nu}}{4} p_\mu \wedge p_\nu\right),
\end{align*}
with $\theta$ a constant antisymmetric matrix. The corresponding star product is
\begin{align}
f(x) \star g(x)&= f(x)g(x)+\left.\sum_{n=1}^\infty\frac{(-1)^n}{2^n n!}\left(\theta^{\mu\nu}\partial_{\mu}^x\partial_{\nu}^y\right)^n f(x)g(y)\right|_{y=x}.
\end{align}
corresponding to the basic commutation relations
\begin{align}
[x^\mu\stackrel{\star}{,}x^\nu]&= i \theta^{\mu\nu},
\end{align}
i.e. a constant $r$ matrix in the vector field representation. Next are noncommutative structures with non-constant $r$ matrices.

\paragraph{Lie-algebraic noncommutative space} Starting with linear coordinate dependence in Cartesian coordinates, we can consider e.g.
\begin{align}
\label{eq:Flambda}
\mathcal{F}_{\lambda}=\exp\left(-\frac{i\lambda}{2}p_3 \wedge M_{12}\right),
\end{align}
and
\begin{align}
\label{eq:Frho}
\mathcal{F}_{\rho}=\exp\left(-\frac{i\rho}{2}p_0 \wedge M_{12}\right).
\end{align}
These twists correspond to nonzero
\begin{equation*}
\scom{x^3}{x^{1/2}}=\pm i\lambda x^{2/1},
\end{equation*}
and
\begin{equation*}
\scom{x^0}{x^{1/2}}=\pm i\rho x^{2/1},
\end{equation*}
respectively, with no higher order corrections. These two examples of so-called Lie-algebraic noncommutative structures are, among other names, known as $\lambda$-Minkowski and $\rho$-Minkowski space, see e.g. \cite{Gutt:1983,Gracia-Bondia:2001ynb,DimitrijevicCiric:2018blz,lizzi:2021dud,Gubitosi:2021itz,Fabiano:2023uhg}, and \cite{Lukierski:2005fc} for details on the twisted structure.

\paragraph{Quadratic (quantum-plane) noncommutative space} A twist with an $r$ matrix quadratic in coordinates leads to noncommutative relations that are of a quantum plane type. Writing $r^{\mu\nu} = (r^{\mu\nu})_{\sigma\rho} x^\rho x^\sigma$, we have
\begin{equation*}
\scom{x^\mu}{x^\nu} = (r^{\mu\nu})_{\sigma\rho} x^\rho x^\sigma.
\end{equation*}
This can be viewed as the leading order expansion of
\begin{equation}
\label{eq:quadraticallorder}
x^\mu \star x^\nu = x^\rho \star x^\sigma R_\sigma{}^{\mu}{}_\rho{}^{\nu}
\end{equation}
where $R_\sigma{}^{\mu}{}_\rho{}^{\nu}$ is the (quantum) $\mathcal{R}$ matrix in the coordinate representation, with the classical $r$ matrix as its first order term.

An example of a twist giving this structure is
\begin{align}
\label{eq:Lorentztwist}
\mathcal{F}=\exp\left(\frac{i\lambda}{2}M_{01}\wedge M_{23}\right),
\end{align}
which we studied before in \cite{Meier:2023kzt}, naming it the Lorentz deformation. The corresponding leading order nonzero commutators are
\begin{equation}
\scom{x^{0/1}}{x^2} = - i \lambda x^{1/0} x^3, \qquad \scom{x^{0/1}}{x^3} =  i \lambda x^{1/0} x^2,
\end{equation}
In all order quantum-plane form we have \eqref{eq:quadraticallorder} with
\begin{align}
\label{eq:LorentztwistRmatrix}
R_\rho{}^\mu{}_\sigma{}^\nu =\cosh(\Lambda^{\mu\nu})\delta^\mu_\rho\delta^\nu_\sigma + i\sinh(\Lambda^{\mu\nu})\epsilon^\mu_\rho\epsilon^\nu_\sigma
\end{align}
with
\begin{align*}
\epsilon^0_1=\epsilon^1_0=\epsilon^2_3=\epsilon^3_2=1
\end{align*}
and $\epsilon^\mu_\nu=0$ otherwise, and no summation over $\mu$ and $\nu$ on the right hand side of equation \eqref{eq:LorentztwistRmatrix}. The matrix $\Lambda$ is given by
\begin{equation}
\label{eq:LorentzDefThetaatrix}
\Lambda^{\mu\nu}=\begin{pmatrix}
0&0&-\lambda&\lambda\\
0&0&-\lambda&\lambda\\
\lambda&\lambda&0&0\\
-\lambda&-\lambda&0&0
\end{pmatrix}^{\mu\nu}.
\end{equation}
The index permutation symmetries of the twist relate the various entries of this $R$ matrix, as mentioned in \cite{Meier:2023kzt}.

A second example that will be relevant below is the light-cone (cousin of the) Lorentz deformation, with twist
\begin{align}
\label{eq:lightconeLorentztwist}
\mathcal{F}=\exp\left(\frac{i\alpha}{2}M_{+1}\wedge M_{+2}\right),
\end{align}
where we use light-cone coordinates
\begin{equation}
x^\pm = x^0 \pm x^3.
\end{equation}
Its noncommutative structure corresponds to the undeformed $R_\mu{}^\rho{}_\nu{}^\sigma = \delta^\rho_\mu \delta^\sigma_\nu$ with additional
\begin{equation}
\begin{aligned}
\label{eq:lightconeLorentztwistRmatrix}
R_-{}^1{}_-{}^2 = i \alpha, \qquad R_-{}^1{}_2{}^+  =  2 i \alpha, \qquad R_1{}^+{}_2{}^+ = 4 i \alpha, \qquad R_-{}^+{}_-{}^+ = - 4 \alpha^2,
\end{aligned}
\end{equation}
plus others related by permuting the first and second pairs of indices, or index values $1$ and $2$, for a sign flip on $\alpha$.

The Lorentz twist and its light-cone cousin are related by an infinite boost in the (03) plane, upon suitable rescaling of the deformation parameter.

\paragraph{Non-abelian $r$ matrices} In parallel to categorizing twists by the Cartesian coordinate dependence of their $r$ matrices, we can also distinguish them in terms of their underlying algebraic structure. From this perspective, all noncommutative structures we gave above are abelian, meaning that the twist is built out of commuting vector fields (generators). We can also consider non-abelian twists. A relevant example is that of almost-abelian twists \cite{Tolstoy:2008zz,Borsato:2016ose,vanTongeren:2016eeb}, which are sequences of noncommuting abelian twists, e.g.
\begin{equation}
\mathcal{F}=\exp\left(\frac{i\beta}{2}p_1 \wedge M_{23}\right)\exp\left(\frac{i\alpha}{2} p_2 \wedge p_3 \right).
\end{equation}
with corresponding
\begin{equation}
\scom{x^2}{x^3}  = -i \alpha, \qquad \scom{x^1}{x^2}  = i \beta x^3,\qquad
\scom{x^1}{x^3}  = -i \beta x^2.
\end{equation}
We will encounter further $r$ matrices and twists of this type below, and give more details on their algebraic structure there. Below we will encounter one further type of non-abelian $r$ matrix, described in section \ref{sec:compatibletwists} and appendix \ref{app:newtwist}.

For abelian $r$ matrices there is always a (local) choice of coordinates in which the $r$ matrix is constant, while this is impossible for non-abelian $r$ matrices.\footnote{In the abelian case, non-constant noncommutativity in Cartesian coordinates still manifests itself through global properties in other coordinates. For instance, the noncommutativity associated to $r = M_{23} \wedge p_1$ becomes constant if we use polar coordinates in the $(2,3)$ plane. However $\theta$ is not globally defined, we should consider $e^{i\theta}$ rather than $\theta$ as an operator before the Weyl map, and $M_{23}(e^{i \theta})$ is not constant.}

\subsection{Twisted differential calculus}
\label{subsec:twisteddifferentialcalculus}

To define field theory actions, in particular gauge theory, we need a differential calculus that is adapted to our twisted algebra. As vector fields, partial derivatives naturally act on star products of functions via the twisted coproduct. We can let the twist act naturally on forms as well via the Lie derivative, leading to a twisted structure there. This twisted differential calculus has been previously discussed in e.g. \cite{Aschieri:2005zs,Aschieri:2009ky}. Here we add to this by keeping track of the appearance of the twist and $\mathcal{R}$ matrices in various representations. We also introduce a generalized Levi-Civita tensor and associated notion of Hodge duality.

\paragraph{Twisted tensor and wedge product} Let us begin by defining the twisted tensor product
\begin{equation}
\label{eq:twistedtensorproduct}
\otimes_\star(\omega,\omega')  = \mu_\otimes \left(\bar{\mathcal{F}}(\omega,\omega')\right) \equiv \bar{f}^\alpha(\omega) \otimes \bar{f}_\alpha(\omega'),
\end{equation}
where $\omega$ and $\omega'$ are differential one forms, $\otimes$ is the usual tensor product of forms, and, here and below, the vector fields in the twist act via Lie derivatives. Next we define the twisted wedge product as\footnote{We hope that the distinction between the tensor and wedge products as they appear in the Hopf algebra, and in the differential calculus, is clear from context.}
\begin{align*}
h\star\omega&=\mu\left(\bar{\mathcal{F}}\, (h,\omega)\right)=\bar{f}^\alpha(h)\bar{f}_\alpha(\omega)\\
\omega\wedge_\star\omega'&=\hat{\mu}\left(\bar{\mathcal{F}}\, (\omega,\omega')\right)\equiv\bar{f}^\alpha(\omega)\wedge\bar{f}_\alpha(\omega'),
\end{align*}
where $h$ is a function, $\omega$ and $\omega'$ are arbitrary degree forms, and $\wedge$ denotes the regular undeformed wedge product of forms. Considering functions to be zero forms, the second formula includes the first of course. Due to the properties of the twist, these products lead to an algebra of forms and functions that carries a representation of the twisted Hopf algebra.

Depending on context it is natural to express coefficients of forms with or without a star product, which we will distinguish by a star superscript, e.g.
\begin{equation}
\label{eq:oneformcomponents}
A = A_\mu  \form{x^\mu}= A_\mu^\star \star \form{x^\mu}.
\end{equation}
It is also helpful to explicitly relate the twisted products between functions and one forms, and between one forms, back to their undeformed counterparts. Since the twist acts via Lie derivatives on a set of basis one forms, acting on a one form simply evaluates the corresponding leg of the twist in the associated representation, picking up a set of indices for each one form. This defines
\begin{equation}
\label{eq:Fwithindicesdef}
\begin{aligned}
\form{x^\mu} \star f(x) & = \form{x^\nu} \bar{F}_\nu{}^\mu f(x),\\
f(x)\star \form{x^\mu} & = \form{x^\nu}(\bar{F}_{op}){}_\nu{}^\mu f(x),\\
\form{x^\mu} \wedge_\star \form{x^\nu} & = \form{x^\sigma} \wedge \form{x^\rho}\bar{F}_\sigma{}^\mu{}_\rho{}^\nu .
\end{aligned}
\end{equation}
The twist still acts on the function moved, via vector fields, i.e. $\bar{F}_\nu{}^\mu$ is a differential operator. In our conventions, a single set of upper and lower indices is associated to the first space acted on, for consistency of placement with double sets of indices. We note that
\begin{equation*}
\bar{F}_\nu{}^\mu(\mathrm{d}x^\kappa) = \bar{F}_\nu{}^\mu{}_\lambda{}^\kappa \mathrm{d}x^\lambda
\end{equation*}
and of course $(\bar{F}_{op})_\sigma{}^\mu{}_\rho{}^\nu = \bar{F}_\rho{}^\nu{}_\sigma{}^\mu$.\footnote{In principle we could consider defining two further $F$s with four indices, associated to $f \star (\mathrm{d}x^\mu \wedge \mathrm{d}x^\nu)$ or $f \star (\mathrm{d}x^\mu \wedge_\star \mathrm{d}x^\nu)$. They are not very natural however, and we will not need them, since they can always be avoided by considering $(f\star \mathrm{d}x^\mu)\wedge_\star \mathrm{d}x^\nu$ instead.} In terms of equation \eqref{eq:oneformcomponents} we then have
\begin{equation}
A_\mu = (\bar{F}_{op}){}_\mu{}^\nu A_\nu^\star.
\end{equation}

As examples, for the Moyal deformation we have $\bar{F}_\nu{}^\mu = \delta^\mu_\nu$ and $\bar{F}_\sigma{}^\mu{}_\rho{}^\nu = \delta^\mu_\sigma \delta^\nu_\rho$, while for the Lorentz deformation of \eqref{eq:Lorentztwist} we have
\begin{equation}
\label{eq:FexampleLorentz}
\begin{aligned}
\bar{F}_\mu{}^\nu &= \tensor{\begin{pmatrix}
\cosh\left(\frac{\lambda}{2}M_{23}\right) & -\sinh\left(\frac{\lambda}{2}M_{23}\right)&0&0\\
-\sinh\left(\frac{\lambda}{2}M_{23}\right) & \cosh\left(\frac{\lambda}{2}M_{23}\right) &0 &0\\
0 & 0 & \cos\left(\frac{\lambda}{2}M_{01}\right) & -\sin\left(\frac{\lambda}{2}M_{01}\right)\\
0 & 0 &  \sin\left(\frac{\lambda}{2}M_{01}\right) & \cos\left(\frac{\lambda}{2}M_{01}\right)
\end{pmatrix}}{_\mu^\nu}\\
\bar{F}_\rho{}^\mu{}_\sigma{}^\nu &=\cosh(\frac{\Lambda^{\mu\nu}}{2})\delta^\mu_\rho\delta^\nu_\sigma+i\sinh(\frac{\Lambda^{\mu\nu}}{2})\epsilon^\mu_\rho\epsilon^\nu_\sigma,
\end{aligned}
\end{equation}
with $\Lambda$ given in equation \eqref{eq:LorentzDefThetaatrix}.

\paragraph{$\mathcal{R}$ matrix} The commutation rules involving forms in our twisted algebra are governed by the $\mathcal{R}$ matrix. In general, moving a function through a basis one form gives
\begin{equation}
\begin{aligned}
\form{x^\mu}\star f(x)
& =\mu(\bar{\mathcal{F}}_{op}\mathcal{R}(\form{x^\mu} , f(x))),\\
& =\mu\left[\bar{\mathcal{F}}_{op}\left(\form{x^\nu}(R^\alpha)_\nu{}^\mu , R_\alpha(f(x)\right)\right],\\
& =R_\alpha(f(x)) \star \form{x^\nu}(R^\alpha)_\nu{}^\mu).
\end{aligned}
\end{equation}
At this point we will make a strong restriction, and consider only twists for which the action of the $\mathcal{R}$ matrix can be suitably factored out of the star product, by assuming that $(R^\alpha)_\nu{}^\mu$ commutes with the $\bar{f}^\beta$ and $\bar{f}_\beta$ governing the star product.\footnote{This holds for any twist with constant $(R^\alpha)_\nu{}^\mu$, i.e. any $r$ matrix built out of vector fields at most linear in coordinates, covering all deformations we will end up considering. At this stage more general solutions are allowed however, e.g. the abelian twist $\exp(i \lambda \, (x^1)^2 \partial_2 \wedge (x^1)^2 \partial_3)$.} We can then write
\begin{equation}
\label{eq:dxfComRule}
\form{x^\mu}\star f(x)=\left(R_\nu{}^\mu f(x)\right)\star\form{x^\nu},
\end{equation}
where $R_\nu{}^\mu = (R^\alpha)_\nu{}^\mu R_\alpha$, which still acts on the function moved via vector fields. Equivalently
\begin{equation}
\label{eq:fdxComRule}
f(x)\star\form{x^\mu}=\form{x^\nu}\star\left(\bar{R}_\nu{}^\mu f(x)\right).
\end{equation}
Conceptually the $R$ matrix of course acts simultaneously on both objects in a product, and it always does for products between general objects, regardless of the choice of twist. At the same time, many twists, in particular all ones based on the Poincar\'e algebra, admit this convenient factorization for products involving basis forms.

The star-anticommutation rules of forms are similarly
\begin{equation}
\label{eq:wedgeComRule}
\form{x^\mu}\wedge_\star\form{x^\nu}=-R^{\enspace \mu\enspace\nu}_{\rho \enspace \sigma} \left(\form{x^\sigma}\wedge_\star\form{x^\rho}\right)=-\bar{R}^{\enspace \nu\enspace\mu}_{\sigma \enspace \rho} \left(\form{x^\sigma}\wedge_\star\form{x^\rho}\right),
\end{equation}
where, again, we assume this can be factored out. For quadratic noncommutative structures, this $R^{\enspace \mu\enspace\nu}_{\rho \enspace \sigma}$ coincides with the one used in the coordinate commutation relations, as in e.g. \eqref{eq:LorentztwistRmatrix}. For forms, $R_{\nu}{}^{\mu}$ and $R^{\enspace \mu\enspace\nu}_{\rho \enspace \sigma}$ are useful objects for general twists, assuming they can be factored out of the star product.

For completeness, the relation \eqref{eq:Rmatrixdef} between the twist and R matrix, becomes
\begin{equation}
\begin{aligned}
R_\nu{}^\mu & = (F_{op}){}_\nu{}^\rho \bar{F}_\rho{}^\mu,\\
R_\rho{}^\mu{}_\sigma{}^\nu & = (F_{op})_\rho{}^\lambda{}_\sigma{}^\zeta \bar{F}_\lambda{}^\mu{}_\zeta{}^\nu,
\end{aligned}
\end{equation}
with $F_{op}=\bar{F}$ for abelian deformations, i.e. $R = \bar{F}^2$ there.

As examples, for the Moyal deformation these $\mathcal{R}$ matrices are trivial: $R_\nu{}^\mu = \delta^\mu_\nu$ and $R_\sigma{}^\mu{}_\rho{}^\nu = \delta^\mu_\sigma \delta^\nu_\rho$. For the Lorentz deformation $R$ is the square of $\bar{F}$ as given in equations \eqref{eq:FexampleLorentz} -- i.e. the same expressions with doubled deformation parameter. $R_\rho{}^\mu{}_\sigma{}^\nu$ of course also matches equation \eqref{eq:LorentztwistRmatrix}. Twists of $\lambda$ or $\rho$-Minkowski type lie between these two, with $R_\nu{}^\mu$ a function of the momenta only, and trivial $R_\rho{}^\mu{}_\sigma{}^\nu$.

\paragraph{Properties of the $\mathcal{R}$ matrix} Since permuting terms in a symmetric star product should do nothing, any $R$ matrix has to satisfy
\begin{equation}
\label{eq:R4indextrivialonsymmetricproduct}
R_\rho{}^\mu{}_\sigma{}^\mu = \delta_\rho^\mu \delta_\sigma^\mu.
\end{equation}
Next, the cocycle condition implies properties for our incarnations of the $\mathcal{R}$ matrix. By associativity of the star product we can move $f$ and $g$ through $\form{x^\mu}$ in $f\star g \star \form{x^\mu}$ either together or consecutively, giving us
\begin{equation}
\label{eq:R2indexdistributionoverstarproduct}
\bar{R}_\mu{}^\nu(f\star g)=\left(\bar{R}_\mu{}^\rho f\right)\star \left(\bar{R}_\rho{}^\nu g\right).
\end{equation}
Similarly, reversing the order of $f \star \form{x^\mu} \star \form{x^\nu}$ by different pairwise permutations gives
\begin{equation}
\label{eq:YB1}
\bar{R}_\rho{}^\zeta{}_\lambda{}^\sigma  \bar{R}_\zeta{}^\nu  \bar{R}_\sigma{}^\mu = \bar{R}_\lambda{}^\sigma  \bar{R}_\rho{}^\zeta \bar{R}_\zeta{}^\nu{}_\sigma{}^\mu,
\end{equation}
i.e. the Yang-Baxter equation for $\mathcal{R}$ evaluated in appropriate representations. For completeness, permuting forms in $\form{x^\mu} \wedge_\star \form{x^\nu}\wedge_\star \form{x^\lambda}$ gives the Yang-Baxter equation
\begin{equation}
\bar{R}^{\enspace \nu \enspace \mu}_{\rho \enspace \sigma}\bar{R}^{\enspace \lambda \enspace \sigma}_{\tau \enspace \omega}\bar{R}^{\enspace \tau \enspace \rho}_{\kappa \enspace \xi}=
\bar{R}^{\enspace \lambda \enspace \nu}_{\rho \enspace \sigma}\bar{R}^{\enspace \rho \enspace \mu}_{\kappa \enspace \tau}\bar{R}^{\enspace \sigma \enspace \tau}_{\xi \enspace \omega}.
\end{equation}

\paragraph{Exterior derivative} In addition to an exterior algebra, we need an exterior derivative. Since the conventional exterior derivative commutes with Lie derivatives, it is not affected by the twist and fulfils the usual Leibniz rule
\begin{equation}
\mathrm{d}(\omega \wedge_\star \chi) = \mathrm{d}\omega \wedge_\star \chi + (-1)^p \omega \wedge_\star d\chi,
\end{equation}
for $p$ and $q$ forms $\omega$ and $\chi$, respectively, including functions as zero forms. We hence use the standard exterior derivative.

When expressing $\mathrm{d}$ in a basis via star products, this introduces a deformed partial derivative $\partial^\star_\mu$,
\begin{align}
\label{eq:extDer}
\mathrm{d}f(x)=\partial_\mu f(x)\form{x^\mu}=\partial^\star_\mu f(x)\star\form{x^\mu},
\end{align}
meaning $\partial^\star_\mu=(F_{op}){}_\mu{}^{\nu}\partial_\nu$. We can then use the Leibniz rule for the exterior derivative to calculate a star-product rule for the partial derivatives, which coincides with the twisted coproduct of the corresponding vector fields (the momenta). In general this takes the form
\begin{equation*}
\partial_\mu (f \star g) = \left(\bar{F}_{op}\right)_\mu{}^\sigma\left[\left((F_{op})_\nu{}^\lambda \partial_\lambda f \right) \star \left(R_\sigma{}^\nu g \right) + f \star \left((F_{op})_\sigma{}^\nu \partial_\nu g\right)\right].
\end{equation*}
Finally, closedness of $\mathrm{d}f$ gives a commutation rule for partial star derivatives
\begin{align*}
\partial^\star_\mu\partial_\nu^\star=\bar{R}_\mu{}^\rho{}_\nu{}^\sigma\partial^\star_\sigma\partial_\rho^\star=R_\nu{}^\sigma{}_\mu{}^\rho\partial^\star_\sigma\partial_\rho^\star.
\end{align*}

\paragraph{Integration} Because the star product is noncommutative, in general the star-wedge product is not antisymmetric
\begin{align*}
\omega\wedge_\star\omega'\neq(-1)^{\text{deg}(\omega)\text{deg}(\omega')}\omega'\wedge_\star\omega.
\end{align*}
It is however desirable to have graded cyclicity for integrated forms, i.e.
\begin{equation}
\int\omega\wedge_\star\omega'=(-1)^{\text{deg}(\omega)\text{deg}(\omega')}\int\omega'\wedge_\star\omega + \mbox{total derivative},
\end{equation}
when $\omega\wedge_\star\omega'$ is a top form of course. This is guaranteed provided the twist satisfies \cite{Aschieri:2009ky}
\begin{equation}
\label{eq:twistunimodularity}
S(\bar{f}^\alpha)\bar{f}_\alpha=1,
\end{equation}
where we recall that the antipode $S$ is extended linearly and anti-multiplicatively to $\mathcal{U}(\mathfrak{g})$, $S(XY) = S(Y)S(X)$.\footnote{This mirrors the structure of integration by parts for multiple derivatives.} In fact, this condition tells us that we can remove a single star product under an integral, i.e.
\begin{equation}
\int\omega\wedge_\star\omega' = \int\omega\wedge\omega'+ \mbox{total derivative}.
\end{equation}
We will come back to this condition in section \ref{sec:compatibletwists}.

\paragraph{Conjugation} Any concrete twist that we consider, will be normalized such that with real deformation parameters we have
\begin{equation}
\label{eq:conjugation}
\overline{\omega \wedge_\star \chi} = (-1)^{pq} \overline{\chi} \wedge_\star \overline{\omega},
\end{equation}
under (conventional) complex conjugation. For functions this means $\overline{f\star g} = \overline{g} \star \overline{f}$.

\section{Hodge duality}
\label{sec:Hodgeduality}

In this section we will introduce our twisted notion of Hodge duality, finding that by insisting on left and right star linearity for our Hodge dual, we have to restrict to twists based on the Poincar\'e algebra. Integral cyclicity restricts us further, leaving only two distinct cases at the level of Hodge duality, working in cartesian coordinates. We will show that in these cases the essential features of Hodge duality carry over to our twisted setting. With this setup established, constructing appropriate Yang-Mills theory actions will prove simple.

\subsection{Definition}

To define Hodge duality we need an appropriate generalization of the Levi-Civita symbol. Based on our experience with the Lorentz deformation \cite{Meier:2023kzt}, we define it via
\begin{align}
\label{eq:generalizedLeviCivita}
\form{x^\mu}\wedge_\star\form{x^\nu}\wedge_\star\form{x^\rho}\wedge_\star\form{x^\sigma}=\epsilon^{\mu\nu\rho\sigma}\form{x^0} \wedge_\star \form{x^1} \wedge_\star \form{x^2} \wedge_\star \form{x^3},
\end{align}
where, importantly, we assume that $\epsilon^{\mu\nu\rho\sigma}$ is star commutative, i.e. $\epsilon \star f = f \star \epsilon = \epsilon f$.\footnote{For a given object $a$, $\scom{a}{f} = 0$ for arbitrary $f$, implies $\xi(a)=0$ for any generator $\xi$ appearing in the $r$ matrix, at leading order in the deformation parameter. This condition is not just necessary, but sufficient for all twists we will end up considering, as discussed in section \ref{subsec:unimodularPoincarermatrices}. There we will also see that $\epsilon$ for us is in fact constant. At this stage, however, we could entertain further options, where e.g. the twist $\exp(i \lambda \, (x^0)^2 \partial_2 \wedge (x^1)^2 \partial_3)$ would give a non-constant but star-commutative $\epsilon$.\label{footnote:starcommutativity}} Our generalized Levi-Civita symbol has two basic properties following from its definition. First, it is totally $R$ antisymmetric, for instance
\begin{equation}
\label{eq:epsilonindexrelations}
\epsilon^{\mu\nu\rho\sigma} = - \bar{R}_\lambda{}^\nu{}_\tau{}^\mu \epsilon^{\lambda\tau\rho\sigma}.
\end{equation}
Second, $\epsilon$ conjugates as
\begin{equation}
\overline{\epsilon^{\mu\nu\rho\sigma}} = \epsilon^{\sigma\rho\nu\mu}.
\end{equation}

We now define the Hodge dual of basis star-forms as in \cite{Meier:2023kzt},
\begin{equation}
\label{eq:hodgestarbasisforms}
* \mathrm{d}x^{\mu_1} \wedge_\star \ldots \wedge_\star \mathrm{d}x^{\mu_k} = \tfrac{(-1)^{\sigma(k)}}{(4-k)!} \epsilon_{\mu_{k+1} \ldots \mu_{4}}{}^{\mu_1\ldots \mu_k} \mathrm{d}x^{\mu_4} \wedge_\star \ldots \wedge_\star \mathrm{d}x^{\mu_{k+1}},
\end{equation}
where $\sigma(p)$ denotes the signature of the reversal of $p$ objects, i.e. $\sigma(1)=\sigma(4)=0$, $\sigma(2)=\sigma(3)=1$, and we raise and lower indices with the usual Minkowski metric\footnote{We can formalize this in star product form, by starting with the usual Minkowski metric
\begin{equation*}
\eta = \eta_{\mu\nu} \form{x^\mu} \otimes \form{x^\nu}
\end{equation*}
and defining our choice of deformed metric as
\begin{equation}
\label{eq:metricdef}
\eta_\star \equiv \eta_{\mu\nu} \form{x^\mu} \otimes_\star \form{x^\nu},
\end{equation}
so that expressed in the general form $\eta_\star = \eta^\star_{\mu\nu} \star \form{x^\mu} \otimes_\star \form{x^\nu}$, we can take $\eta^\star_{\mu\nu} = \eta_{\mu\nu}$.}. As we will soon see, cyclically related choices of index ordering are admissible, but the reversed ordering of the indices on the dual form is essential. We would like to, and will, extend this left and right star linearly to arbitrary forms. This turns out to restrict us to twists based on the Poincar\'e algebra.

\subsection{Star linearity and Poincar\'e twists}

If we assume the Hodge star operation is left and right star linear, we can consider e.g. $*\mathrm{d}x^\mu \star f_\mu$, and bracket and permute objects appropriately to find
\begin{equation}
\begin{aligned}
*(\form{x^\mu} \star f_\mu) & = (*\form{x^\mu}) \star f_\mu \\
& = \frac{1}{6} \epsilon_{\rho \sigma \tau}{}^\mu \form{x^\tau} \wedge_\star \form{x^\sigma} \wedge_\star \form{x^\rho} \star f_\mu\\
& = \frac{1}{6} \epsilon_{\rho \sigma \tau}{}^\mu  R_\nu{}^\tau R_\xi {}^\sigma R_\zeta{}^\rho f_\mu \star \form{x^\nu} \wedge_\star \form{x^\xi} \wedge_\star \form{x^\zeta},
\end{aligned}
\end{equation}
as well as
\begin{equation}
\begin{aligned}
*(\form{x^\mu} \star f_\mu) & = *(R_\nu{}^\mu f_\mu \star \form{x^\nu})\\
& = R_\nu{}^\mu f_\mu \star *(\form{x^\nu})\\
& = \frac{1}{6}\epsilon_{\zeta\xi\rho}{}^\nu R_\nu{}^\mu f_\mu \star \form{x^\rho} \wedge_\star \form{x^\xi} \wedge_\star \form{x^\zeta}.
\end{aligned}
\end{equation}
By consistency, since $f$ is arbitrary, we find
\begin{equation}
\epsilon^{\tau\kappa\zeta\phi}\tensor{R}{^\rho_\zeta}\tensor{R}{^\nu_\kappa}\tensor{R}{^\mu_\tau} = \epsilon^{\mu\nu\rho\sigma} \tensor{R}{_\sigma^\phi}.
\end{equation}
We can derive similar relations starting from other degree forms, a two form giving, e.g.
\begin{equation}
\epsilon^{\tau\kappa\chi\phi}\tensor{R}{^\nu_\kappa}\tensor{R}{^\mu_\tau} = \epsilon^{\mu\nu\psi\sigma} \tensor{R}{_\psi^\chi}\tensor{R}{_\sigma^\phi}.
\end{equation}
These two equations are related by the left action of $\tensor{R}{^\rho_\chi}$, since we assume $\epsilon$ is star commutative. Compatibility then requires
\begin{equation}
\label{eq:RraiselowerRtransposeis1}
\tensor{R}{_\sigma^\mu} \tensor{R}{^\sigma_\nu} = \delta^\mu_\nu,
\end{equation}
i.e.
\begin{equation}
\label{eq:Rmatrixinversebyraisinglowering}
R^\mu{}_\nu = \bar{R}_\nu{}^\mu,
\end{equation}
implying that $R$ is a (vector field valued) element of the Poincar\'e group. It is convenient to recast this condition as having an \emph{uncharged metric}
\begin{equation}
f \star \eta_\star = \eta_\star \star f,
\end{equation}
where $\eta_\star = \eta_{\mu\nu} dx^\mu \otimes_\star dx^\nu$ as in equation \eqref{eq:metricdef}, and $f$ is an arbitrary function. In components this directly gives
\begin{equation}
\label{eq:metricinvarianttensor}
\eta_{\lambda \rho} R_\mu{}^{\lambda}R_\nu{}^{\rho} = \eta_{\mu\nu},
\end{equation}
which is equivalent to equation \eqref{eq:Rmatrixinversebyraisinglowering}, and makes the metric tensor an invariant of the $\mathcal{R}$ matrix. Put differently, starting from $f \star \eta_\star = \eta_\star \star f$, recalling that the twist acts via the Lie derivative $\mathcal{L}$ on forms, with $r = r^{ij} \xi_i \wedge \xi_j$, at leading order we immediately get
\begin{equation*}
r^{ij} \xi_i(f) \mathcal{L}_{\xi_j} (\eta) = 0,
\end{equation*}
with $\eta$ the undeformed metric tensor, i.e. $\mathcal{L}_{\xi_i} (\eta) = 0$, meaning the vector fields in the twist must be Killing vector fields for Minkowski space, hence that we are dealing with twists based on the Poincar\'e algebra. We note that this restriction to the Poincar\'e algebra does not directly rely on the specific form we have chosen for our deformed Levi-Civita symbol, and hence would apply for any other (star-commutative) choice of $\epsilon$ used in \eqref{eq:hodgestarbasisforms}.

If we now introduce the volume form
\begin{equation}
\mathrm{d}_\star^4x = \form{x^0} \wedge_\star \form{x^1} \wedge_\star \form{x^2} \wedge_\star \form{x^3},
\end{equation}
we can view the remaining consistency conditions on star-linearity of Hodge duality as having an \emph{uncharged volume form}
\begin{equation}
f \star\mathrm{d}_\star^4x = \mathrm{d}_\star^4x \star f,
\end{equation}
where $f$ is an arbitrary function. Namely, by eqn. \eqref{eq:generalizedLeviCivita}, and the fact that $\epsilon^{\mu\nu\rho\sigma}\epsilon_{\mu\nu\rho\sigma}$ is star commutative, this directly implies
\begin{equation}
\label{eq:epsilonRinvariant}
\epsilon^{\tau\kappa\zeta\phi}\tensor{R}{_\tau^\mu}\tensor{R}{_\kappa^\nu}\tensor{R}{_\zeta^\rho}\tensor{R}{_\phi^\sigma} = \epsilon^{\mu\nu\rho\sigma}.
\end{equation}
which is the consistency condition for star-linearity of Hodge duality for four forms. Together with property \eqref{eq:RraiselowerRtransposeis1}, suitable left multiplication by $R$s gives the required relations for other degree forms. Eqn. \eqref{eq:epsilonRinvariant} also states that our generalized Levi-Civita tensor is an invariant of the $R$ matrix.\footnote{Our generalized Levi-Civita symbol either depends on the deformation parameter -- like $R$ -- or is undeformed, so there is no contradiction with the known invariant tensors of the Lorentz group.} With this perspective it is clear that all consistency conditions are automatically satisfied for any twist built on the Poincar\'e algebra, since for such twists $\mathrm{d}_\star^4x$ is a star-commutative (constant) multiple of the standard Poincar\'e invariant volume form $\mathrm{d}^4x$.

We will now check that our twisted notion of Hodge duality has the usual features expected of Hodge duality, by evaluating it for concrete twists.

\subsection{Concrete generalized Levi-Civita symbols}

Many twists based on the Poincar\'e algebra actually lead to an undeformed Levi-Civita symbol. At the level of the $r$ matrix, only terms built on the Lorentz algebra, i.e. of the form $M \wedge M$, contribute nontrivially to star-wedge products between general basis forms, so that only cases with quadratic noncommutativity have a deformed Levi-Civita symbol. Since Lorentz generators act as linear operators on basis forms, these deformed Levi-Civita symbols are constant. Taking into account integral cyclicity, as discussed in detail in the next section, there are effectively only two cases to consider. The first of these is the Lorentz twist \eqref{eq:Lorentztwist}
\begin{align}
\mathcal{F}=\exp\left(\frac{i\lambda}{2}M_{01}\wedge M_{23}\right),
\end{align}
with an associated $\epsilon$ symbol that is graded cyclic, and has 32 nonzero components given by \cite{Meier:2023kzt}
\begin{equation}
\begin{aligned}
\epsilon^{0123} & = -\epsilon^{0132}=\epsilon^{0231}=-\epsilon^{0321} =1,\\
\epsilon^{1212} & = -\epsilon^{0202}=\epsilon^{1313}= -\epsilon^{0303}=i \sinh{\lambda},\\
\epsilon^{0312} & = -\epsilon^{0213}=\cosh{\lambda},
\end{aligned}
\end{equation}
plus others related by graded cyclicity.

The second is the light-cone version of the Lorentz twist
\begin{align}
\mathcal{F}=\exp\left(\frac{i\alpha}{8}M_{+1}\wedge M_{+2}\right),
\end{align}
with an associated $\epsilon$ symbol that is similarly graded cyclic, and, in light cone coordinates, has 32 nonzero components, 24 of which are the usual components of the undeformed $\epsilon$ symbol. The remaining nonzero components are
\begin{equation}
\label{eq:lightconelorentzepsilon}
\epsilon^{+++-}  = - 2 i \alpha ,\quad  \epsilon^{+1+1}=\epsilon^{+2+2}= i \alpha,
\end{equation}
plus others related by graded cyclicity.

In both cases the volume form, in our conventions, is undeformed
\begin{equation}
\mathrm{d}^4_\star x =\mathrm{d}^4 x.
\end{equation}

\subsection{Contractions and projectors}

A deformed Hodge star operation similar to ours was discussed in \cite{Meyer:1994wi} for $q$-Minkowski space, and generalized to other braided geometries in \cite{Majid:1994mh}. While we found our Hodge star independently starting from the natural relation \eqref{eq:generalizedLeviCivita}, are working in a twisted setting, and our concrete generalized Levi-Civita symbols and normalization factors differ from the one for $q$-Minkowski space, most of the general structure discussed in \cite{Meyer:1994wi,Majid:1994mh} applies in our setting as well. In particular, similarly to how contractions of the undeformed Levi-Civita symbol give projectors onto totally antisymmetric tensors, contractions of our generalized Levi-Civita symbol give projectors onto totally $R$-antisymmetric tensors. Concretely, we have
\begin{equation}
\label{eq:epsiloncontractions}
\begin{aligned}
\epsilon^{\rho_1\dots\rho_k\mu_1\dots\mu_{4-k}}\epsilon_{\nu_1\dots\nu_{4-k}\rho_k\dots\rho_1}&=-k!(4-k)!\tensor{P}{^{\mu_1\dots\mu_{4-k}}_{\nu_{4-k}\dots\nu_1}},
\end{aligned}
\end{equation}
where $\tensor{P}{^{\mu_1\dots\mu_{4-k}}_{\nu_1\dots\nu_{4-k}}}$ is the projector onto $R$-antisymmetric $4-k$ forms, with $\tensor{P}{^{\mu\ldots \nu}_{\kappa\ldots \zeta}}\tensor{P}{^{\kappa\ldots \zeta}_{\rho \ldots \sigma}}=\tensor{P}{^{\mu \ldots \nu}_{\rho \ldots \sigma}}$. Following \cite{Majid:1994mh}, these projectors can be nicely expressed via $R$ matrices as
\begin{equation}
\begin{aligned}
\tensor{P}{}= & 1 \\
\tensor{P}{^{\mu}_\nu} =& \delta^\mu_\nu P\\
\tensor{P}{^{\mu_1\mu_2}_{\nu_1\nu_2}} = &\frac{1}{2}\delta^{\tau_1}_{\nu_1}\tensor{P}{^{\tau_2}_{\nu_2}}\left(\delta^{\mu_1}_{\tau_1} \delta^{\mu_2}_{\tau_2} - \tensor{R}{_{\tau_2}^{\mu_1}_{\tau_1}^{\mu_2}}\right) \\
\tensor{P}{^{\mu_1\mu_2\mu_3}_{\nu_1\nu_2\nu_3}} = &\frac{1}{3}\delta^{\tau_1}_{\nu_1} \tensor{P}{^{\tau_2\tau_3}_{\mu_2\mu_3}} \left(\delta^{\mu_1}_{\tau_1} \delta^{\mu_2}_{\tau_2} \delta^{\mu_3}_{\tau_3} -\tensor{R}{_{\tau_2}^{\mu_1}_{\tau_1}^{\mu_2}} \delta^{\mu_3}_{\tau_3} + \tensor{R}{_{\tau_2}^{\mu_1}_{\tau_1}^{\sigma}} \tensor{R}{_{\tau_3}^{\mu_2}_{\sigma}^{\mu_3}} \right)\\
\tensor{P}{^{\mu_1\mu_2\mu_3\mu_4}_{\nu_1\nu_2\nu_3\nu_4}} = &\frac{1}{4} \delta^{\tau_1}_{\nu_1} \tensor{P}{^{\tau_2\tau_3\tau_4}_{\nu_2\nu_3\nu_4}} \left( \delta^{\mu_1}_{\tau_1}\delta^{\mu_2}_{\tau_2}\delta^{\mu_3}_{\tau_3}\delta^{\mu_4}_{\tau_4} -\tensor{R}{_{\tau_2}^{\mu_1}_{\tau_1}^{\mu_2}}\delta^{\mu_3}_{\tau_3}\delta^{\mu_4}_{\tau_4}\right.\\
& \quad + \left.\tensor{R}{_{\tau_2}^{\mu_1}_{\tau_1}^{\sigma}}\tensor{R}{_{\tau_3}^{\mu_2}_{\sigma}^{\mu_3}}\delta^{\mu_4}_{\tau_4} -\tensor{R}{_{\tau_2}^{\mu_1}_{\tau_1}^{\sigma_2}} \tensor{R}{_{\tau_3}^{\mu_2}_{\sigma_2}^{\sigma_1}} \tensor{R}{_{\tau_4}^{\mu_3}_{\sigma_1}^{\mu_4}}\right).
\end{aligned}
\end{equation}
This also allows us to express our generalized Levi-Civita symbols via $R$ matrices, as
\begin{equation}
\begin{aligned}
\epsilon^{\mu\nu\rho\sigma}&=4!\tensor{P}{^{\mu\nu\rho\sigma}_{3210}}\\
\epsilon_{\mu\nu\rho\sigma}&=-4!\tensor{P}{^{0123}_{\sigma\rho\nu\mu}}.
\end{aligned}
\end{equation}

\subsection{Properties of the twisted Hodge star}

Our twisted Hodge duality has a number of properties that will be essential in the following sections. First of all, it is important to note that, while we restricted ourselves to the Poincar\'e algebra as a consistency requirement on star linearity of our Hodge star, so far our definition of the Hodge star is not constructively star linear. In principle there could be undesirable tension with the conventional differential calculus underpinning our twisted calculus. Fortunately, for the standard and both nontrivial cases, the Hodge star commutes with Lie derivatives along Poincar\'e vector fields when evaluated on basis star forms, as can be verified by explicit computation. This means that the Hodge star commutes with the associated star product on basis star forms, and can hence be naturally extended left and right star linearly, in a way that is compatible with regular linearity. By this we mean that star linearity of the Hodge star is equivalent to regular linearity of tensor contractions in conventional differential calculus, i.e. that we can evaluate the Hodge dual of, say, $f_\mu \star \mathrm{d}x^\mu$ as $f_\mu \star *\mathrm{d}x^\mu$ (star linear) or as $F_\nu{}^\mu(f) *\mathrm{d}x^\nu$ (linear), making the two manifestly compatible.
By regular linearity of our twisted Hodge star, and the conventional product rule for Lie derivatives, Poincar\'e Lie derivatives then commute with our Hodge star in general\footnote{For basis star forms we have
\begin{equation}
[\mathcal{L}_\xi,*]\mathrm{d}x = 0,
\end{equation}
for $\xi \in \mathcal{P}$, where $\mathrm{d}x$ schematically denotes an arbitrary basis star $p$ form. This means that, again schematically,
\begin{equation}
[\mathcal{L}_\xi,*] f^\star \star \mathrm{d}x  = [\mathcal{L}_\xi,*] F(f^\star)\mathrm{d}x  = 0
\end{equation}
where the second equality follows by the regular linearity of our twisted Hodge star, and the conventional product rule for Lie derivatives. Here $F$ denotes the twist with the required indices to represent the star product in the sense of eqs. \eqref{eq:Fwithindicesdef}.}
\begin{equation}
\label{eq:hodgecommuteswithPoincare}
[*,\mathcal{L}_\xi] = 0, \quad \xi \in \mathcal{P},
\end{equation}
in other words, our Hodge star is Poincar\'e equivariant. This is also essential for twisted symmetry, discussed in section \ref{sec:twistedsymmetry}.  Algebraically, the compatibility of (left and right) regular and star linearity, amount to
\begin{equation}
[1 \otimes *, \bar{\mathcal{F}}]=[* \otimes 1, \bar{\mathcal{F}}]=0,
\end{equation}
which clearly hold given eqn. \eqref{eq:hodgecommuteswithPoincare}.

With linearity settled, we can move on to the other properties of our Hodge star. The first two of these follow from the general definition of $\epsilon$. Namely, our Hodge star maps star forms to star forms, preserving their symmetry properties thanks to the total $R$-antisymmetry of $\epsilon$, and it preserves reality of star forms by the conjugation properties of $\epsilon$. Next, the contraction identities \eqref{eq:epsiloncontractions} show that for equal-degree $p$ forms $\omega$ and $\chi$
\begin{equation}
\label{eq:equalformshodgeviavolumeform1}
\omega \wedge_\star *\chi = (-1)^{\sigma(p)+1} p! \omega^\star_{\mu \ldots \nu} \star R_{\kappa}{}^\mu \ldots R_{\rho}{}^\nu \chi^{\star \rho \ldots \kappa} \,\form{{}^4x},
\end{equation}
which is nothing but the defining property of the conventional Hodge star, dressed by appropriate star products and $R$ matrices. We also have
\begin{equation}
\label{eq:equalformshodgeviavolumeform2}
* \omega \wedge_\star \chi  = (-1)^{p+\sigma(p)+1} p! \omega^\star_{\mu \ldots \nu} \star \bar{R}^\mu{}_{\kappa} \ldots \bar{R}^\nu{}_{\rho} \chi^{\star \rho \ldots \kappa} \,\form{{}^4x},
\end{equation}
so that by eqn. \eqref{eq:Rmatrixinversebyraisinglowering},
\begin{equation}
\label{eq:Hodgestarflip}
\omega \wedge_\star *\chi =  (-1)^{p} \, * \omega \wedge_\star \chi,
\end{equation}
and our Hodge star gives the usual symmetric inner product
\begin{equation}
\langle \omega, \chi \rangle \equiv \int \omega \wedge_\star *\chi = \int \chi \wedge_\star * \omega = \langle \chi , \omega \rangle ,
\end{equation}
by the graded cyclicity of our star product under the integral.

Finally, the contraction identities \eqref{eq:epsiloncontractions} show that applying the Hodge star twice is the same as applying a projection operator, up to a possible sign. Applying a projection operator to an already $R$-antisymmetric form does nothing, and as a result we have
\begin{equation}
\label{eq:hodgestarsquared}
** \omega = -(-1)^p \omega,
\end{equation}
for a (star) $p$ form $\omega$, meaning that our Hodge star provides an actual duality.

The Lorentz twist and its light-cone cousin are the only twists that both fit our constraints, and result in a nonstandard $\epsilon$ in Cartesian coordinates.\footnote{For abelian twists we could locally work in coordinates that trivialize the action of the twist on basis forms, to get an undeformed Hodge star. Such a choice of coordinates is not always desirable however, and, more importantly, is impossible for nonabelian twists.} As we will discuss in the next section, all other twists we will consider, either have a standard, undeformed Levi-Civita symbol, or extend the Lorentz twist or its light-cone cousin by terms that do not further deform $\epsilon$. As a result, our twisted Hodge star has the above properties in all cases, keeping in mind that $R_\mu{}^\nu$ is affected by $M\wedge p$ type terms if they are present in a given twist, which should be taken into account in eqs. \eqref{eq:equalformshodgeviavolumeform1} and \eqref{eq:equalformshodgeviavolumeform2}.

To finish our discussion of Hodge duality, let us note that while we introduced our choice of generalized Levi-Civita symbol by an intuitive relation, the projectors above concretely demonstrate that there is only one totally $R$-antisymmetric symbol in each of our cases, up to an overall function. This function is fixed by the defining property of the Hodge star, making our choice essentially unique. Moreover, while we arrived at our Hodge star and its Poincar\'e equivariance by direct construction, the fact that the undeformed Hodge star is a Poincar\'e-equivariant map, means that exactly for Poincar\'e twists it should admit a formal quantization that continues to be equivariant (and hence star linear), see e.g. part III of \cite{Schenkel:2011biz}.\footnote{We thank Richard Szabo for discussions on this point.} Both this formal argument, as well as the uniqueness of the undeformed Hodge star in its Poincar\'e equivariance combined with our ability to express our star basis forms invertibly via regular basis forms, suggest that our deformed Hodge star can be expressed via the undeformed one. Indeed, computationally our Hodge star is equivalent to expressing star basis forms in terms of regular basis forms, applying the regular Hodge star, and then mapping back to star forms.

Interestingly, our discussion also shows that for twists beyond the Poincar\'e algebra but otherwise fitting our assumptions, attempts to find a Hodge star that retains some notion of star linearity, in general will require further structure, or suitably relaxing the properties of the Hodge star. For example, any twist leading to a trivial star-wedge product necessarily has an undeformed Levi-Civita symbol, if we demand the usual properties for the Hodge star. If we then assume this twist to be abelian, left star linearity is equivalent to right star linearity by flipping the sign of the deformation parameter. By our earlier analysis, it then follows that one can have neither left or right star linearity beyond the Poincar\'e algebra, for this type of twist.\footnote{Two nontrivial examples of such twists are the ones associated to $r = x^i \partial_i \wedge \partial_\mu$ for any choice of $\mu$, or $r = x^1 \partial_2 \wedge p_3$.}

Our deformed Hodge star can be used to define further structure such as a codifferential and a deformed Laplacian. The latter is directly relevant for us, and we will come back to it below.

\section{Compatible twists}
\label{sec:compatibletwists}

Above we indicated how star linearity of our Hodge star operation restricts us to twists based on the Poincar\'e algebra. In this section we will discuss all such twists, for which we also have graded cyclic integration of forms.

\subsection{Integral cyclicity and unimodularity}

In section \ref{subsec:twisteddifferentialcalculus} we noted that if we impose the constraint \eqref{eq:twistunimodularity}, i.e.
\begin{equation}
\label{eq:Sff=1local}
S(\bar{f}^\alpha)\bar{f}_\alpha=1,
\end{equation}
our star product will be graded cyclic upon integration. Interestingly, this condition implies that the corresponding $r$ matrix is unimodular. To see this, first we note that any twist satisfying eqn. \eqref{eq:Sff=1local} is automatically $r$ symmetric, where we recall that this means that its leading order expansion starts with the $r$ matrix, $\mathcal{F}_{12} = -\tfrac{r_{12}}{2}$. This follows because any twist can be put in $r$-symmetric form by the transformation $\mathcal{F} \rightarrow v^{-1} \otimes v^{-1} \mathcal{F} \Delta(v)$ with $v = (S(\bar{f}^\alpha)\bar{f}_\alpha)^{-1/2}$ \cite{Tolstoy:2008zz}, i.e. by doing nothing when eqn. \eqref{eq:Sff=1local} is satisfied.\footnote{For completeness, there are $r$-symmetric twists which do not satisfy eqn. \eqref{eq:Sff=1local}, in particular jordanian ones, as discussed in e.g. section 5 of \cite{Dimitrijevic:2014dxa}.} Then, starting from a twist $\bar{\mathcal{F}} = 1\otimes 1 + \lambda/2\, r + \ldots$, with $r= r^{ij} \xi_i \wedge \xi_j$, we find
\begin{equation}
\label{eq:unimodularity}
S(\bar{f}^\alpha)\bar{f}_\alpha = 1 - \tfrac{\lambda}{2} r^{ij} [\xi_i, \xi_j] + \mathcal{O}(\lambda^2),
\end{equation}
so that condition \eqref{eq:Sff=1local} implies
\begin{equation*}
r^{ij} [\xi_i, \xi_j] = 0.
\end{equation*}
This means that $r$ is unimodular, i.e. the quasi-Frobenius algebra corresponding to $r$ is a unimodular Lie algebra.\footnote{Constant solutions of the classical Yang-Baxter equation over $\mathfrak{g}$ correspond one-to-one to quasi-Frobenius subalgebras $\mathfrak{f} \subset \mathfrak{g}$ \cite{STOLIN1999285}. This subalgebra $\mathfrak{f}$ corresponds to the support of the $r$ matrix, or, equivalently, given a nondegenerate bilinear form $\langle\,,\rangle$ on $\mathfrak{g}$ it corresponds to the image of $\langle r, \cdot \rangle_2$. Unimodularity now means $\Tr{\mbox{ad}_\xi} = 0, \forall \xi \in \mathfrak{f}$. See e.g. \cite{Borsato:2016ose} for details, and \cite{Hoare:2021dix} for a pedagogical discussion, not including unimodularity however.}

While the constraint \eqref{eq:Sff=1local} was presented as a sufficient condition in \cite{Aschieri:2009ky}, the unimodularity that it implies at leading order, is a necessary condition. Namely, demanding graded cyclicity of $\int \omega \wedge_\star \chi$ for general $\omega$ and $\chi$, at leading order requires
\begin{equation}
\int r^{ij} \mathcal{L}_{\xi_i}(\omega) \wedge \mathcal{L}_{\xi_j}(\chi) =0,
\end{equation}
up to a total derivative. Writing this as
\begin{equation}
\int r^{ij} \mathcal{L}_{\xi_i}(\omega \wedge \mathcal{L}_{\xi_j}(\chi)) - r^{ij} \omega \wedge  \mathcal{L}_{\xi_i} \mathcal{L}_{\xi_j}\chi = 0,
\end{equation}
by Cartan's formula the first term is a total derivative. The second term vanishes (including up to a total derivative) if and only if
\begin{equation}
r^{ij} \mathcal{L}_{\xi_i} \mathcal{L}_{\xi_j} = \tfrac{1}{2} r^{ij} [\mathcal{L}_{\xi_i}, \mathcal{L}_{\xi_j}] = 0,
\end{equation}
i.e. if and only if $r$ is unimodular.

We are not aware of a result for the converse direction, i.e. whether every unimodular $r$ matrix admits a twist which satisfies eqn. \eqref{eq:Sff=1local}. We will simply consider this question in the restricted setting of the Poincar\'e algebra, where we can show that this is the case.

\subsection{Overview of twists}

\label{subsec:unimodularPoincarermatrices}

We find ourselves considering twists based on the Poincar\'e algebra, with unimodular $r$ matrix. Any abelian $r$ matrix is automatically unimodular, but for nonabelian $r$ matrices this is a nontrivial constraint that for example excludes basic jordanian $r$ matrices such as $r = M_{+-} \wedge p_+$. Fortunately, homogeneous $r$ matrices for the Poincar\'e algebra are classified in \cite{zakrzewski:1997}, and we can readily check which of them are unimodular. We present all unimodular $r$ matrices for the Poincar\'e algebra in Table \ref{tab:rmatrices}, up to automorphisms.

\begin{table}[h]
{\small
\begin{center}
\begin{tabular}{c|c|c|c}
r matrix &
$\tilde{r}$ & N in \cite{zakrzewski:1997}\\
\hline
\hline
$r_1$ & $p_1 \wedge p_+$ &  19\\
$r_2$ & $p_1 \wedge p_2$ & 20\\
$r_3$ & $p_+ \wedge p_- + \alpha p_1 \wedge p_2$ & 21\\
\hline
$r_4$&$ M_{03} \wedge M_{12} + \left(\alpha p_+ \wedge p_- +\tilde{\alpha} p_1 \wedge p_2\right)$ & 1\\
$r_5$& $ p_+ \wedge M_{12} + M_{+2} \wedge M_{+1}$ & 2 with $\beta_1=0$\\
$r_6$&$M_{+1} \wedge M_{+2} + p_+ \wedge p_1$  & 3 with $\beta=0$\\
$r_7$&$M_{+1} \wedge M_{+2} + p_+ \wedge \left(p_1 + \alpha p_2\right)$ & 4 with $\beta=0$\\
$r_8$&$p_2 \wedge \left(M_{+1}+ p_-\right) + \alpha p_+ \wedge p_1$ & 11\\
$r_9$&$p_0 \wedge \left(M_{12} + \alpha_1 p_-\right) + \alpha_2 p_1 \wedge p_2$ & 13\\
$r_{10}$&$p_3\wedge \left(M_{12} + \alpha_1 p_-\right) + \alpha_2 p_1 \wedge p_2$ & 14\\
$r_{11}$&$p_+ \wedge \left(M_{12} +  p_-\right) + \alpha p_1 \wedge p_2$ & 15\\
$r_{12}$& $p_1 \wedge \left(M_{+-} + \alpha_2 p_2 \right) + \alpha_1 p_+ \wedge p_-$ & 16\\
$r_{13}$& $ M_{+1}\wedge p_+ + (2 \alpha_1 p_1 + \alpha_2 p_+) \wedge p_2 + (M_{+2} + \alpha_1 (p_- + 2 p_2)) \wedge p_1 $& 10\\
\hline
$r_{14}$& $p_+ \wedge M_{+1} + p_- \wedge \left( \alpha p_+ + \alpha_1 p_1\right) +\tilde{\alpha} p_+ \wedge p_2$ & 12\tablefootnote{This case, as presented in \cite{zakrzewski:1997}, contains a $p_- \wedge p_2$ term that breaks the classical Yang-Baxter equation unless it is set to zero, as also noted in \cite{Tolstoy:2008zz}.}
\end{tabular}
\end{center}
}
\caption{Homogeneous unimodular r matrices for the Poincar\'e algebra, based on the classification of \cite{zakrzewski:1997}.  The first three $r$ matrices are abelian and correspond to the Groenewold-Moyal deformation(s). The next ten $r$ matrices have (can be given) an almost-abelian structure, as discussed in the main text. The last $r$ matrix is not of almost-abelian type. For reference, the right-most column refers to the classification in Table 1 of \cite{zakrzewski:1997}, up to some inconsequential relative numerical factors. The $\alpha$s denote free parameters, leading to multi-parameter deformations. We recall $x^\pm = x^0 \pm x^3$.}
\label{tab:rmatrices}
\end{table}

In our construction we need a twist, however, not just an $r$ matrix. Fortunately, as we will now discuss, we know how to construct a suitable twist for each of the above cases.

In the discussion below we refer to the rank and the symmetry algebra of an $r$ matrix. The rank of an $r$ matrix is the dimension of its associated quasi-Frobenius algebra. This is also twice the minimal number of wedge terms used to express a given $r$ matrix. The symmetry algebra of an $r$ matrix is the algebra of generators $\xi\in \mathcal{P}$ for which
\begin{equation*}
\Delta(\mbox{ad}_\xi)(r) = 0.
\end{equation*}

\paragraph{Abelian twists} Cases $r_{1-3}$ are abelian $r$ matrices, and thereby automatically unimodular. They correspond to abelian twists, with $r$-symmetric $\mathcal{F} = e^{-i\lambda /2\, r}$. Twists of this form satisfy  \eqref{eq:Sff=1local} to all orders. This class includes the Groenewold-Moyal and Lorentz deformations, as well as e.g. the $\lambda$ and $\rho$-Minkowski ones mentioned in section \eqref{subsec:twistexamples}. In our presentation, the latter three appear as special cases of our next class of twists.

\paragraph{Almost-abelian twists} With the exception of $r_{14}$, the nonabelian $r$ matrices in Table \ref{tab:rmatrices} are almost abelian, by which we mean that they are ordered sums of abelian terms, with each new term constructed out of symmetries of the previous terms. I.e. an almost abelian $r$ matrix is of the form $r = \hat{r} + \ldots + \check{r} + \tilde{r}$ with each term abelian and constructed out of symmetries of the terms to its left, e.g. $\tilde{r}$ is constructed out of symmetries of $\hat{r} + \ldots + \check{r}$. The associated almost-abelian twists \cite{Tolstoy:2008zz,vanTongeren:2016eeb}, are products of abelian twists of the form $\mathcal{F} = \tilde{\mathcal{F}}\check{\mathcal{F}} \ldots \hat{\mathcal{F}}$, where the terms do not commute. The cocycle condition is satisfied nontrivially thanks to the ordered symmetry structure, also known as a subordinate structure \cite{Tolstoy:2008zz}.  As each abelian term satisfies eqn. \eqref{eq:twistunimodularity}, any twist of this product form satisfies eqn. \eqref{eq:Sff=1local} thanks to the anti-multiplicative nature of the antipode. In Table \ref{tab:rmatrices} we indicate the almost abelian structure of our $r$ matrices by the left-to-right ordering of the sum. The subordinate structure of the homogeneous $r$ matrices of \cite{zakrzewski:1997}, including jordanian (nonunimodular) cases, is discussed in detail in \cite{Tolstoy:2008zz}.

\paragraph{A non-factorized rank four twist} We have have covered all $r$ matrices in Table \ref{tab:rmatrices} except $r_{14}$. This case is actually closely related to $r_{13}$, which itself is a special case, because it can be factorized in an almost abelian form only once we effectively extend its support, as indicated in Table \ref{tab:rmatrices}. Put differently, $r_{13}$ is rank four, but its almost abelian factorization uses a five-dimensional algebra. Still, as discussed in Appendix \ref{app:newtwist}, we can rewrite the almost abelian twist for $r_{13}$ in a form using only the generators appearing in its support. This gives a twist of a type we have not encountered in the literature, which we believe to be the first example of a fundamentally rank four twist, i.e. one defined purely in terms of the support of the $r$ matrix, while not admitting a factorization in rank two terms. Importantly, this also allows us to construct a twist for $r_{14}$, because its fundamental algebraic structure is the same as that of $r_{13}$, as we discuss in Appendix \ref{app:newtwist}.\footnote{This covers the one missing case in \cite{Tolstoy:2008zz}, completing the explicit construction of twists for the Poincar\'e algebra.} In contrast to $r_{13}$, $r_{14}$ does not admit an almost abelian factorization, however, due to its different embedding in the Poincar\'e algebra. These twists satisfy eqn. \eqref{eq:Sff=1local}, since they are algebraically equivalent to the the almost-abelian form of the twist for $r_{13}$, which satisfies it.

\paragraph{Star commutativity} In section \ref{sec:Hodgeduality}, in particular footnote \ref{footnote:starcommutativity}, we indicated that star commutativity of an object $a$ -- $\scom{a}{f}=0$ for all $f$ --  requires it to be uncharged under the generators in the support of the $r$ matrix associated to the twist under consideration. As we now have an explicit twist for each $r$ matrix, constructed out of the generators in its support, we see that being uncharged with respect to the support of an $r$ matrix is also a sufficient condition for star commutativity for our twists.

\subsection{Deformed Levi-Civita symbols, Hodge star, and simplifications}
\label{subsec:Laplaciandiscussion}

Only the twists for $r_{4-7}$ result in nontrivially deformed Levi-Civita symbols. Of these, $r_4$ gives an extended version of the Lorentz deformation, while $r_{5-7}$ involve its light-cone cousin. In each case the extra terms do not affect the generalized Levi-Civita symbol, meaning all cases in Table \ref{tab:rmatrices} have either an undeformed Levi-Civita symbol, or the one for the Lorentz deformation or its light-cone cousin, so that our discussion of Hodge duality in the previous section, covers all deformations based on Poincar\'e twists with unimodular $r$ matrices.\footnote{Dropping the unimodularity constraint would allow for a further case with a nontrivial generalized Levi-Civita symbol, associated to the jordanian $r = M_{+-} \wedge M_{+1}$.}

With a concrete list of twists, we can also discuss the explicit form of the deformed Laplacian, which we will encounter below. We define it in the usual way, as
\begin{equation}
\Delta f \equiv * \mathrm{d} * \mathrm{d} f.
\end{equation}
By the properties of our Hodge star and definition of $\partial^\star_\mu$, in components this gives
\begin{equation}
\Delta f =  \partial^{\star\mu} \partial^\star_\mu f.
\end{equation}
However, it turns out that for our class of deformations, the Laplacian is not actually deformed, i.e.
\begin{equation}
\Delta f = \partial^\mu \partial_\mu f.
\end{equation}
The reason for this is intimately tied to the commutativity of the Hodge star and the star product. Namely, given the latter, we have
\begin{equation}
\begin{aligned}
\mathrm{d}x^\mu \wedge_\star * \mathrm{d}x^\nu = & \mathrm{d}x^\alpha \wedge \bar{F}_\alpha{}^\mu(* \mathrm{d}x^\nu)\\
= & \left(\mathrm{d}x^\alpha \wedge *\mathrm{d}x^\beta\right) \bar{F}_\alpha{}^\mu{}_\beta{}^\nu.
\end{aligned}
\end{equation}
Upon multiplying by $F$ and using property \eqref{eq:equalformshodgeviavolumeform1} of the twisted Hodge star, gives
\begin{equation}
\begin{aligned}
\mathrm{d}x^\mu \wedge * \mathrm{d}x^\nu = &\left(\mathrm{d}x^\alpha \wedge_\star *\mathrm{d}x^\beta\right) F_\alpha{}^\mu{}_\beta{}^\nu
 = &  F_\alpha{}^\mu{}^\alpha{}^\nu \mathrm{d}^4 x.
 \end{aligned}
\end{equation}
Now, for abelian twists with $\bar{F} = R |_{\lambda\rightarrow\lambda/2}$, by the triviality of the $R$ matrix for symmetric products, which results in eqn. \eqref{eq:R4indextrivialonsymmetricproduct}, this gives
\begin{equation}
\label{eq:dxhodgedx_with_without_star_product}
\mathrm{d}x^\mu \wedge_\star * \mathrm{d}x^\nu = \mathrm{d}x^\mu \wedge * \mathrm{d}x^\nu = \eta^{\mu\nu} \mathrm{d}^4 x.
\end{equation}
The same applies for the nonabelian twists in Table \ref{tab:rmatrices}, since only $M\wedge M$ terms contribute nontrivially to the twisted wedge product of basis forms, and these terms are all abelian.\footnote{Similarly we can see that $\eta_\star = \eta$ in our setting.} This allows us to evaluate the Laplacian, using regular linearity of the Hodge star, as
\begin{equation}
\label{eq:laplacian}
\begin{aligned}
\Delta f = &*\mathrm{d}(*\partial_\mu f \mathrm{d}x^\mu) \\
= & *\mathrm{d}(\partial_\mu f)\wedge * \mathrm{d}x^\mu\\
= & * (\partial_\nu \partial_\mu f (\mathrm{d}x^\nu \wedge *\mathrm{d}x^\mu))\\
= & \partial_\nu\partial_\mu f (\eta^{\nu\mu} *\mathrm{d}^4x)\\
= & \partial^\mu \partial_\mu f.
\end{aligned}
\end{equation}
Eqn. \eqref{eq:dxhodgedx_with_without_star_product} shows the close relationship between graded cyclicity of the generalized Levi-Civita symbol, and having an undeformed Laplacian. Since contractions with $\epsilon$ are invertible on star forms, eqn. \eqref{eq:dxhodgedx_with_without_star_product} implies graded cyclicity of $\epsilon$. Of course, alternatively we could directly evaluate $\partial_\mu^\star \partial^{\star\mu}=\tensor{\left(F_{op}\right)}{_\mu^\rho}\partial_\rho \tensor{\left(F_{op}\right)}{^\mu_\sigma}\partial^\sigma$, which reduces to the undeformed expression using $\tensor{\left(F_{op}\right)}{^\mu_\sigma}=\tensor{\left(\bar{F}_{op}\right)}{_\sigma^\mu}$ and $\left[\tensor{\left(F_{op}\right)}{_\mu^\rho},\partial_\rho\right]=0$, which hold for our Poincar\'e twists.

\section{Noncommutative Yang Mills theory}

Now that we have systematically set up our differential calculus, and found a suitable Hodge star, we can straightforwardly write down actions for noncommutative Yang-Mills theory on the noncommutative versions of Minkowski space described by the twists discussed above. In fact, with our machinery in place, we can simply mimic the well-known construction for the Groenewold-Moyal deformation, as reviewed in \cite{Szabo:2001kg}.

\paragraph{Gauge transformations}
Local gauge transformations are functions on spacetime, and hence affected by the twist, making it natural to consider star-gauge transformations. This means that e.g. a fundamental field $\Phi$ should transform as
\begin{align*}
\delta_\varepsilon\Phi(x)&=i\varepsilon(x)\star\Phi(x),
\end{align*}
under a gauge transformation with parameter $\varepsilon \in \mathfrak{h}$, where $\mathfrak{h}$ is the Lie algebra of the gauge group $H$. One immediate concern is whether such transformations close in the Lie algebra, which is in fact not automatically the case. If we consider
\begin{equation}
\label{eq:gaugeClose}
\begin{aligned}
\com{\delta_{\varepsilon_1}}{\delta_{\varepsilon_2}}\Phi&=\delta_{\scom{\varepsilon_1}{\varepsilon_2}}\Phi\\
\scom{\varepsilon_1}{\varepsilon_2}&=\left(\varepsilon_1^a\star\varepsilon_2^b-\varepsilon_2^a\star\varepsilon_1^b\right)T^aT^b,
\end{aligned}
\end{equation}
where $T^a$ denote the generators of the gauge algebra, we see that in general, the product of two generators is not an element of the gauge algebra, since the right hand side has contributions symmetric in $a$ and $b$. Since our star product is suitably hermitian, however, as in eqs. \eqref{eq:conjugation}, we can consistently demand anti-hermiticity, since
\begin{equation}
\scom{\varepsilon_1}{\varepsilon_2}^\dagger = - \scom{\varepsilon_1}{\varepsilon_2}
\end{equation}
In other words, we can directly consider gauge transformations in the fundamental representation of $\mathfrak{u}(N)$, but not other algebras.\footnote{Of course $\mathfrak{gl}(n)$ is always an option where closure is concerned, but this is undesirable for other reasons.} Other gauge algebras can be considered by letting gauge transformations live in the universal enveloping algebra, as discussed in \cite{Jurco:2001rq}. This guarantees that the right-hand side in \eqref{eq:gaugeClose} is in the same algebra, so that the gauge transformations close. However, the universal enveloping algebra formally has infinitely many degrees of freedom. Still, these additional degrees can be reduced to the finitely many degrees of freedom of the undeformed gauge algebra, at least perturbatively in the deformation parameter, by making use of the Seiberg-Witten map \cite{Seiberg:1999vs}. Given our intended applications in AdS/CFT, we are primarily interested in $\mathfrak{u}(N)$.\footnote{There has also been recent work on an alternative, braided approach to noncommutative gauge theory \cite{Ciric:2021rhi,Giotopoulos:2021ieg}, where closure is automatic for any gauge group. This realizes gauge symmetry differently from the present star-gauge form, however, and its relation to string theory, and hence AdS/CFT, is currently unclear.}

\paragraph{Covariant derivative} To define a covariant derivative, we normally consider the partial derivative of a given field and its behavior under gauge transformations. In our noncommutative setting this gives
\begin{equation}
\begin{aligned}
\partial_\mu\left(\delta_\varepsilon\Phi\right)&=i\mu_\mathcal{F}(\Delta_\mathcal{F}(\partial_\mu)(\varepsilon,\Phi))\\
&=\left(\bar{F}_{op}\right)_\mu{}^\sigma\left[\left((F_{op})_\nu{}^\lambda \partial_\lambda \varepsilon \right) \star \left(R_\sigma{}^\nu \Phi \right) + \varepsilon \star \left((F_{op})_\mu{}^\nu \partial_\nu \Phi\right)\right].
\end{aligned}
\end{equation}
We see that the partial derivative, as a vector field, does not act locally on products, due to the deformed product rule. As such, a locally transforming gauge field cannot complete the partial derivative to a covariant derivative \cite{Dimitrijevic:2011jg}. We can avoid this obstacle by working in the exterior algebra and its deformation -- where the product rule for the exterior derivatives is not deformed -- by expressing every quantity in terms of forms. We then treat these forms as the fundamental objects to assign gauge transformations. This allows us to define a gauge covariant exterior derivative, which in turn defines the component covariant derivative.

We start with the transformation of exterior derivatives of a fundamental field $\Phi$,
\begin{align*}
\mathrm{d}\left(\delta_\varepsilon\Phi\right)&=\mathrm{d}\left(i\varepsilon\star\Phi\right)\\
&=i\mathrm{d}\varepsilon\star\Phi+i\varepsilon\star\mathrm{d}\Phi.
\end{align*}
To define a covariant exterior derivative $\mathrm{D}$, we introduce a gauge one form such that we get the desired transformation. I.e. we take
\begin{equation}
\label{eq:covDerPhi}
\mathrm{D}\Phi=\mathrm{d}\Phi+iA\star\Phi,
\end{equation}
with
\begin{equation}
\delta_\varepsilon A=\mathrm{d} \varepsilon+i\scom{\varepsilon}{A}, \qquad \delta_\varepsilon\left(\mathrm{D}\Phi\right)=i \varepsilon \star \mathrm{D}\Phi.
\end{equation}
In index free notation this is identical to the construction for the Groenewold-Moyal case, but in (Cartesian) components all other cases pick up new nontrivial R matrix factors, as
\begin{equation}
\mathrm{D} \Phi = \mathrm{D}_\mu^\star \Phi \star \form{x^\mu} = (\partial^\star_\mu \Phi  + i A^\star_\nu \star \bar{R}_\mu{}^\nu \Phi )\star \form{x^\mu}.
\end{equation}

\paragraph{Gauge invariant noncommutative Yang-Mills theory}

The gauge one form $A$ has an associated field strength tensor\footnote{We use $G$ to avoid confusion with the twist function.}
\begin{equation}
G=\mathrm{d}A-iA\wedge_\star A,
\end{equation}
which transforms star covariantly,
\begin{equation}
\delta_\varepsilon G=i\scom{\varepsilon}{G},
\end{equation}
and is in general not Lie algebra valued. By the left and right star linearity of our Hodge star, we can also immediately write down the dual field strength tensor $*G$, which similarly transforms star covariantly
\begin{equation}
\delta_\varepsilon (*G)=i\scom{\varepsilon}{*G}.
\end{equation}
The simplest star-product compatible action for Yang-Mills theory with the correct undeformed limit is then\footnote{We normalize the generators of the gauge group such that $\tr\left(T^aT^b\right)=\frac{1}{2}\delta^{ab}$.}
\begin{equation}
\label{eq:YMAction}
S_{\text{NC-YM}}= -\frac{1}{2g_{\text{\tiny{YM}}}^2}\int\tr~ G\wedge_\star *G,
\end{equation}
which is gauge invariant thanks to the cyclicity of the star product under integration, guaranteed by imposing eqn. \eqref{eq:twistunimodularity}. In components, gauge invariance of the action follows by the properties \eqref{eq:Rmatrixinversebyraisinglowering} and \eqref{eq:epsilonRinvariant}, as discussed for the Lorentz deformation in \cite{Meier:2023kzt}.

\paragraph{$\theta$ term and self duality}

We can add a noncommutative version of the $\theta$ term
\begin{equation}
\label{eq:YMActiontheta}
S_{\text{NC-YM-$\theta$}}=-\frac{\theta}{8\pi^2} \int\tr~ G\wedge_\star G,
\end{equation}
which is similarly gauge invariant by graded cyclicity of our star-wedge product under the integral. As in the commutative case, it is exact,
\begin{equation}
\int\tr~ G\wedge_\star G = \int\tr~ \mathrm{d} (A \wedge_\star d A - i \frac{2}{3} A \wedge_\star A \wedge_\star A - X),
\end{equation}
where
\begin{equation}
\int \tr ~\mathrm{d} X = \int \tr ~A\wedge_\star A \wedge_\star A \wedge_\star A
\end{equation}
identically vanishes in the commutative setting, and is now a total derivative since our star-wedge product is graded cyclic under an integral, up to total derivative terms.

In Euclidean signature, as for the commutative case, our noncommutative Yang-Mills action can be written as
\begin{equation}
\label{eq:YMActionEuclideanrewriting}
S_{\text{NC-YM}}= \frac{1}{4g_{\text{\tiny{YM}}}^2}\int\tr~ (G\pm *G)\wedge_\star *(G \pm *G) \,\mp \frac{1}{2g_{\text{\tiny{YM}}}^2} \int \tr~G\wedge_\star G .
\end{equation}
Here of course we consider the Euclidean analogue of our deformations, where only the Lorentz deformation has a Hermitian Euclidean counterpart with a deformed Hodge star, associated to the unique antisymmetric $r$ matrix solving the CYBE for $\mathfrak{so}(4)$.\footnote{The analytic continuation to Euclidean signature can be straightforwardly implemented in all purely algebraic properties. In particular, it does not affect the property \eqref{eq:Hodgestarflip}, but gives the usual extra overall sign in eqn. \eqref{eq:hodgestarsquared}. Other properties are unchanged, up to modification of the star product and $R$ matrices of course.} Using eqs. \eqref{eq:equalformshodgeviavolumeform1} and \eqref{eq:equalformshodgeviavolumeform2} to write out the first term then gives
\begin{equation}
\frac{1}{4g_{\text{\tiny{YM}}}^2}\int\tr~ (G\pm *G)\wedge_\star *(G \pm *G) = \frac{1}{2g_{\text{\tiny{YM}}}^2}\int \tr~ (G^\star_{\mu\nu} \pm *G^\star_{\mu\nu}) \star \bar{R}^\mu{}_\rho \bar{R}^\nu{}_\sigma (G^{\star\sigma\rho} \pm *G^{\star\sigma\rho}) \mathrm{d}^4x.
\end{equation}
We can use graded cyclicity (unchargedness of the volume form) to discard the star product between the two terms, and then integrate by parts on one of the $R$s to find
\begin{equation}
\frac{1}{4g_{\text{\tiny{YM}}}^2}\int\tr~ (G\pm *G)\wedge_\star *(G \pm *G) = \frac{1}{2g_{\text{\tiny{YM}}}^2}\int \tr~ \bar{R}^{\rho\mu}(G^\star_{\mu\nu} \pm *G^\star_{\mu\nu}) \bar{R}^{\nu\sigma} (G^{\star}_{\sigma\rho} \pm *G^{\star}_{\sigma\rho}) \mathrm{d}^4x,
\end{equation}
i.e. the first term in the action \eqref{eq:YMActionEuclideanrewriting} is positive definite, while the second gives a boundary contribution, as in the commutative setting. Leaving the investigation of noncommutative instanton contributions as a relevant open question, we note that this second contribution is presumably discrete, being a deformation of the usual winding number. As a result, in each sector the action would be minimized on solution of the noncommutative (anti-)self-dual Yang-Mills equations
\begin{equation}
G = \pm * G,
\end{equation}
which are deformed by the star product at second order in the gauge field, and overall via our twisted Hodge star.

\section{Matter Fields}

Now that we have a noncommutative Yang-Mills action, we would like to couple it to matter fields in the fundamental and adjoint representation of the gauge group. Scalar matter can be directly incorporated using our twisted notion of Hodge duality. To include fermionic matter, in particular in the adjoint representation, we will use a suitable index-free notation involving ``half form'' analogues of the basis forms used for gauge and scalar fields.

\subsection{Scalar fields}

\paragraph{Fundamental representation}

A scalar field and its conjugate in the fundamental representation transform as
\begin{align}
\delta_\varepsilon\phi&=i\varepsilon\star\phi & \delta_\varepsilon\phi^\dagger&=-i\phi^\dagger\star\varepsilon.
\end{align}
The corresponding covariant exterior derivative is given by
\begin{align}
\mathrm{D}\phi&=\mathrm{d}\phi - i A \star \phi & \mathrm{D}\phi^\dagger&=\mathrm{d}\phi^\dagger + i \phi^\dagger \star A,
\end{align}
and we can construct a gauge invariant kinetic action as
\begin{align}
S_{\text{NC-}\phi}&=-\int~\left(\mathrm{D}\phi\right)^\dagger\wedge_\star*\mathrm{D}\phi,
\end{align}
where the Hodge dual term transforms covariantly by star linearity of the Hodge star. Any scalar interactions built out of $\phi^\dagger\star\phi$ blocks are automatically gauge invariant.

\paragraph{Adjoint representation}
A scalar field in the adjoint representation transforms as
\begin{equation}
\delta_\varepsilon\phi=i\scom{\varepsilon}{\phi},
\end{equation}
leading to a covariant derivative
\begin{equation}
\mathrm{D}\phi=\form{\phi}-i\scom{A}{\phi}.
\end{equation}

We then define a star-product-compatible action as
\begin{equation}
\label{eq:scalarAction}
S_{\text{NC-}\phi}=-\int\tr~\mathrm{D}\phi^\dagger\wedge_\star*\mathrm{D}\phi,
\end{equation}
which is gauge invariant similarly to the gauge field action, since also $*\mathrm{D}\phi$ transforms covariantly, by star linearity of the Hodge star. We can add scalar interaction terms obtained by replacing products by star products in usual commutative interaction terms. These will be gauge invariant since we are free to commute the gauge parameter through the volume form, by star commutativity of the latter for our class of twists.

\subsection{Fermionic fields and spinors}

The action of our twists can be extended to spinors via the spinor Lie derivative, which for a vector field $X=X^\mu\partial_\mu$, acting on a Dirac spinor $\tilde{\psi}$, is given by \cite{SpinorLie}
\begin{equation}
\mathcal{L}_X(\tilde{\psi}) = X(\tilde{\psi}) -\frac{1}{8}\left(\partial_\mu X_\nu-\partial_\nu X_\mu\right)\gamma^\mu \gamma^\nu \tilde{\psi},
\end{equation}
where the $\gamma^\mu$ are the Dirac gamma matrices. Rather than working in components, we want to mimic our construction for gauge and scalar fields involving basis forms, and manifest the nature of the Lie derivative of a spinor field. To do so, we split Dirac spinors into left and right-handed Weyl spinors, and introduce the Grassmann-odd left-handed basis Weyl spinor $s^\alpha$, $\alpha=1,2$, and the right-handed $\bar{s}^{\dot{\alpha}}$, $\dot{\alpha}=1,2$. We then write
\begin{equation}
\begin{aligned}
\psi &= s^\alpha \psi_\alpha &= s^\alpha \star \psi^\star_\alpha,\\
\bar{\psi} &= \bar{\psi}_{\dot{\alpha}} \bar{s}^{\dot{\alpha}} &= \bar{\psi}^\star_{\dot{\alpha}} \star \bar{s}^{\dot{\alpha}},
\end{aligned}
\end{equation}
where $\psi_\alpha$ and $\bar{\psi}^{\dot{\alpha}}$ are the Grassmann-odd left-handed and right-handed component spinors respectively. We define our associated Grassmann integrals as
\begin{equation}
\int d^2 s ~s^\alpha s^\beta = \epsilon^{\alpha\beta}, \qquad \int d^2 \bar{s} ~ \bar{s}^{\dot{\alpha}} \bar{s}^{\dot{\beta}} = - \epsilon^{\dot{\alpha}\dot{\beta}},
\end{equation}
such that
\begin{equation}
\int d^2 s ~\psi ~\chi = \psi^\alpha\chi_\alpha, \qquad \int d^2 \bar{s} ~\bar{\psi} ~\bar{\chi} = \bar{\psi}_{\dot{\alpha}} \bar{\chi}^{\dot{\alpha}},
\end{equation}
where we raise and lower indices by the two dimensional (undeformed) Levi-Civita symbol, with $\epsilon^{12} = \epsilon^{\dot{1}\dot{2}}= 1$. We will refer to the Grassmann-even $\psi$ and $\bar{\psi}$ as half forms. Finally, we define a Grassmann-even one form $\sigma$, related to the Pauli $\sigma$ matrices as
\begin{equation}
\sigma=s^\alpha\tensor{\sigma}{_\mu_{\alpha\dot{\alpha}}}\bar{s}^{\dot{\alpha}}\form{x^\mu}=s^\alpha \star \tensor{\sigma}{^\star_\mu_{\alpha\dot{\alpha}}} \star \bar{s}^{\dot{\alpha}} \star \form{x^\mu},
\end{equation}
where $\sigma_{\mu\alpha\dot{\alpha}}=\eta_{\mu\nu} \sigma^\nu_{\alpha\dot{\alpha}}$ which are defined by the following properties
\begin{equation}
\begin{aligned}
\sigma^\mu_{\alpha\dot{\alpha}} \sigma^{\nu\dot{\alpha}\beta} + \sigma^\nu_{\alpha\dot{\alpha}} \sigma^{\mu\dot{\alpha}\beta}=2\eta^{\mu\nu}\delta^\beta_\alpha\\
\sigma^{\mu\dot{\alpha}\alpha} \sigma^\nu_{\alpha\dot{\beta}} + \sigma^{\nu\dot{\alpha}\alpha} \sigma^\mu_{\alpha\dot{\beta}} = 2 \eta^{\mu\nu} \delta^{\dot{\alpha}}_{\dot{\beta}}.
\end{aligned}
\end{equation}
Here, spinor indices are lowered and raised by the levi civita symbol
\begin{equation}
\sigma^{\mu\dot{\alpha}\alpha}=\sigma^\mu_{\beta\dot{\beta}}\epsilon^{\alpha\beta} \epsilon^{\dot{\alpha}\dot{\beta}}.
\end{equation}
The $\sigma$ matrices fulfill the Fierz identity
\begin{equation}
\sigma^\mu_{\alpha\dot{\alpha}}\sigma_{\mu}^{\dot{\beta}\beta}=2\delta_\alpha^\beta \delta_{\dot{\alpha}}^{\dot{\beta}}.
\end{equation}

The algebra of basis spinors behaves similarly to the algebra of forms with regard to star-commutation rules. Correspondingly, we can evaluate the twist function with one or both legs acting our basis spinors, whereby it picks up appropriate spinor indices. This defines
\begin{equation}
\label{eq:Fwithspinorindices}
\begin{aligned}
f(x) \star s^\alpha&=s^\beta \left(\bar{F}_{op}\right)_\beta^{\enspace\alpha} f(x) &&&
f(x) \star \bar{s}^{\dot{\alpha}}&=\bar{s}^{\dot{\beta}} \left(\bar{F}_{op}\right)_{\dot{\beta}}^{\enspace\dot{\alpha}} f(x)\\
\bar{s}^{\dot{\alpha}}\star s^\alpha&=\bar{F}^{\enspace\dot{\alpha}\enspace \alpha}_{\dot{\beta}\enspace\beta} \bar{s}^{\dot{\beta}} s^\beta &&&
s^\alpha \star \bar{s}^{\dot{\alpha}}&=\bar{F}^{\enspace \alpha \enspace \dot{\alpha}}_{\beta \enspace \dot{\beta}} s^\beta \bar{s}^{\dot{\beta}}.\\
s^{\alpha}\star s^\beta&=\bar{F}^{\enspace\alpha\enspace \beta}_{\gamma\enspace\delta} \bar{s}^{\gamma} s^\delta &&&
\bar{s}^{\dot{\alpha}} \star \bar{s}^{\dot{\beta}}&=\bar{F}^{\enspace \alpha \enspace \beta}_{\gamma \enspace \delta} \bar{s}^{\dot{\gamma}} \bar{s}^{\dot{\delta}}
\end{aligned}
\end{equation}
We can now also move functions and basis one forms through the basis half forms, which introduces $\mathcal{R}$ matrices with appropriate spinor indices as\footnote{As above, the $\mathcal{R}$ matrix can be factored out in this fashion since we are dealing with vector fields at most linear in coordinates.}
\begin{equation}
\begin{aligned}
s^\alpha \star f(x)&=R_\beta~^\alpha f(x)\star s^\beta&&&
\bar{s}^{\dot{\alpha}} \star f(x)&=R_{\dot{\beta}}~^{\dot{\alpha}} f(x)\star \bar{s}^{\dot{\beta}}\\
\form{x^\mu} \star s^\alpha&=R^{\enspace\mu\enspace\alpha}_{\nu\enspace\beta} s^\beta \star \form{x^\nu}&&&
\form{x^\mu} \star \bar{s}^{\dot{\alpha}}&=R^{\enspace\mu\enspace\dot{\alpha}}_{\nu\enspace\dot{\beta}} \bar{s}^{\dot{\beta}} \star \form{x^\nu}.
\end{aligned}
\end{equation}
Finally, commuting two basis half forms reads
\begin{equation}
\begin{aligned}
\bar{s}^{\dot{\alpha}}\star s^\alpha&=-R^{\enspace\dot{\alpha}\enspace \alpha}_{\dot{\beta}\enspace\beta} s^\beta \star \bar{s}^{\dot{\beta}}&&&
s^\alpha \star \bar{s}^{\dot{\alpha}}&=-R^{\enspace \alpha \enspace \dot{\alpha}}_{\beta \enspace \dot{\beta}} \bar{s}^{\dot{\beta}} \star s^\beta\\
s^\alpha\star s^\beta&=-R^{\enspace\alpha\enspace \beta}_{\gamma\enspace\delta} s^\delta \star s^\beta&&&
\bar{s}^{\dot{\alpha}} \star \bar{s}^{\dot{\beta}}&=-R^{\enspace \dot{\alpha} \enspace \dot{\beta}}_{\dot{\gamma} \enspace \dot{\delta}} \bar{s}^{\dot{\delta}} \star \bar{s}^{\dot{\gamma}}.
\end{aligned}
\end{equation}
This defines the noncommutative structure upon which we want to define our gauge theory. Similar to the gauge transformation for the gauge field, we will treat the full spinors $\psi$ and $\bar{\psi}$ as the fundamental objects to assign gauge transformations.

\paragraph{Fundamental fermionic fields}
Fundamental fermionic fields $\psi$ and $\bar{\psi}$ transform under gauge transformations as
\begin{equation}
\delta_\varepsilon\psi = i\varepsilon \star \psi, \qquad \delta_\varepsilon\bar{\psi} = -i\bar{\psi} \star \varepsilon,
\end{equation}
leading to the covariant derivatives
\begin{equation}
\mathrm{D}\psi = \mathrm{d}\psi - i A \star \psi, \qquad \mathrm{D}\bar{\psi} = \mathrm{d}\bar{\psi} + i \bar{\psi} \star A.
\end{equation}
We can immediately construct a star-product-compatible gauge-invariant kinetic action for fermionic fields as\footnote{To avoid excessive integral signs, here and below we use a single integral sign for all Grassmann integrals, with a second integral sign for a spacetime integral.}
\begin{equation}
S_{\text{NC-}\psi}=\int\mathrm{d}^2s\mathrm{d}^2\bar{s}~\int~\bar{\psi} \star \mathrm{D}\psi \wedge_\star*\sigma.
\end{equation}
Together with an adjoint scalar field $\phi$, we can consider Yukawa-like interactions such as
\begin{equation}
\int\mathrm{d}^2s~\int~*(\psi \star \phi \star \psi),
\end{equation}
which are gauge invariant by construction.

\paragraph{Adjoint fermionic fields}
Adjoint fermionic fields transform as
\begin{equation}
\delta_\varepsilon\psi=i\scom{\varepsilon}{\psi},
\end{equation}
with covariant derivative
\begin{equation}
\mathrm{D}\psi=\form{\psi}-i\scom{A}{\psi},
\end{equation}
and the same for $\bar{\psi}$. As a natural action with correct undeformed limit we can take
\begin{align*}
\int\mathrm{d}^2s\mathrm{d}^2\bar{s}~\int~\tr\left[\bar{\psi} \star \sigma \wedge_\star \left(*\mathrm{D}\psi\right)\right],
\end{align*}
which is gauge invariant provided the Grassmann-even one form $\sigma$ is star commutative, i.e.
\begin{equation}
f \star \sigma = \sigma \star f,
\end{equation}
for any function $f$. This means we need $\sigma_{\mu\alpha\dot{\alpha}}s^\alpha\bar{s}^{\dot{\alpha}}$ to transform as a basis vector with lower index $\mu$ with respect to the twist. In terms of $\mathcal{R}$ matrices the constraint reads
\begin{equation}
\label{eq:sigmaconstraint}
\sigma_{\mu\alpha\dot{\alpha}}R_\beta~^\alpha R_{\dot{\beta}}~^{\dot{\alpha}}R_\nu{}^\mu=\sigma_{\nu\beta\dot{\beta}}.
\end{equation}
This is automatically satisfied, since restricted to the Poincar\'e algebra, the Lie derivative of a spinor gives the spin representation of the Poincar\'e algebra, which the Pauli matrices convert to the fundamental representation, compensating the transformation of the one basis form. Like star linearity of the Hodge star and star commutativity of the volume form, star commutativity of $\sigma$ highlights the special role played by the Poincar\'e algebra in our construction.

We can consider Yukawa-like interactions such as
\begin{equation}
\int\mathrm{d}^2s~\int~*\tr\left(\psi \star \phi \star \psi\right),
\end{equation}
which are again gauge invariant by star linearity of the Hodge star.

\section{A planar equivalence theorem for Feynman diagrams}
\label{sec:filk1}

Our noncommutative deformations affect the Feynman rules and diagrams of our theory, in general drastically affecting the resulting amplitudes and correlation functions. Concretely, the noncommutativity makes the Feynman rules for vertices sensitive to the ordering of propagators entering a given vertex. At the same time, the cyclicity of the star product under integration results in cyclic vertices, reminiscent of ordinary for Yang-Mills theory, or single trace interactions for general matrix valued fields. This structure allows us to introduce a notion of planarity for diagrams. For the Groenewold-Moyal deformation this structure was used by Filk \cite{Filk:1996dm} to describe planar diagrams of the deformed quantum field theory in terms of the corresponding diagrams of the undeformed theory, showing that the two only differ by star product factors between the external lines, and hence that they have the same type of divergences. We will refer to this as the planar equivalence theorem. In this section we will prove a generalization of this theorem for all deformations covered by our construction, for theories involving scalars, fermions, and gauge fields, whose undeformed counterparts have Poincar\'e symmetry.\footnote{The original proof by Filk, see also e.g. section 3.1 of the review \cite{Szabo:2001kg}, is in momentum space, where the effect of the Groenewold-Moyal twist is particularly simple. Our proof is its position space analogue, generalized to arbitrary Poincar\'e-based twists, and general field content.} We will start by discussing some relevant simplifications that happen for our class of deformations.

\subsection{Simplifications and notation}

\paragraph*{Cyclic scalar interactions} Because for our class of twists, star wedge products of forms are graded cyclic under integration, and the volume form is star commutative, star products of functions become cyclic under integration, i.e.
\begin{equation}
\label{eq:cyclicIntegral}
\int \mathrm{d}^4_\star x\star \left(\phi \star \chi\right) = \int \mathrm{d}^4 x~ \left(\phi \star \chi\right) = \int \mathrm{d}^4 x~ \left(\chi \star \phi\right) = \int \mathrm{d}^4 x~ \phi\, \chi,
\end{equation}
for any functions $\phi$ and $\chi$. This means that any scalar interactions are manifestly cyclic. For matrix valued fields this requires single trace interactions.

\paragraph*{Propagators} The kinetic terms of our theories can be simplified using the undeformed laplacian \eqref{eq:laplacian} together with integration by parts. For a scalar field, for instance,
\begin{equation}
S=-\int~\mathrm{d}\bar{\phi} \wedge_\star \left(*\mathrm{d}\phi\right)=\int~\bar{\phi} \star \mathrm{d}\left(*\mathrm{d}\phi\right)=\int~\bar{\phi} (*\Delta\phi)=-\int \mathrm{d}^4x~\partial_\mu\bar{\phi} \partial^\mu \phi,
\end{equation}
meaning that the kinetic term is undeformed. Mass terms are similarly undeformed, since we can freely remove the single star product in the associated quadratic term. Consequently, scalar propagators are not deformed. By similar reasoning, the kinetic action and propagator of the component gauge fields can be shown to remain undeformed, as discussed in \cite{Meier:2023kzt} for the case of the Lorentz twist. Finally, for the fermionic kinetic action, we find
\begin{equation}
S_{\bar{\psi}\psi}=\int\int\mathrm{d}^2s\mathrm{d}^2\bar{s}~\bar{\psi} \star \sigma \wedge_\star *d\psi=\int\int\mathrm{d}^2s\mathrm{d}^2\bar{s}~\bar{\psi} \left(\sigma \wedge_\star *d\psi\right)=\int\mathrm{d}^4x\int\mathrm{d}^2s\mathrm{d}^2\bar{s}~\bar{\psi} \sigma^\mu \partial_\mu \psi,
\end{equation}
which is the undeformed kinetic action, where \eqref{eq:dxhodgedx_with_without_star_product} and the star commutativity of $\sigma$ were used. In summary, propagators are undeformed.

\begin{figure}[h]
\begin{center}
\includegraphics[width=0.48\textwidth]{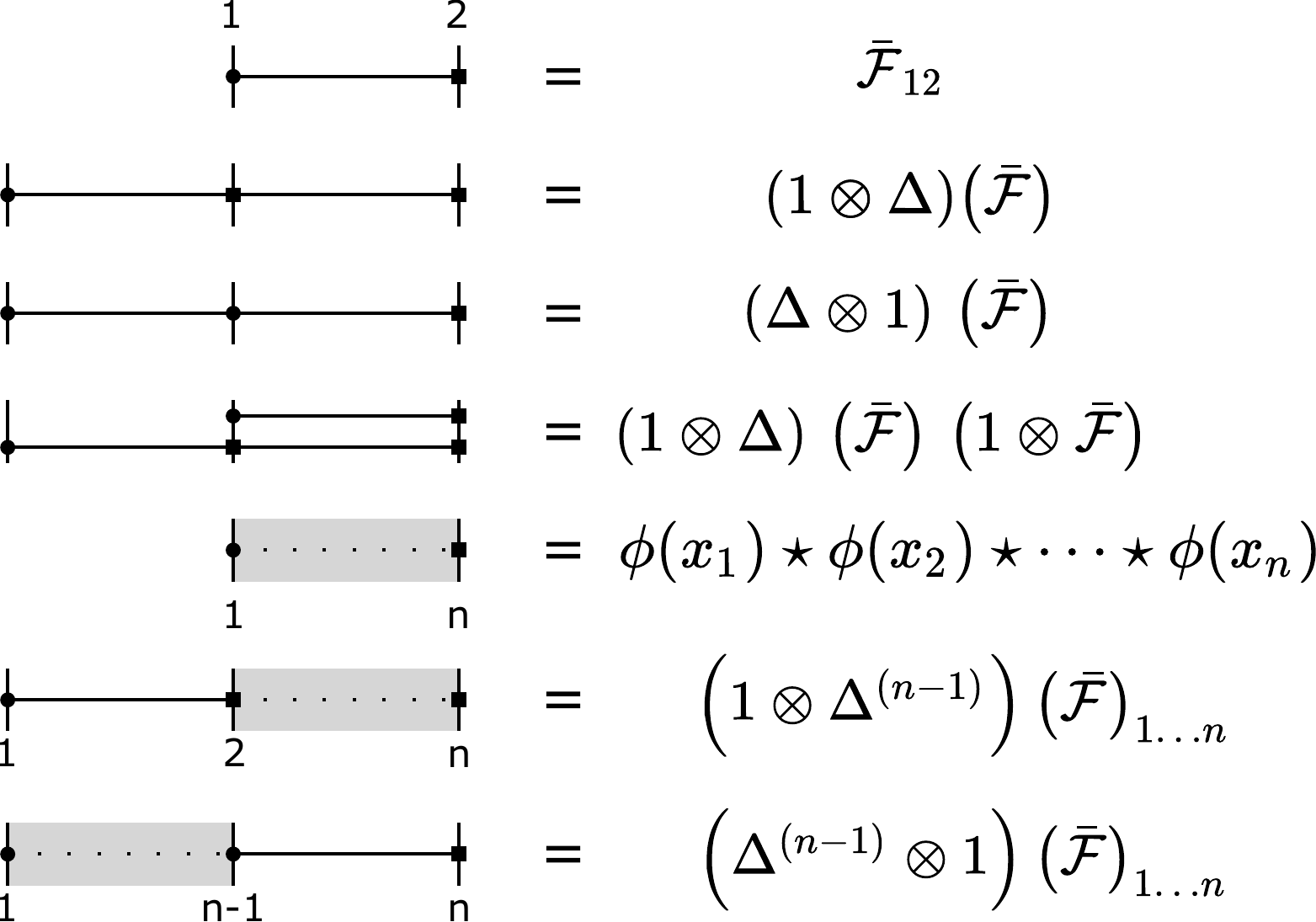}
\caption[Diagrammatic twist function]{Feynman rules of star products between legs of a diagram. The vertical lines here denote legs in a hypothetical Feynman diagram, while the horizontal lines correspond to twist functions, and will be called twist lines. In actual diagrams the distinction between legs and twist lines will be clear from context. A dot indicates the action of $\bar{f}^\alpha$, while a square indicates the action of $\bar{f}_\alpha$, extended by coproducts to multiple legs, as in the first three equations. Grey areas represent arbitrary many legs in between the first and last leg, as in the last three equations.  Since our twist functions in general do not commute, there is an intrinsic ordering, accounted for in the current orientation by reading a diagram from the top down, as in the fourth equation. For twist lines appearing in actual diagrams, with legs connected to a vertex, we apply twists reading outward from the vertex.}
\label{fig:DefineDiagrammTwist}
\end{center}
\end{figure}

\begin{figure}[h]
\begin{center}
\includegraphics[width=0.48\textwidth]{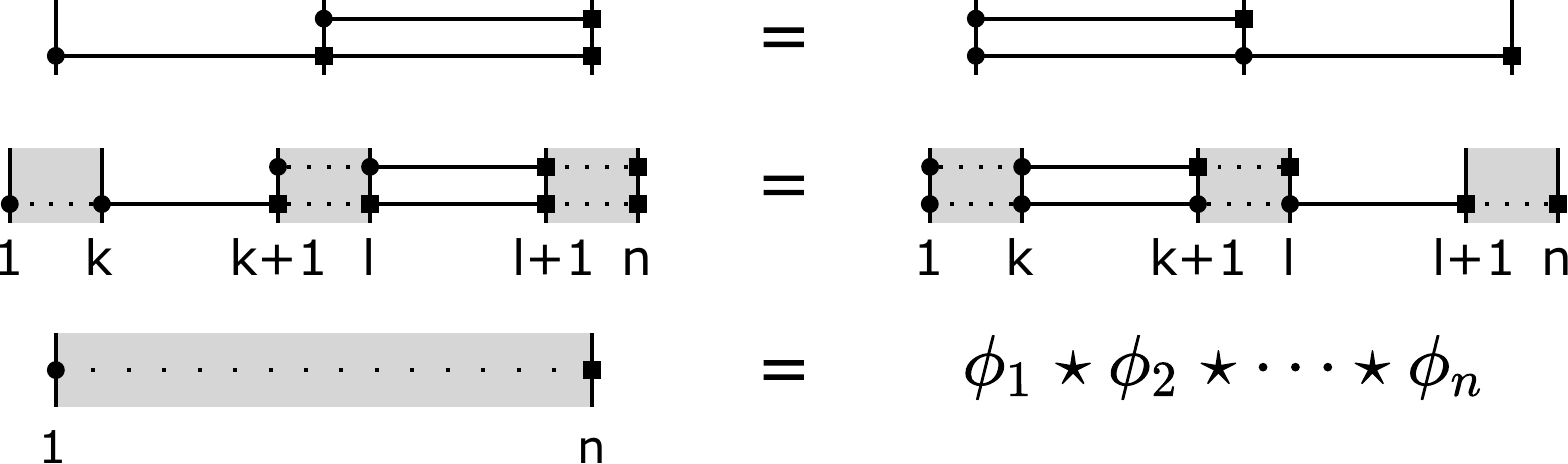}
\caption{The cocycle condition (associativity) for twist lines. The first line directly corresponds to the cocycle condition $\left(1\otimes\Delta\right)\left(\mathcal{F}^{-1}\right)\left(1\otimes\mathcal{F}^{-1}\right)=\left(\Delta\otimes1\right)\left(\mathcal{F}^{-1}\right)\left(\mathcal{F}^{-1}\otimes1\right)$, while the second equation will appear for star products between arbitrary many fields and represents the equation $\left(\Delta^{(k)}\otimes\Delta^{(l-k)}\otimes \Delta^{(n-k-l)}\right) \left[\left(1\otimes\Delta\right)\left(\mathcal{F}^{-1}\right)\left(1\otimes\mathcal{F}^{-1}\right)\right]= \left(\Delta^{(k)}\otimes\Delta^{(l-k)}\otimes \Delta^{(n-k-l)}\right)\left[\left(\Delta\otimes1\right)\left(\mathcal{F}^{-1}\right)\left(\mathcal{F}^{-1}\otimes1\right)\right]$. Because star products are associative, we can collect multiple star products, and consistently denoting them by a dashed twist line starting with a circle and finishing with a rectangle, as in the last line.}
\label{fig:cocycle}
\end{center}
\end{figure}

\paragraph*{Notation} To account for the star products (twists) appearing in our Feynman rules through the interaction terms, we will dress our Feynman diagrams with extra lines, indicating twists acting between legs coming out of vertices. To distinguish the spaces in which the twist $\bar{\mathcal{F}} = \bar{f}^\alpha \otimes \bar{f}_\alpha \in \mathcal{U(P)}\otimes \mathcal{U(P)}$ acts, we will denote the action of the twist in the first space ($\bar{f}^\alpha$) by circles and the second ($\bar{f}_\alpha$)by rectangles, extended to multiple lines by the corresponding coproducts, as indicated in figure \ref{fig:DefineDiagrammTwist}.
We can collect multiple star products, and correspondingly their twists, in a single twist line, as indicated in figure \ref{fig:cocycle}. This is well-defined since the star product is associative, following directly from the cocycle condition for the twist. Moreover, by this associativity, any set of twist lines corresponding to a given bracketing of star products, can be replaced by another set of twist lines, corresponding to an alternate choice of bracketing, as also indicated in figure \ref{fig:cocycle}.

We are now ready to discuss the planar equivalence theorem, starting with scalar field theories. Our proof will then readily generalize to fermionic and gauge fields.

\subsection{Scalar field theories}
\label{sec:scalarFilk}
The Feynman rules for vertices in our noncommutative theories, introduce star products acting on propagators in any given Feynman diagram. To have any hope to relate these diagrams to their undeformed counterparts, we will need to manipulate these twists to cancel any of them acting on internal propagators. For this we need several identities involving twists on scalar propagators, which will be used in our proof of the planar equivalence theorem below.

\begin{figure}[h]
\begin{center}
\includegraphics[width=0.7\textwidth]{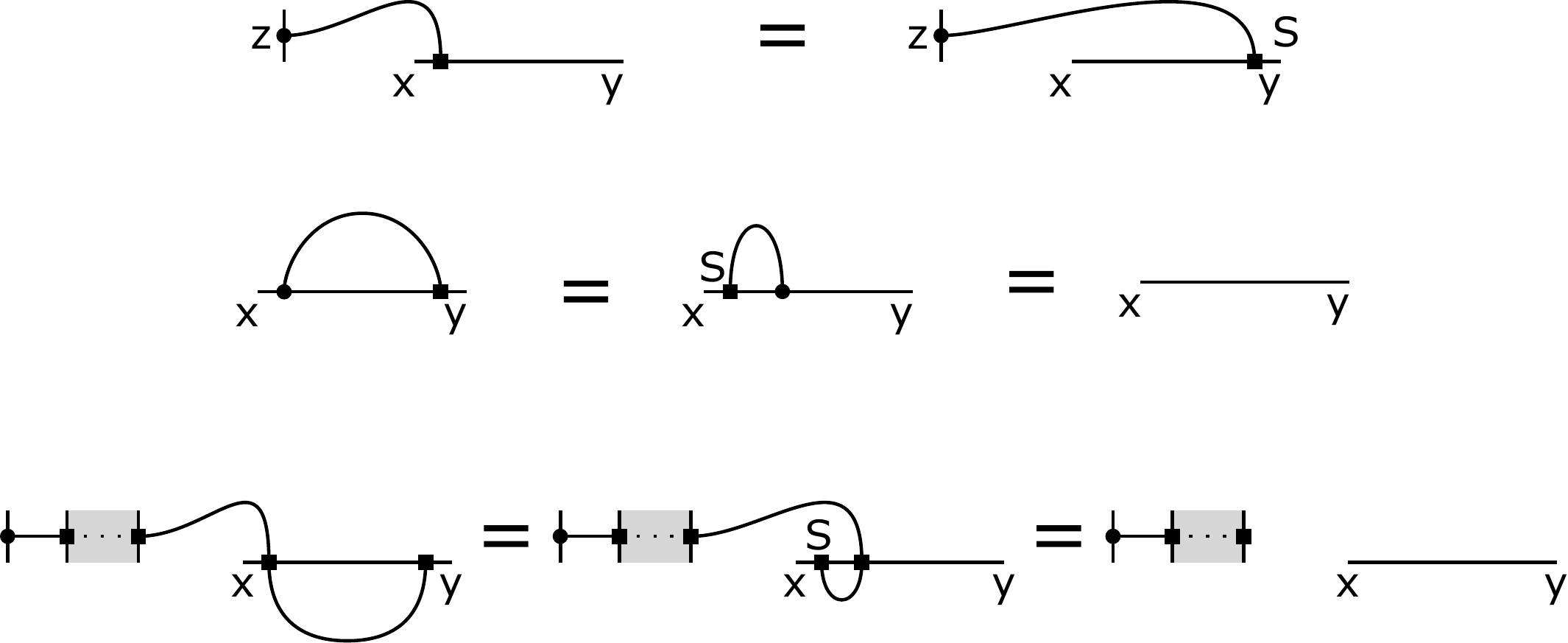}
\caption{Diagrammatic representation of \eqref{eq:TwistOnProp}, \eqref{eq:TwistOnProp2} and \eqref{eq:TwistOnProp3}.}
\label{fig:PropRelations}
\end{center}
\end{figure}

\paragraph{Twist-propagator identities} Since the scalar propagator is undeformed, it takes the usual form
\begin{align*}
\Delta_\phi(x-y)&=\int\mathrm{d}^4p~\frac{1}{p^2+m^2+i\epsilon}e^{-i p\left(x-y\right)},
\end{align*}
reducing to
\begin{align*}
\Delta_{\phi,m=0}(x-y)&=\frac{1}{(x-y)^2}
\end{align*}
for massless fields. Regardless of mass, this propagator is Poincar\'e invariant, hence for $X \in \mathcal{P}$
\begin{equation*}
\begin{aligned}
\left(X_x+X_y\right)\Delta_\phi(x-y)&=0,
\end{aligned}
\end{equation*}
where the subscript denotes the position the vector field $X$ acts on. By anti-linearity of the antipode, this extends to arbitrary elements of the universal enveloping algebra as
\begin{equation}
\label{eq:invarianceOfPropagator}
\begin{aligned}
\left(X_x-S\left(X_y\right)\right)\Delta_\phi(x-y)&=0~~~~\forall X\in\mathcal{U(P)}.
\end{aligned}
\end{equation}
Starting from this invariance, we are able to map generic elements of $\mathcal{U(P)}$ from one end of a propagator to the other. Since, $\mathcal{F}\in\mathcal{U(P)}\otimes \mathcal{U(P)}$ for all cases in this work, we can map twist functions and hence star products from one end of a scalar propagator to the other. In particular, \eqref{eq:invarianceOfPropagator} implies
\begin{equation}
\label{eq:TwistOnProp}
\begin{aligned}
\mathcal{\bar{F}}_{xz}\left(\Delta_\phi(x-y)\right)&=\bar{f}^\alpha_x\left(\Delta(x-y)\right)\bar{f}_{z\alpha}\\
&=S(\bar{f}^\alpha_y)\left(\Delta_\phi(x-y)\right)\bar{f}_{z\alpha}\\
&=\left(S\otimes 1\right)\left(\bar{\mathcal{F}}\right)_{yz}\Delta_\phi(x-y),
\end{aligned}
\end{equation}
where the subscripts denote the objects on which the twist is acting. In the case of an abelian twist this reduces to the simpler
\begin{equation*}
\mathcal{\bar{F}}_{xz}\left(\Delta_\phi(x-y)\right)=\mathcal{F}_{yz}\left(\Delta_\phi(x-y)\right).
\end{equation*}
Next, a twist which acts on both ends of the propagator is trivial
\begin{equation}
\label{eq:TwistOnProp2}
\begin{aligned}
\mathcal{\bar{F}}_{xy}\left(\Delta_\phi(x-y)\right)&=\bar{f}^\alpha_x\bar{f}_{y\alpha}\left(\Delta_\phi(x-y)\right)\\
&=\bar{f}^\alpha_x S(\bar{f}_{x\alpha})\Delta_\phi(x-y)\\
&=\Delta_\phi(x-y),
\end{aligned}
\end{equation}
where we use the fact that our twists satisfy \eqref{eq:twistunimodularity}.
Finally, from \eqref{eq:invarianceOfPropagator} it directly follows that for any $X\in \mathcal{U(P)}$
\begin{equation}
\Delta(X)_{xy}\Delta_\phi(x-y)=\nabla\left[\left(1\otimes S\right)\Delta(X)\right]_x\Delta_\phi(x-y)=\epsilon(X)\Delta_\phi(x-y)=\left\lbrace\begin{matrix}
\Delta_\phi(x-y)~&X=1\\
0&\text{otherwise}
\end{matrix}\right.,
\end{equation}
where $\nabla:\mathcal{U}(\mathfrak{g})\otimes\mathcal{U}(\mathfrak{g})\rightarrow \mathcal{U}(\mathfrak{g}), \nabla(X\otimes Y)=XY$ denotes the product of our algebra. Here we used the Hopf algebra relation $\nabla\left(1\otimes S\right)\Delta= \epsilon$. Hence, for a twist hitting each end of the propagator with the same twist leg, we have
\begin{equation}
\label{eq:TwistOnProp3}
\begin{aligned}
&\left(1\otimes\Delta^{(i+1)}\right)\left(\bar{\mathcal{F}}\right)_{x y_1 \dots y_{i+1}}\Delta_\phi(y_i-y_{i+1})\\
=& \left(1^{i\otimes}\otimes\Delta\right)\left(1\otimes\Delta^{(i)}\right)\left(\bar{\mathcal{F}}\right)_{x y_1 \dots y_{i+1}}\Delta_\phi(y_i-y_{i+1})\\
=&\left(1^{i\otimes}\otimes\nabla\right)\left(1^{i\otimes}\otimes1 \otimes S\right)\left(1^{i\otimes}\otimes\Delta\right)\left(1\otimes\Delta^{(i)}\right)\left(\bar{\mathcal{F}}\right)_{x y_1 \dots y_{i}}\Delta_\phi(y_i-y_{i+1})\\
=&\left(1^{i\otimes}\otimes\epsilon\right)\left(1^{(i-1)\otimes}\otimes\Delta\right)\left(1\otimes\Delta^{(i-1)}\right)\left(\bar{\mathcal{F}}\right)_{x y_1 \dots y_{i-1}}\Delta_\phi(y_i-y_{i+1})\\
=&\left(1\otimes\Delta^{(i-1)}\right)\left(\bar{\mathcal{F}}\right)_{x y_1 \dots y_{i-1}}\Delta_\phi(y_i-y_{i+1}),
\end{aligned}
\end{equation}
where in the third equality, we used that $\epsilon$ is a counit and hence satisfies $\left(1\otimes\epsilon\right)\Delta=1$. For an abelian twist the derivation simplifies and amounts to
\begin{equation}
\mathcal{\bar{F}}_{xy_{i}}\mathcal{\bar{F}}_{xy_{i+1}}\Delta_\phi(y_i-y_{i+1})=\mathcal{\bar{F}}_{xy_{i}}\mathcal{F}_{xy_{i}}\Delta_\phi(y_i-y_{i+1})=\Delta_\phi(y_i-y_{i+1}).
\end{equation}
as the coproduct in the first term generates several commuting twist, where just two of them hit the propagator and need to be considered. We can diagrammatically represent the identities \eqref{eq:TwistOnProp}, \eqref{eq:TwistOnProp2} and \eqref{eq:TwistOnProp3} as in fig. \ref{fig:PropRelations}.

\begin{figure}[h]
\begin{center}
\includegraphics[width=0.9\textwidth]{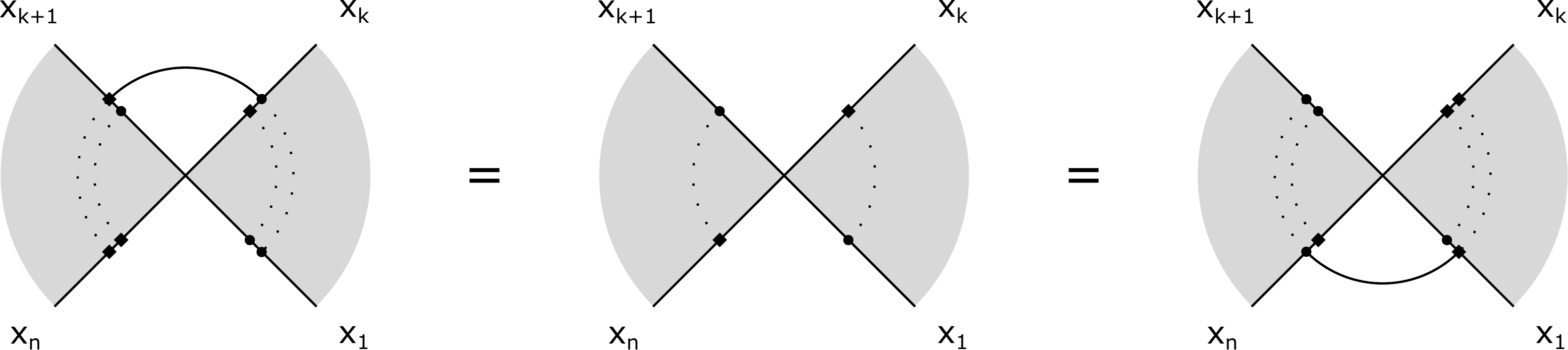}
\caption{The cyclicity relation for a single vertex corresponding to \eqref{eq:VertInv}.}
\label{fig:VertexCyclic}
\end{center}
\end{figure}

\begin{figure}[h]
\begin{center}
\includegraphics[width=0.6\textwidth]{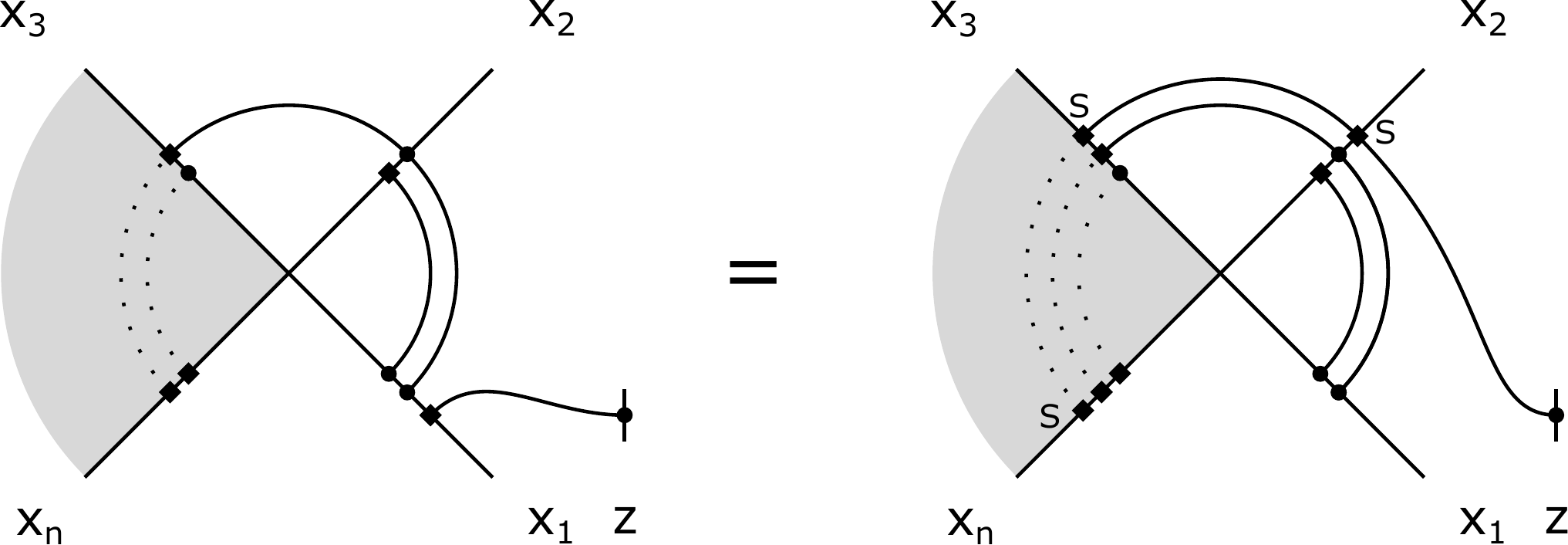}
\caption{The Poincar\'e invariance relation for a single vertex corresponding to \eqref{eq:VertInv2}.}
\label{fig:VertexInv}
\end{center}
\end{figure}

\paragraph{Properties of interaction vertices} Beyond properties of propagators, we also need to discuss fundamental properties of interaction vertices. Consider a general scalar interaction with $n$ fields entering a vertex. Its corresponding interaction term in the deformed action is given by
\begin{align*}
\label{eq:scalarVert}
\int\mathrm{d}^4x~\phi \star \dots \star \phi &=\int\mathrm{d}^4x~\left(\overset{n}{\underset{i=1}{\large{\bigstar}}}\phi(x_i)\right)_{x_k=x},
\end{align*}
where we define
\begin{equation}
\overset{n}{\underset{i=1}{\large{\bigstar}}}\phi(x_i)=\phi (x_1) \star \phi (x_2) \star \dots \star \phi (x_n).
\end{equation}
Here we separate the positions of the fields entering the vertex from each other to be able to track Wick contractions later on. Cyclicity under integration gives
\begin{equation}
\begin{aligned}
\label{eq:VertInv}
\int\mathrm{d}^4x~\left(\overset{n}{\underset{k=1}{\large{\bigstar}}}\phi(x_k)\right)_{x_i=x}&=\int\mathrm{d}^4x~\left(\overset{k}{\underset{i=1}{\large{\bigstar}}}\phi(x_i)\overset{n}{\underset{j=k+1}{\large{\bigstar}}}\phi(x_j)\right)_{x_i=x}\\
&=\int\mathrm{d}^4x~\left(\overset{n}{\underset{j=k+1}{\large{\bigstar}}}\phi(x_j)\star\overset{k}{\underset{i=1}{\large{\bigstar}}}\phi(x_i)\right)_{x_i=x}&~~,0\le k\le n,
\end{aligned}
\end{equation}
diagramatically represented in figure \ref{fig:VertexCyclic}.
Next, by Poincar\'e invariance of the volume element, integration by parts tells us
\begin{equation}
\label{eq:VertInv2}
\int\mathrm{d}^4x~\left(\left[\bar{\mathcal{F}}_{zx}\phi(x)\right]\star\overset{n-1}{\underset{k=1}{\large{\bigstar}}}\phi(x)\right)=\int\mathrm{d}^4x~\phi(x)\star\left(\left(1\otimes S \right)\left(\bar{\mathcal{F}}\right)_{zx}\overset{n-1}{\underset{k=1}{\large{\bigstar}}}\phi(x)\right),
\end{equation}
which can be diagramatically represented as in figure \ref{fig:VertexInv}.
Since this is related to Poincar\'e invariance, we will refer to it as the Poincar\'e invariance relation.

\paragraph{Structure of undeformed Feynman diagrams} Before considering deformed Feynman integrals we want to express undeformed Feynman integrals in a suitable way. For this, we first strip off external propagators. In other words, the fields to be Wick contracted to external fields in a proper Feynman diagram, are left uncontracted. To evaluate the diagrams we finally need to perform these missing Wick contractions, but for now we are interested in the structure of integrands of Feynman diagrams in our deformed theories. In what follows, we will call these uncontracted fields external.

In this setting, a general, planar, undeformed Feynman diagram can be expressed as
\begin{align}
\label{eq:undeformedDiagram}
a_n\left(I_{n};x_1\dots x_n\right)&=I_{n} \times \prod_{k=1}^n \phi(x_k),
\end{align}
where $I_{n}$ collects all internal propagators and integrations associated to the vertices of the diagram. The remaining fields are external.

\subsubsection*{The planar equivalence theorem}

\begin{figure}[h]
\begin{center}
\includegraphics[width=0.3\textwidth]{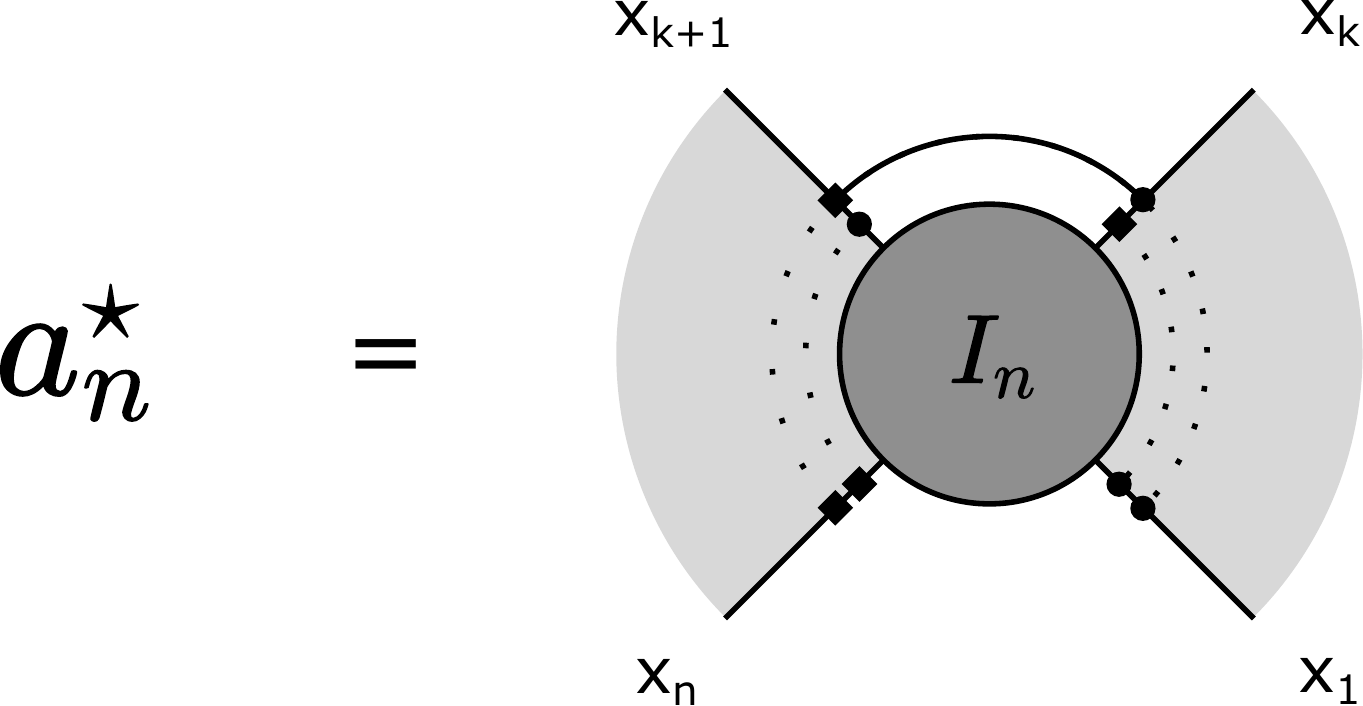}
\caption{The planar equivalence structure. A deformed planar diagram is equal to its undeformed counterpart, up to star products between external legs.}
\label{fig:Filk1}
\end{center}
\end{figure}

\begin{figure}[h]
\begin{center}
\includegraphics[width=0.8\textwidth]{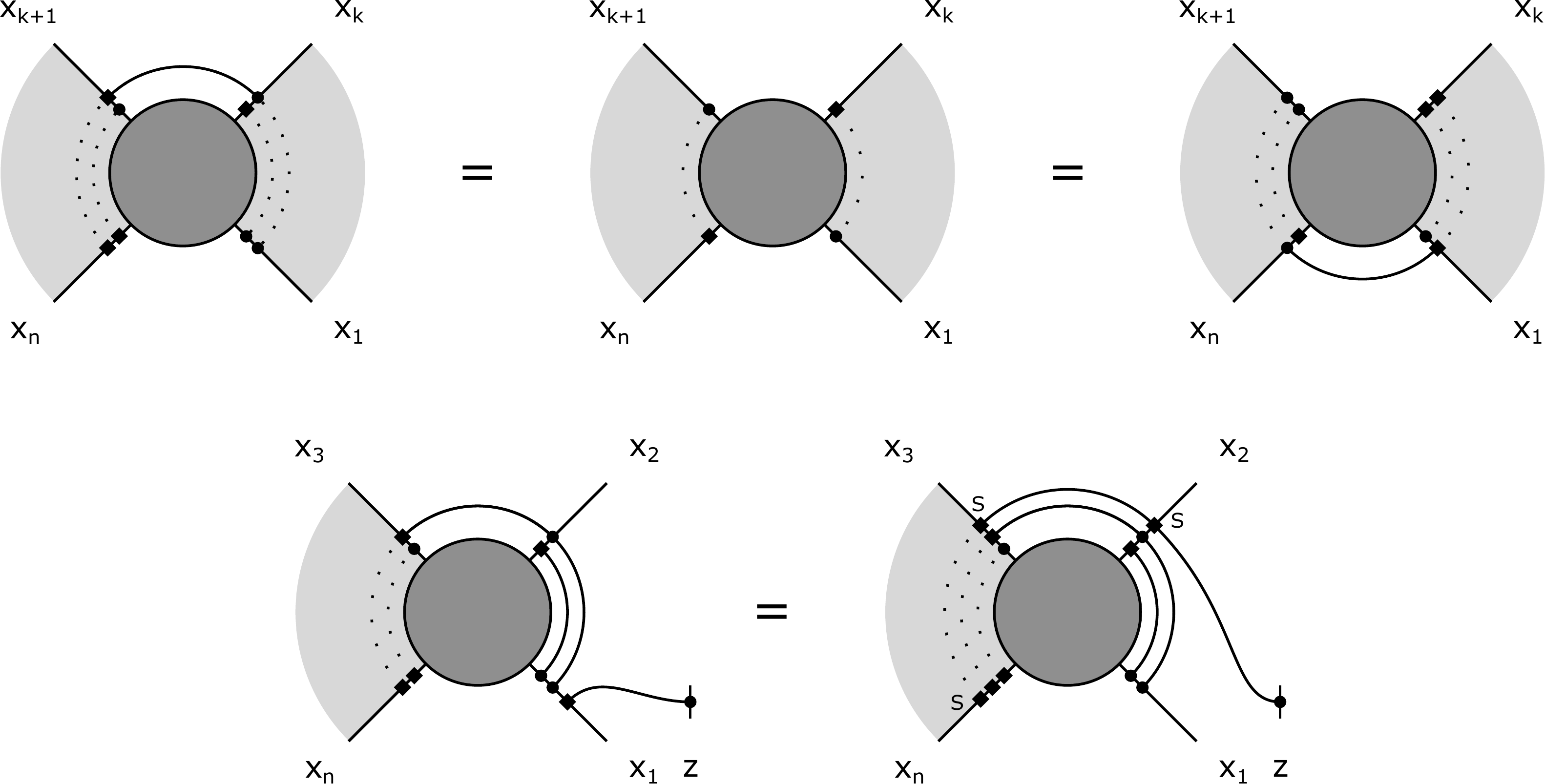}
\caption{The cyclicity and Poincar\'e invariance relation for planar diagrams corresponding to \eqref{eq:ampAss2}.}
\label{fig:Filk2}
\end{center}
\end{figure}

We say that a deformed diagram corresponds to an undeformed one, if the undeformed limit of the former is given by the latter, i.e. they have the same internal structure up to potential star products. The planar equivalence theorem for Feynman diagrams then firstly states that the deformed planar diagram corresponding to the undeformed one of eqn. \eqref{eq:undeformedDiagram}, has the structure
\begin{align}
\label{eq:ampAss1}
a^\star_n\left(I_n\right)=I_{n} \times \overset{n}{\underset{k=1}{\large{\bigstar}}} \phi(x_k)=\underset{i=1}{\overset{n-1}{\overleftarrow{\prod}}}\left(\Delta^{(i)}\otimes 1\right)\left(\bar{\mathcal{F}}\right)_{x_1\dots x_{i+1}}a_n\left(I_n;x_1\dots x_n\right),
\end{align}
represented diagramatically in fig. \ref{fig:Filk1}.
In other words, the internal structure of the deformed diagram is identical to the undeformed one, while their difference lies in the added star products between the external fields. The second part of the planar equivalence theorem states that these diagrams have the same cyclicity and Poincar\'e invariance relations as the deformed interaction vertices, i.e.
\begin{equation}
\begin{aligned}
\label{eq:ampAss2}
I_{n}\times\overset{n}{\underset{k=1}{\large{\bigstar}}}\phi(x_k)&=I_{n}\times \overset{k}{\underset{i=1}{\large{\bigstar}}}\phi(x_i)\overset{n}{\underset{j=k+1}{\large{\bigstar}}}\phi(x_j)\\
&=I_{n}\times \overset{n}{\underset{j=k+1}{\large{\bigstar}}}\phi(x_j)\star\overset{k}{\underset{i=1}{\large{\bigstar}}}\phi(x_i)\\
I_{n}\times\left[\bar{\mathcal{F}}_{x_1z}\phi(x_1)\right]\star\overset{n}{\underset{k=2}{\large{\bigstar}}}\phi(x_k)&=I_{n}\times\phi(x_1)\star\left(\left(S\otimes1\right)\left(\bar{\mathcal{F}}\right)_{xz}\overset{n}{\underset{k=2}{\large{\bigstar}}}\phi(x_k)\right),
\end{aligned}
\end{equation}
as illustrated in fig. \ref{fig:Filk2}.
These three identities are automatically satisfied by diagrams containing exactly one vertex. In other words, our basic interaction vertices fit the requirements of the planar equivalence theorem.

\subsubsection*{Proof of the planar equivalence theorem}

Our proof of the planar equivalence theorem will be diagrammatic and inductive, and consist of two parts. First, starting from single-vertex diagrams, we will add additional vertices by connecting a new basic vertex via one propagator to an arbitrary external field of the existing diagram. This allows us to construct arbitrary diagrams with tree like structure, which we will then prove to fit the planar equivalence theorem. Second, we will include planar loops by contracting neighboring fields of a given diagram with each other, proving that the result fits the planar equivalence theorem. By combining these two steps we can generate arbitrary planar diagrams starting from our basic interaction vertices, completing the proof.

\begin{figure}[h]
\begin{center}
\includegraphics[width=0.9\textwidth]{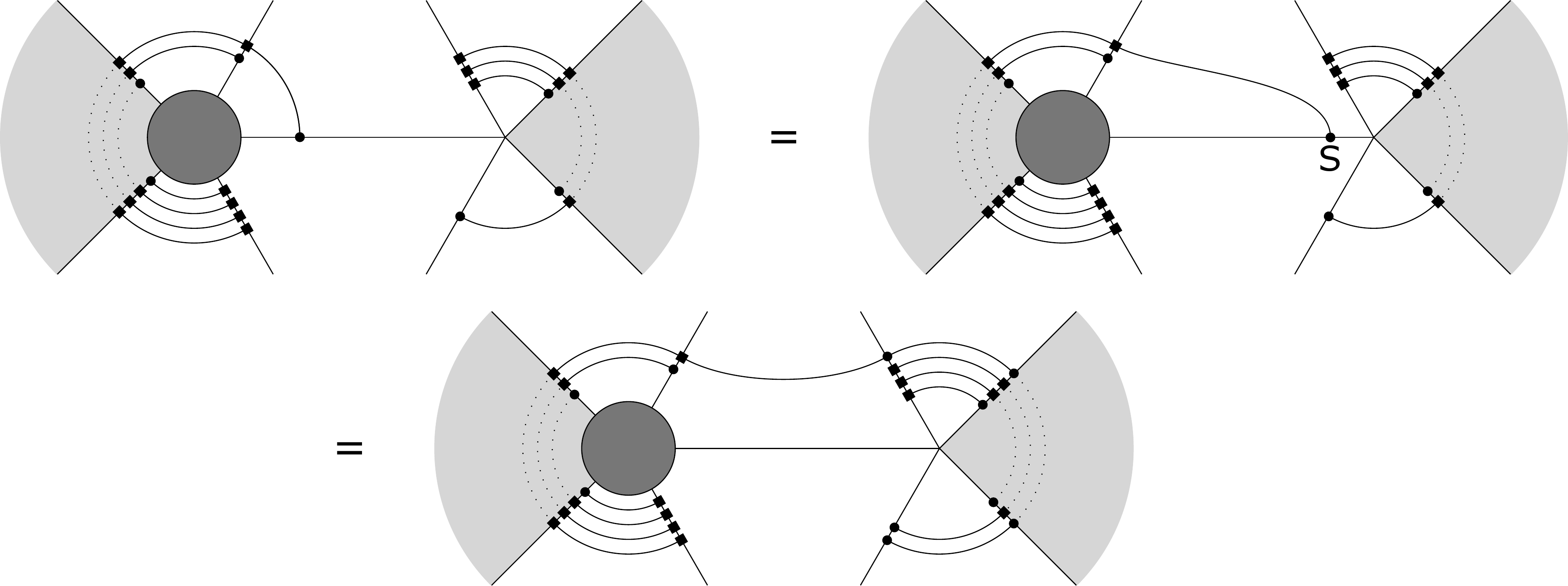}
\caption{Adding vertices recursively to a given planar diagram. By the cyclicity condition for the two subdiagrams we chose to start with the top left configuration. The first equality follows from (the properties illustrated in) fig. \ref{fig:PropRelations}, the second from fig. \ref{fig:VertexInv}.}
\label{fig:addVertex}
\end{center}
\end{figure}

\begin{figure}[h]
\begin{center}
\includegraphics[width=0.9\textwidth]{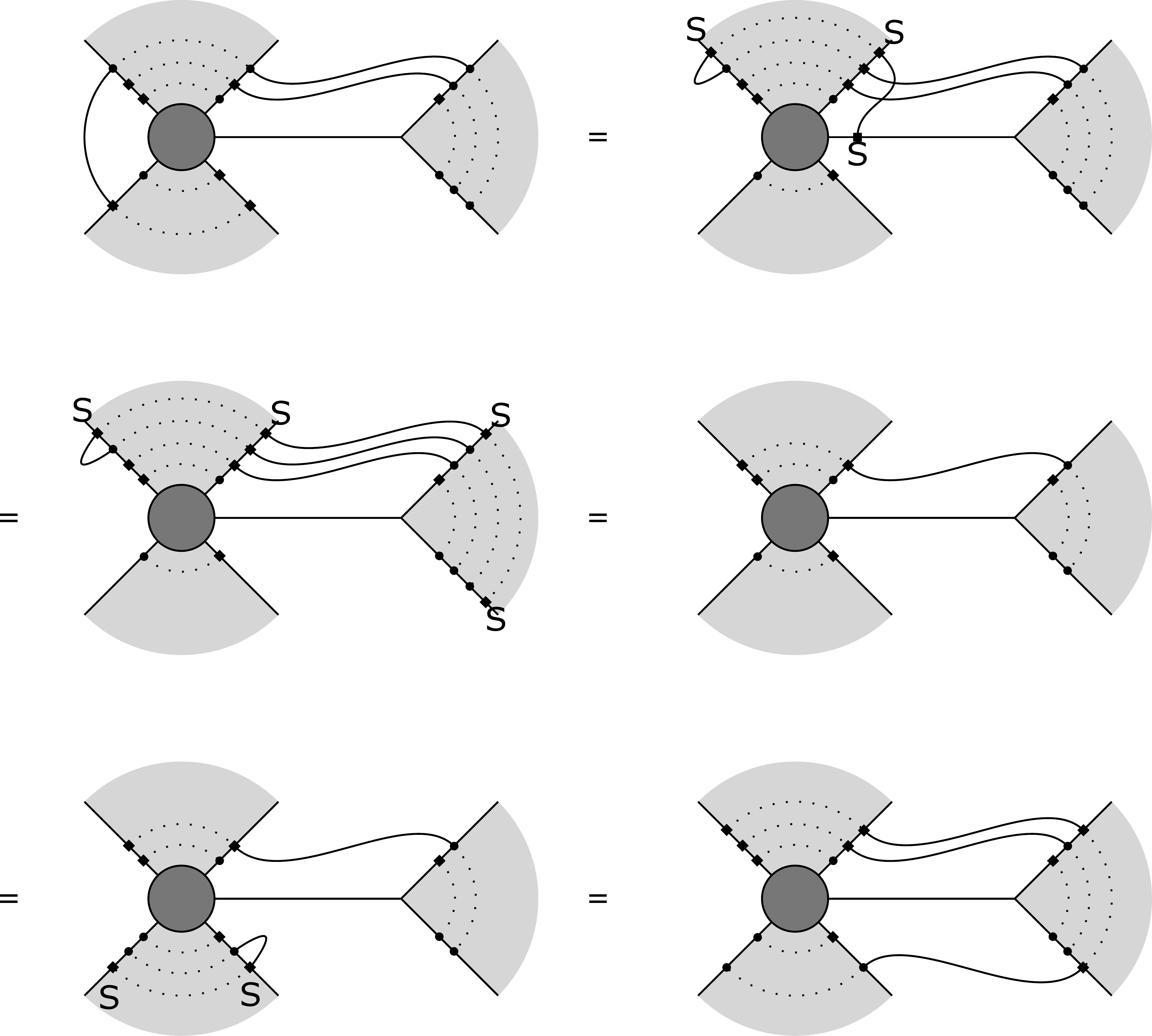}
\caption{Proving cyclicity of the external star products after adding a new vertex. Starting from the last diagram in fig. \ref{fig:addVertex}, we use the cocycle condition of fig. \ref{fig:cocycle} to consider the top left configuration of twists. We then use the Poincar\'e invariance relation in the first step, and eqn. \eqref{eq:TwistOnProp} and the Poincar\'e invariance relation again in the second step. For the third and fourth steps we use $S(\bar{f}^\alpha)\bar{f}_\alpha=1$.}
\label{fig:addVertexCyc}
\end{center}
\end{figure}

\paragraph*{Adding a vertex} Consider a given diagram $a^\star_n$ and a set of basic interaction vertices, all having the planar equivalence properties. We want to extend our diagram $a^\star_n$ by successively connecting the new vertices to it, connecting each via only one leg. Since, by assumption, $a^\star_n$ and the vertices have the cyclicity property \eqref{eq:ampAss2} we can connect $a^\star_n$ and a vertex via their first legs, without loss of generality, as in the left diagram of fig. \ref{fig:addVertex}. The remainder of this figure shows how the resulting diagram has the required external star product structure \eqref{eq:ampAss1}.\\
Next, we can prove that the star product structure of the resulting diagram is cyclic, following the steps of fig. \ref{fig:addVertexCyc}. By similar considerations, the Poincar\'e invariance relation follows as well.
This proves the planar equivalence theorem for tree diagrams.

\begin{figure}[h]
\begin{center}
\includegraphics[width=0.7\textwidth]{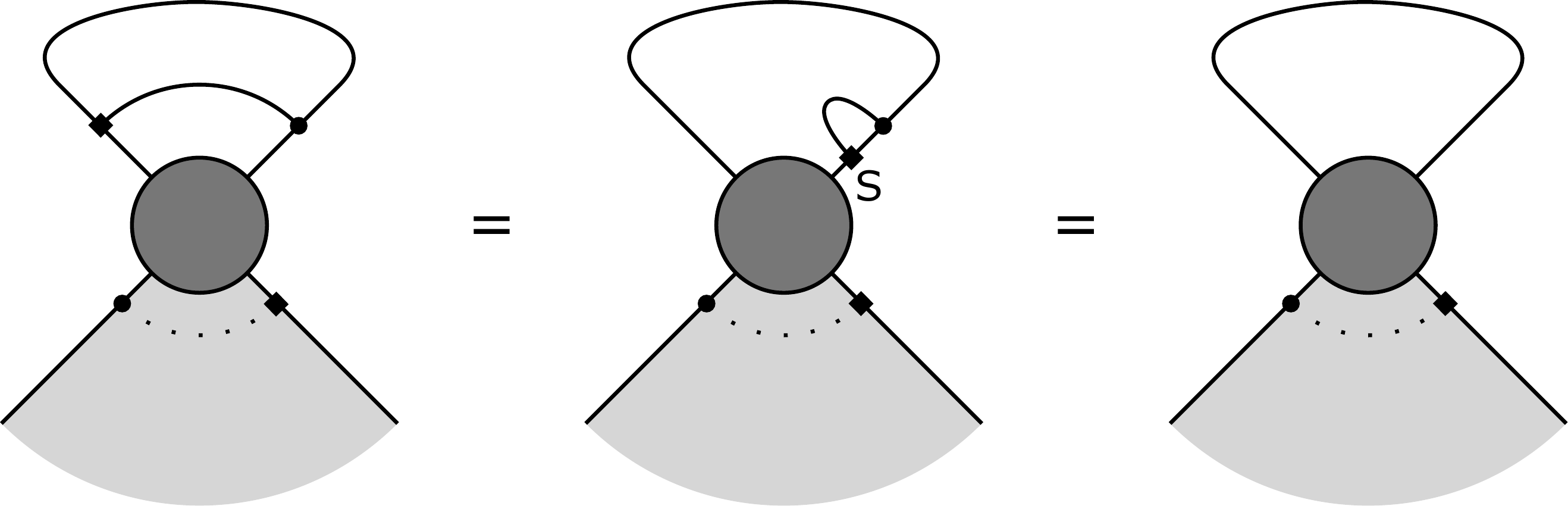}
\caption{Generating closed planar loops in a general planar diagram by connecting two neighboring legs with each other. Due to cyclicity of the external star products, we can start from the left configuration. The right diagram is of the desired structure of fig. \ref{fig:Filk1} again.}
\label{fig:closeLoop}
\end{center}
\end{figure}

\begin{figure}[h]
\begin{center}
\includegraphics[width=0.9\textwidth]{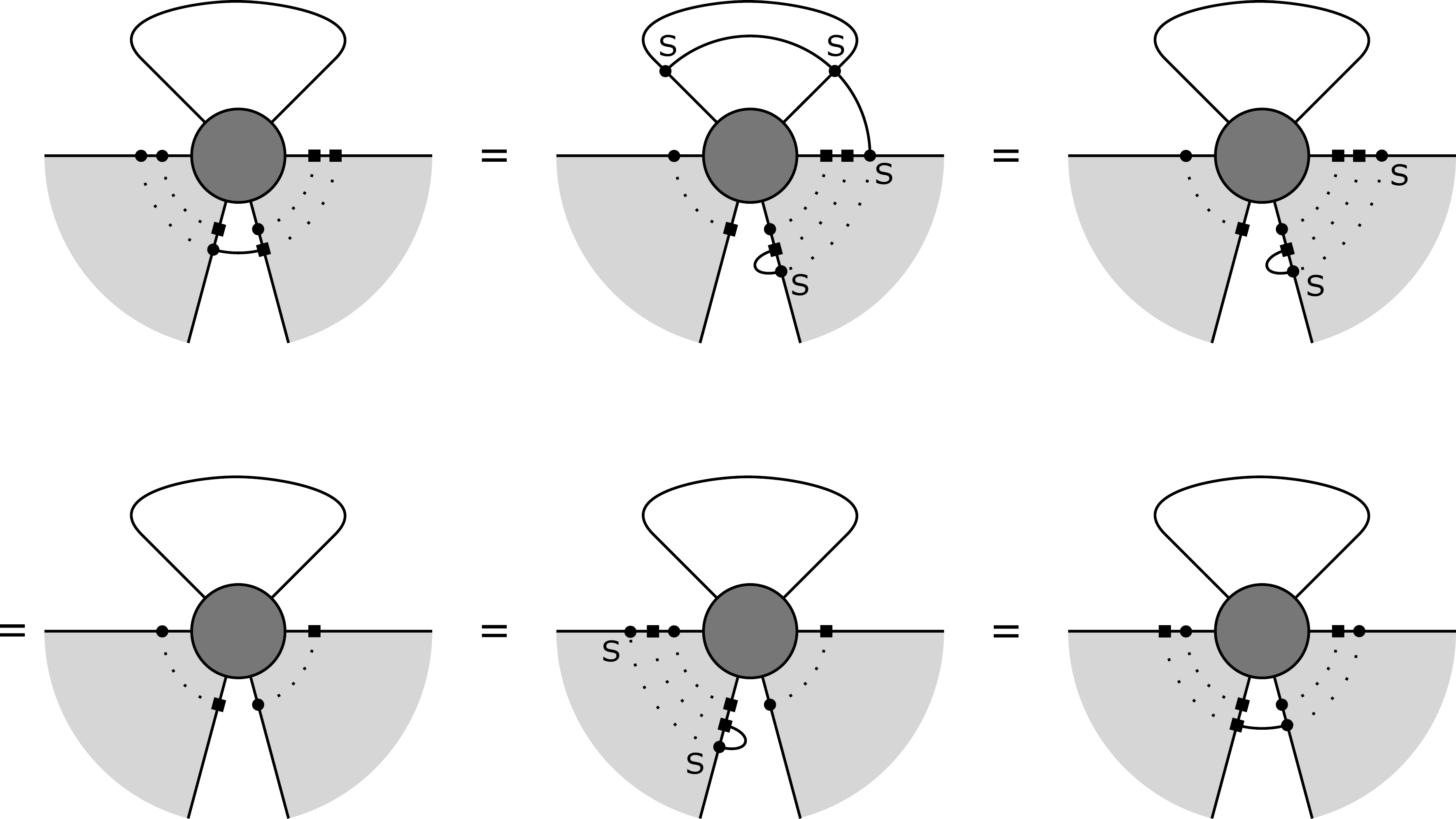}
\caption{Proving the cyclicity condition for diagrams with closed loops. In the first equality, we use the Poincar\'e invariance relation for the formerly uncontracted diagram. In the second step we use eqn. \eqref{eq:TwistOnProp3}, while in the third and fourth equality we use $S(\bar{f}^\alpha)\bar{f}_\alpha=1$.}
\label{fig:closeLoopCyc}
\end{center}
\end{figure}

\paragraph*{Closing loops} To describe arbitrary planar diagrams we need to be able to close loops in tree diagrams. I.e. we want to connect external lines of the diagram in fig. \ref{fig:Filk1} with each other. Restricting ourselves to planar diagrams, however, means that we only need to connect neighboring fields. By the cyclicity property of the original diagram, the resulting diagram automatically has the desired star product structure, as shown in fig. \ref{fig:closeLoop}.
The cyclicity condition for the resulting diagram now follows by the steps indicated in fig. \ref{fig:closeLoopCyc}.
The Poincar\'e invariance condition follows similarly. In summary, all planar diagrams have the desired star product structure of fig. \ref{fig:Filk1} and fulfill the cyclicity and  Poincar\'e invariance relations of fig. \ref{fig:Filk2}. To get the final structure of proper Feynman diagrams, we will now contract them with external fields.

\subsubsection*{Contraction with external fields}

To get proper planar Feynman diagrams, we take a given $a^\star_n$ and perform the missing Wick contractions with the external fields $\phi(y_i)$. This leads to
\begin{equation}
\label{eq:diagramContract}
\begin{aligned}
A_n^\star\left(y_1,\dots,y_n\right)&=I_{n}\times \underset{i=1}{\overset{n}{\bigstar}_x}\Delta_\phi(x_i-y_i)\\
&=I_{n}\times \left[\mathcal{\bar{F}}_{x_1x_2}\Delta_\phi(x_1-y_1)\Delta_\phi(x_2-y_2)\right]\star_x\left(\underset{i=3}{\overset{n}{\bigstar}_x}\Delta_\phi(x_i-y_i)\right)\\
&=I_{n}\times \left[\left(S\otimes S\right)\left(\mathcal{\bar{F}}\right)_{y_1y_2}\Delta_\phi(x_1-y_1)\Delta_\phi(x_2-y_2)\right]\star_x\left(\underset{i=3}{\overset{n}{\bigstar}_x}\Delta_\phi(x_i-y_i)\right),
\end{aligned}
\end{equation}
where we used associativity of the star product in the second line and eqn. \eqref{eq:TwistOnProp} in the third. For the remaining twists still acting on internal positions,we follow the same idea to map them to the external positions, using the commutativity of the antipode and the coproduct, $\left(S\otimes S\right)\circ\Delta=\Delta\circ S$, of our cocommutative Hopf algebra. This results in
\begin{equation}
A_n^\star\left(y_1,\dots,y_n\right)=I_{n}\times \underset{i=1}{\overset{n-1}{\overrightarrow{\prod}}}\left[\left(\Delta^{(i)} \otimes 1\right) \left(S\otimes S\right)\left(\mathcal{F}\right)\right]_{y_1\dots y_{i+1}}\prod_{j=1}^n\Delta_\phi(x_j-y_j).
\end{equation}
Finally, for our class of twists we have $\left(S\otimes S\right)(\mathcal{\bar{F}})=\mathcal{F}_{op}$,\footnote{This follows immediately for abelian twists, where $\left(S\otimes S\right)(\mathcal{\bar{F}})=\mathcal{\bar{F}}=\mathcal{F}_{op}$. For almost abelian twists, e.g. a rank four twist $\mathcal{\bar{F}}=\bar{\hat{\mathcal{F}}}\bar{\tilde{\mathcal{F}}}$ with $\bar{\tilde{\mathcal{F}}}$ and $\bar{\hat{\mathcal{F}}}$ abelian, the antimultiplicativity of the antipode leads to
\begin{equation*}
\left(S\otimes S\right)(\bar{\mathcal{F}})=\left(S\otimes S\right)(\bar{\tilde{\mathcal{F}}})\left(S\otimes S\right)(\bar{\hat{\mathcal{F}}})=\tilde{\mathcal{F}}_{op}\hat{\mathcal{F}}_{op}=\mathcal{F}_{op},
\end{equation*}
which naturally extends to higher rank twists. For the twists for $r_{13}$ and $r_{14}$, the same follows since they are algebraically identical, and the one for $r_{13}$ admits an almost-abelian factorization. It can also be directly verified using the explicit form given in Appendix \ref{app:newtwist}.} meaning we can express our deformed planar diagrams via undeformed ones with opposite twists between the external positions,
\begin{equation}
A_n^\star\left(y_1,\dots,y_n\right)=I_{n}\times \underset{i=1}{\overset{n-1}{\overrightarrow{\prod}}}\left[\left(\Delta^{(i)} \otimes 1\right)\left(\mathcal{F}_{op}\right)\right]_{y_1\dots y_{i+1}}\prod_{j=1}^n\Delta_\phi(x_j-y_j).
\end{equation}
For abelian deformations, in particular the Groenewold-Moyal deformation, the opposite twists reduce to star products between the external positions, so that the above coincides with Filk's original planar equivalence theorem \cite{Filk:1996dm}. Our planar equivalence theorem generalizes this to all twists of the Poincar\'e algebra, that give to cyclic star products under integration.

\subsection{Generalization to gauge and fermionic fields}

Our proof of the planar equivalence theorem for scalar fields is based on two properties: the cyclicity of the fields in the vertices, and the invariance of the propagators under the generators appearing in the twist function, i.e. their Poincar\'e invariance. However, the last property in particular, is unique for scalar fields. The massless fermionic propagator for example,
\begin{align*}
\Delta_{\alpha\dot{\alpha}}(x-y)&=\frac{(x-y)_\mu}{(x-y)^4}\sigma^\mu_{\alpha\dot{\alpha}}
\end{align*}
is not Poincar\'e invariant on its own. A further complication is that, written in component fields, interactions which include non-scalar fields contain several R-matrices with indices, which would potentially act on external fields as well. Fortunately, both these complications disappear if we use the index free notation introduced earlier. In index-free form, the interactions are (graded) cyclic and consequently behave similar to the pure scalar case, i.e. they fulfill a graded version of the cyclicity relation \eqref{eq:VertInv} and the Poincar\'e invariance relation \eqref{eq:VertInv2}, with the grading of course identical to that of the corresponding undeformed vertices. Moreover, in index-free notation, the natural propagators of the gauge field and fermionic fields are given by
\begin{align*}
\Delta_A(x-y)=\Braket{A(x)A(y)}&=\int\mathrm{d}^4p~\frac{\eta_{\mu\nu}}{p^2}e^{-ip(x-y)}\form{x^\mu}\form{y^\nu}\\
\Delta_\Psi(x-y)=\Braket{\bar{\Psi}(x)\Psi(y)}&=\int\mathrm{d}^4p \frac{-ip_\mu \Gamma^\mu+M}{p^2+m^2+i\epsilon}e^{-ip(x-y)}
\end{align*}
where
\begin{equation*}
\begin{aligned}
\Psi&=\begin{pmatrix}
\psi_L^\alpha s_\alpha\\ \left.\bar{\psi}_R\right._{\dot{\alpha}} \bar{s}^{\dot{\alpha}}
\end{pmatrix}
\end{aligned}
\end{equation*}
is a Dirac spinor field written in index free notation with the left-handed two spinor $\psi_L$ and the right-handed two spinor $\psi_R$. Here we defined
\begin{equation*}
\begin{aligned}
\Gamma^\mu&\equiv\begin{pmatrix}
0 & \sigma^\mu_{\alpha\dot{\alpha}}s_x^\alpha \bar{s}_y^{\dot{\alpha}}\\
\sigma^{\mu\dot{\alpha}\alpha}\bar{s}_{x\dot{\alpha}} s_{y\alpha}
\end{pmatrix}\\
M&\equiv m\begin{pmatrix}
s_x^\alpha s_{y\alpha}&0\\
0&\bar{s}_{x\dot{\alpha}}\bar{s}_y^{\dot{\alpha}}
\end{pmatrix}.
\end{aligned}
\end{equation*}
For a massless fermionic field, the two components decouple and we have the massless propagator for a two-component spinor field
\begin{equation*}
\begin{aligned}
\Delta_{\psi}(x-y)&=\Braket{\bar{\psi}(x)\psi(y)}&=\frac{(x-y)^\mu}{(x-y)^4}\sigma_\mu^{\dot{\alpha}\alpha}\bar{s}_{x\dot{\alpha}} s_{y\alpha}.
\end{aligned}
\end{equation*}
These propagators are obtained by contracting the component field propagators by the appropriate basis forms or spinors for the fields determining the propagator. By construction, these propagators are Poincar\'e invariant, and hence have the properties \eqref{eq:TwistOnProp} and \eqref{eq:TwistOnProp2}.

Since both the vertices and propagators now behave as their scalar counterparts, the proof of section \ref{sec:scalarFilk} directly generalizes to diagrams written in index-free notation but containing all types of fields. In other words, the planar equivalence theorem also applies to any gauge theory including matter in the fundamental or adjoint representation, allowing us to relate all planar Feynman diagrams to their undeformed counterparts with additional opposite twists acting between the external fields. It is important to emphasize that the planar equivalence theorem only applies to index-free diagrams. In components, the twists appearing in the planar equivalence theorem explicitly mix the component diagrams\footnote{Proper (component) Feynman diagrams can be understood and read off, as the components with respect to a basis of forms and spinors for the external legs.} involving any external gauge fields or spinors.

\section{Twisted symmetry}
\label{sec:twistedsymmetry}

Noncommutative field theories do not have conventional Poincar\'e symmetry. For example, the constant matrix $\theta$ for the Groenewold-Moyal deformation picks out fixed directions in spacetime, clearly breaking Lorentz invariance. For the Groenewold-Moyal deformation, Poincar\'e symmetry is not completely gone however, but instead is realized in a twisted, nonlocal fashion \cite{Chaichian:2004za,Wess:2003da,Chaichian:2004yh}. The same applies quite generally to other noncommutative deformations based on Drinfel'd twists, and to our gauge theories. Let us recall the concept of twisted symmetries using the simple example of the Groenewold-Moyal deformation of $\phi^3$ theory, and then discuss a general framework that demonstrates that our noncommutative gauge theories all have twisted Poincar\'e symmetry, and even twisted conformal or supersymmetry in appropriate cases.

\subsection{Groenewold-Moyal deformed scalar fields}

To recall how twisted symmetry arises, consider $\phi^3$ theory, with undeformed action
\begin{equation}
S_{\phi^3} = \int \tfrac{1}{2} \partial_\mu \phi \partial^\mu \phi - \tfrac{m}{2} \phi^2 - g \phi^3  \mathrm{d}^4 x,
\end{equation}
This action is Poincar\'e invariant, with the Poincar\'e algebra acting via Lie derivatives on individual fields. For instance, under a rotation in the $(1,3)$ plane, its interaction term transforms as
\begin{equation}
\delta_{M_{13}} S_{\mbox{\tiny int}} \sim \int \left((\mathcal{L}_{M_{13}} \phi) \phi^2 + \phi (\mathcal{L}_{M_{13}} \phi) \phi + \phi^2 (\mathcal{L}_{M_{13}} \phi) \right)  \mathrm{d}^4 x = \int \mathcal{L}_{M_{13}}(\phi^3) \mathrm{d}^4 x \approx 0,
\end{equation}
where we collected terms by the product rule to get a total derivative, giving a symmetry of the action. If we now consider the Groenewold-Moyal deformation with, say, $\theta^{12} = \theta$, and zero otherwise, vary the individual fields as before, and consider the variation of its interaction term $\phi \star \phi \star \phi$, we find
\begin{equation}
\delta_{M_{13}} S^\star_{\mbox{\tiny int}} \sim \int \left((\mathcal{L}_{M_{13}}\phi) \star \phi \star \phi+ \phi \star (\mathcal{L}_{M_{13}}\phi) \star \phi + \phi \star \phi \star (\mathcal{L}_{M_{13}}\phi)\right) \mathrm{d}^4 x.
\end{equation}
This is no longer a total derivative since the star product prevents us from collecting the terms into a single overall Lie derivative. As a result, this deformation breaks Poincar\'e invariance to the symmetry generated by the subset of generators that commute with the twist, in this case $M_{03},M_{12}$ and the $p_\mu$.\footnote{These are, of course, also the symmetries of the corresponding $r$ matrix in the sense of section \ref{subsec:unimodularPoincarermatrices}.}

If the Lie derivative instead acted ``after'' taking the star product, in a suitable sense, in the above variation, there would be no problem in showing invariance of the action. To formalize this idea, consider that in the commutative setting the local action of $\xi\in \mathcal{P}$ on individual fields can be written via coproducts as
\begin{equation}
\delta_\xi (\phi^n) = \mu(\Delta^{(n)}(\xi)(\phi,\ldots,\phi)),
\end{equation}
Because of the product rule we have
\begin{equation}
\mu(\Delta^{(n)}(\xi)(\phi,\ldots,\phi)) = \xi(\phi^n),
\end{equation}
meaning
\begin{equation}
\delta_\xi (\phi^n) = \mu(\Delta^{(n)}(\xi)(\phi, \ldots , \phi))= \xi(\phi^n),
\end{equation}
which we used above to find invariance for the interaction term in the action. In our twisted setting we instead deal with star products, which have a twisted product rule of eqn. \eqref{eq:twistedprodrule},
\begin{equation}
\xi (f\star g) =\mu_\mathcal{F}\left(\Delta_\mathcal{F}(\xi)(f, g)\right).
\end{equation}
This means that if we let the Poincar\'e algebra act in a twisted fashion on star products of fields as
\begin{equation}
\delta^\star_\xi (\phi^{\star n}) \equiv \mu_\mathcal{F} ( \Delta_\mathcal{F}^{(n)}(\xi) (\phi,\ldots,\phi)),
\end{equation}
we have
\begin{equation}
\delta^\star_\xi (\phi^{\star n}) = \xi (\phi^{\star n}),
\end{equation}
which gives a total derivative under an integral, and hence invariance of the corresponding term in the action. As an example of the nonlocality of this action of the Poincar\'e algebra, for instance
\begin{equation}
\delta^\star_{M_{13}}(\phi \star \phi) = (\mathcal{L}_{M_{13}}\phi) \star \phi +   \phi  \star(\mathcal{L}_{M_{13}}\phi) +  \theta (p_{[3} \phi) \star (p_{2]} \phi).
\end{equation}
where square brackets indicate antisymmetrization, with a factor of $1/2$. It is this twisted action of the Poincar\'e algebra that is a symmetry of our example noncommutative $\phi^3$ theory.

\subsection{Twisted Poincar\'e symmetry}

The considerations above directly extend to any twist-noncommutative field theory, provided that all products between fields are star products, and that the local operations on single fields are compatible with the action of the symmetry generators. What we mean by the latter is perhaps best illustrated by example. All deformations except the Groenewold-Moyal one, deform the differential calculus involving forms, making it natural to work in index-free notation. In this language, Poincar\'e symmetry of the action of e.g. a free (undeformed) scalar field
\begin{equation}
\int \mathrm{d} \phi \wedge *\mathrm{d}\phi,
\end{equation}
follows as for the interaction term discussed earlier, provided $\mathrm{d}\phi$ and $*\mathrm{d} \phi$ transform as $\phi$, i.e. if we have
\begin{equation}
[\mathcal{L}_\xi,\mathrm{d}] \phi = 0,
\end{equation}
and
\begin{equation}
[\mathcal{L}_\xi,*]\mathrm{d} \phi = 0.
\end{equation}
The first of these is just the well-known commutativity of the Lie and exterior derivative which holds for any vector field $\xi$. The second, however, does not hold in general, but does for $\xi$ in the Poincar\'e algebra.

In our twisted context, the exterior derivative is undeformed and hence continues to commute with the Lie derivative, but the commutativity of the twisted Hodge star with the action of the Poincar\'e algebra is a nontrivial requirement in principle. Fortunately, as discussed in section \ref{sec:Hodgeduality}, it continues to hold for our twisted Hodge star.

This means that for our class of twists, any action constructed out of star products of fields, their exterior derivatives, and twisted Hodge duals -- in particular our Yang-Mills action -- has twisted Poincar\'e symmetry. The detailed form of this twisted Hopf algebra of course depends on the deformation under consideration, as discussed in detail in \cite{Lukierski:2005fc} for the Lorentz twist for example.

For theories with spacetime symmetries beyond the Poincar\'e algebra, i.e. conformal or supersymmetric theories, we get further requirements if we want the full symmetry algebra to be twisted.

\subsection{Twisted conformal symmetry}

For conformal theories, we need the action of the dilatation and special conformal generators on exterior derivatives and Hodge duals of individual fields, to be isomorphic before and after deformation. If this is the case, twisted conformal symmetry follows as above, since the twisted coproduct essentially reduces the question of symmetry to the same question for the undeformed action. For the Hodge star this is again a nontrivial requirement. For the dilatation generator, $D$, in the untwisted setting we have
\begin{equation}
[\mathcal{L}_D,*] \omega = (4-2p)i * \omega,
\end{equation}
for a $p$ form $\omega$. This extends directly to the twisted setting, as we still have
\begin{equation}
[\mathcal{L}_D,*] \mathrm{d}x = (4-2p)i * \mathrm{d}x,
\end{equation}
where $\mathrm{d}x$ again schematically denotes a basis star $p$ form, since the twist does not change the degree of forms. The same then applies for arbitrary star forms, by regular linearity of our twisted Hodge star, and the conventional product rule for the Lie derivative. While less obvious, we have verified that also for the special conformal generators $k_\mu$, $[\mathcal{L}_{k_\mu},*]$ has the same action on basis star forms as it has on regular basis forms in the undeformed setting, so that our twisted Hodge star is compatible with the action of the conformal algebra on individual fields. As a result, our twist-noncommutative deformation of a conformal theory, such as for instance maximally supersymmetric Yang-Mills theory discussed below, will have twisted conformal symmetry. At the classical level the same applies to pure Yang-Mills theory, but there we know this will inevitably be broken at the quantum level.

\subsection{Twisted supersymmetry}

The final possible extension of Poincar\'e symmetry affected by spacetime twists is supersymmetry. In contrast to Poincar\'e symmetry, for supercharges we are dealing with field variations producing other fields, and various nontrivial terms canceling against one another, rather than the individual variations combining into a single total derivative.\footnote{It would be interesting to investigate twisted superspace in detail, where we would find total derivatives. There is no superspace formulation manifesting all supersymmetry of SYM, however, which is presently our main aim.} In this case the coproduct does not formalize a product rule that we can use to show invariance of the action, instead it merely indicates how the variation is actually implemented. To show invariance of a twist-noncommutative action under twisted supersymmetry then, requires more work.

We could compute the effect of the twisted coproduct of supercharges on the component action, and attempt to show that this is a symmetry by brute force. Fortunately this turns out not to be necessary, since we can treat supersymmetry in a suitable index-free fashion, whereby twisted supersymmetry of a deformed action follows straightforwardly from regular supersymmetry of its undeformed counterpart. Our strategy will be to show that, for our class of twists, the coproduct of index-free supercharges acts on products of fields in a way that is formally unchanged by twisting, so that by introducing suitable star products in our field variations, the entire variation is formally unchanged, and any algebraic cancellations are immediate. This then leaves cancellations involving the kinetic terms, which as we shall see, still work out for our twisted differential structure. We will discuss this using the simple example of the free Wess-Zumino model as illustration.

To set the stage, let us recall the standard supersymmetry of the undeformed free Wess-Zumino model in components. The action of the undeformed free Wess-Zumino model is given by
\begin{equation}
\label{eq:WZMcomp}
S_{WZ} =\int \mathrm{d}^4x~ \partial_\mu\bar{\phi}\partial^\mu\phi-i\bar{\psi}_{\dot{\alpha}} \sigma^{\mu\alpha\dot{\alpha}} \partial_\mu \psi_\alpha.
\end{equation}
Under the supersymmetry transformations
\begin{equation}
\label{eq:basicsusyfieldvariations}
\begin{aligned}
\delta_{\varepsilon,\bar{\varepsilon}}\phi&=\varepsilon^\alpha \psi_\alpha &&& \delta_{\varepsilon,\bar{\varepsilon}} \bar{\phi} &= \bar{\psi}_{\dot{\alpha}} \bar{\varepsilon}^{\dot{\alpha}}\\
\delta_{\varepsilon,\bar{\varepsilon}}\psi_\alpha &= i\sigma^\mu_{\alpha\dot{\alpha}}\bar{\varepsilon}^{\dot{\alpha}} \partial_\mu \phi &&& \delta_{\varepsilon,\bar{\varepsilon}} \bar{\psi}_{\dot{\alpha}} &= -i\varepsilon^{\alpha} \sigma^\mu_{\alpha\dot{\alpha}} \partial_\mu \bar{\phi},
\end{aligned}
\end{equation}
its variation, upon integration by parts, is given by
\begin{equation}
\label{eq:SuSyWZM}
\begin{aligned}
\delta S_{WZ} = &
\int\mathrm{d}^4x~ \frac{1}{2} \bar{\psi}_{\dot{\alpha}} \left(\sigma^{\mu\dot{\alpha}\alpha} \sigma^\nu_{\alpha\dot{\beta}} + \sigma^{\mu\dot{\alpha}\alpha} \sigma^\nu_{\alpha\dot{\beta}} \right) \bar{\varepsilon}^{\dot{\beta}} \partial_\mu \partial_\nu \phi - \bar{\varepsilon}_{\dot{\alpha}} \bar{\psi}^{\dot{\alpha}} \partial_\mu \partial^\mu \phi\\
&\quad +
\int\mathrm{d}^4x~ \frac{1}{2}\varepsilon^\alpha \left( \sigma^\mu_{\alpha\dot{\alpha}} \sigma^{\nu\dot{\alpha}\beta} + \sigma^\nu_{\alpha\dot{\alpha}} \sigma^{\mu\dot{\alpha}\beta} \right) \psi_\beta  \partial_\mu \partial_\nu \bar{\phi} - \varepsilon^\alpha \psi_\alpha \partial_\mu \partial^\mu \bar{\phi}.
\end{aligned}
\end{equation}
This vanishes because of the standard Pauli matrix property,
\begin{equation}
\label{eq:CliffAlgindex}
\sigma^\mu_{\alpha\dot{\alpha}} \sigma^{\nu\dot{\alpha}\beta} + \sigma^\nu_{\alpha\dot{\alpha}} \sigma^{\mu\dot{\alpha}\beta} = 2\eta^{\mu\nu} \delta_\alpha^\beta,
\end{equation}
and similarly for the other index contraction, giving supersymmetry of the action \eqref{eq:WZMcomp}.

As a first step towards twisting this, let us discuss it again, in index-free notation. The action becomes
\begin{equation}
S_{WZ} = \int \mathrm{d}\bar{\phi}\wedge*\mathrm{d}\phi - i\int \mathrm{d}^2s\mathrm{d}^2\bar{s} ~ \int ~\bar{\psi}\sigma\wedge*\mathrm{d}\psi,
\end{equation}
using our Grassmann even spinors $\bar{\psi}=\bar{\psi}_{\dot{\alpha}} \bar{s}^{\dot{\alpha}}$ and $\psi = s^\alpha \psi_\alpha$. We now define index-free supercharges $Q$ and $\bar{Q}$ that act on the covariant fields $\phi$, $\psi$ and $\bar{\psi}$ as
\begin{equation}
\label{eq:freeWZvariationsindexfree}
\begin{aligned}
Q\phi &= \psi &&& \bar{Q} \bar{\phi} &= \bar{\psi}\\
Q\bar{\psi} &= -i *\left( \sigma \wedge  *\mathrm{d}\bar{\phi} \right) &&& \bar{Q} \psi &= i *\left( \sigma \wedge *\mathrm{d}\phi \right),
\end{aligned}
\end{equation}
These reproduce the variations \eqref{eq:basicsusyfieldvariations} above, via
\begin{equation}
\delta_{\varepsilon,\bar{\varepsilon}}(\cdot)=\int\mathrm{d}^2s~ \varepsilon Q~(\cdot) + \int\mathrm{d}^2\bar{s}~ \bar{Q} \bar{\varepsilon}~(\cdot),
\end{equation}
where
\begin{equation}
\varepsilon = \varepsilon^\alpha s_\alpha, \qquad \bar{\varepsilon} = \bar{\varepsilon}_{\dot{\alpha}} \bar{s}^{\dot{\alpha}}
\end{equation}
and we can write
\begin{equation}
Q = s^\alpha Q_\alpha, \qquad \bar{Q}=\bar{Q}_{\dot{\alpha}} \bar{s}^{\dot{\alpha}},
\end{equation}
so that
\begin{equation}
\int \mathrm{d}^2 s \,\epsilon\, Q = \epsilon^\alpha Q_\alpha, \qquad \int \mathrm{d}^2 \bar{s} \, \bar{Q}\,\bar{\epsilon} = \bar{\epsilon}^{\dot{\alpha}} \bar{Q}_{\dot{\alpha}}.
\end{equation}
Finally, property \eqref{eq:CliffAlgindex} can be nicely recast as
\begin{equation}
\label{eq:CliffAlg}
\int \mathrm{d}^2 \bar{s} (\sigma^\mu\sigma'^\nu + \sigma^\nu \sigma'^\mu) = 2 \eta^{\mu\nu} s_\alpha s'^\alpha 
\end{equation}
where we use primes to distinguish the various Grassmann variables, i.e. $\sigma'=\sigma'_\mu \mathrm{d}x^\mu  = \sigma_{\mu\alpha\dot{\alpha}}s'^\alpha \bar{s}^{\dot{\alpha}}\mathrm{d}x^\mu$, and there is a similar relation with an integral over $s$ instead. The variation of the action now becomes
\begin{equation}
\begin{aligned}
\delta S_{WZ} =&
\int \mathrm{d}^2s\int \varepsilon \,\mathrm{d} \bar{\phi} \wedge * \mathrm{d} \psi + \int\mathrm{d}^2s\mathrm{d}^2\bar{s}\mathrm{d}^2s'\int \varepsilon' *(\sigma' \wedge *\mathrm{d} \bar{\phi})(\sigma \wedge *\mathrm{d}\psi),\\
\end{aligned}
\end{equation}
where $\varepsilon'=\varepsilon^\alpha s_{\alpha}'$, plus a similar term proportional to $\bar{\varepsilon}$. Integrating by parts and flipping the Hodge dual we find
\begin{equation}
\label{eq:SWZundefvariation2}
\begin{aligned}
\delta S_{WZ} =&
\int \mathrm{d}^2s\int  \varepsilon \left( \mathrm{d}*\mathrm{d}\bar{\phi} \right) \psi
+
\int\mathrm{d}^2s\mathrm{d}^2\bar{s}\mathrm{d}^2s' \int\varepsilon' \left(\sigma \wedge *\mathrm{d}* \left(\sigma' \wedge *\mathrm{d}\bar{\phi} \right)\right) \psi ,
\end{aligned}
\end{equation}
Now, for the undeformed Hodge star we have
\begin{equation}
\label{eq:sigmawedgehodgeexterior}
*(\sigma \wedge  *\mathrm{d}\phi) =  \sigma^\mu \partial_\mu \phi,
\end{equation}
which, applied to the second term, gives
\begin{equation}
\begin{aligned}
\int\mathrm{d}^2s\mathrm{d}^2\bar{s}\mathrm{d}^2s'\int \varepsilon' \sigma \wedge *\mathrm{d}* \left(\sigma' \wedge *\mathrm{d}\bar{\phi} \right) \psi = & \int \mathrm{d}^2s\mathrm{d}^2\bar{s}\mathrm{d}^2s'\int \varepsilon' \left(*\sigma^\mu \sigma'^\nu \partial_\mu \partial_\nu \bar{\phi}\right) \psi\\
= & \int\mathrm{d}^2s\mathrm{d}^2s'\int \varepsilon' s_\alpha s'^\alpha \left(*\partial^\mu \partial_\mu \bar{\phi}\right) \psi\\
= & - \int \mathrm{d}^2s\int \varepsilon (\mathrm{d}*\mathrm{d} \bar{\phi}) \psi,
\end{aligned}
\end{equation}
where the last equality uses the definition of the Laplacian, as discussed in section \eqref{subsec:Laplaciandiscussion}. This clearly cancels against the first term, establishing invariance.

%

Now we can consider the twist-noncommutative free Wess-Zumino action
\begin{equation}
\label{eq:WZMdef}
\int \mathrm{d}\bar{\phi}\wedge_\star*\mathrm{d}\phi -i \int \mathrm{d}^2s\mathrm{d}^2\bar{s} ~ \bar{\psi}\star\sigma\wedge_\star*\mathrm{d}\psi.
\end{equation}
We expect this action to be invariant under twisted supersymmetry, acting on products of fields via the twisted coproduct.\footnote{Strictly speaking, in components this quadratic action can be reduced the undeformed one, which of course has regular supersymmetry. We prefer to not clutter our discussion by adding interaction terms. Due to coassociativity, quadratic terms suffice to illustrate the general argument.} Focussing on the $\varepsilon^\alpha$ variation, for the product two fields we then should have
\begin{equation}
\delta_{\epsilon^\alpha Q_\alpha} (f\star g) = \mu_\mathcal{F}(\Delta_{\mathcal{F}}(\epsilon^\alpha Q_\alpha)(f,g)),
\end{equation}
where, by abuse of notation, we use $\Delta_{\mathcal{F}}(\epsilon^\alpha Q_\alpha)$ as a shorthand for $\Delta_{\mathcal{F}}(Q_\alpha)$ suitably contracted with $\epsilon^\alpha$, keeping track of Grassmann signs,\footnote{The position of $\epsilon$ is important when acting on fermionic terms.} i.e.
\begin{equation}
\Delta_{\mathcal{F}}(\epsilon^\alpha Q_\alpha) = \mathcal{F} \left(\epsilon^\alpha Q_\alpha \otimes 1 + 1 \otimes \epsilon^\alpha Q_\alpha \right) \mathcal{F}^{-1}.
\end{equation}
As discussed in appendix \ref{app:twistedsusycoproduct}, the twisted coproduct can be written as
\begin{equation}
\Delta_{\mathcal{F}}(Q_\alpha) = Q_\beta \otimes F^\beta{}_\alpha + (F_{op})^\beta{}_\alpha \otimes Q_\beta.
\end{equation}
where $F^\beta{}_\alpha $ and $(F_{op})^\beta{}_\alpha$ are (inverses of) the $F$s appearing in eqs. \eqref{eq:Fwithspinorindices}. We want to rephrase this in an index-free fashion, in terms of $Q$. Despite appearances, $Q$ has a nontrivial twisted coproduct, since $s^\alpha$ is not an algebra element. At the same time, it is a Lorentz scalar if we transform it via the Lie bracket and the Lie derivative simultaneously, which has convenient consequences. Namely, as discussed in appendix \ref{app:twistedsusycoproduct},  the twisted action of $Q_\alpha$ on a star product, effectively appears untwisted in terms of $Q$, meaning
\begin{equation}
\delta_Q (f \star g) = \mu_{\mathcal{F}}(\Delta_\mathcal{F}(Q) (f,g)) = Q(f) \star g + f \star Q(g),
\end{equation}
where $\Delta_\mathcal{F}(Q)=\Delta_\mathcal{F}(s^\alpha Q_\alpha)$ is defined by the same abuse of notation as $\Delta_\mathcal{F}(\epsilon^\alpha Q_\alpha)$. In terms of $Q$, supercharges act on the various products of fields in the action exactly as in the commutative setting, just replacing products with star products. Put differently, the action of $Q$ on single particle states (effectively) commutes with taking star products, just as we saw earlier for our Hodge star and exterior derivative. To reintroduce $\epsilon^\alpha$ and get a supersymmetry variation by $\epsilon^\alpha Q_\alpha$, we simply act with
\begin{equation}
\int d^2 \tilde{s} \,\epsilon \,(\cdot).
\end{equation}
Since it is Grassmann even, we can place this integral operator all the way to the left in any variation -- outside any star products -- as in the undeformed setting. A similar discussion applies to $\bar{Q}$.

At this stage, the supersymmetry variation of a twist-noncommutative action is identical in form to that of its commutative counterpart, only the product differs, also when including interaction terms. To complete this picture at the level of the individual field variations, we note that we can freely add star products in eqs. \eqref{eq:freeWZvariationsindexfree}, to get
\begin{equation}
\label{eq:freeWZvariationsindexfreewithstar}
\begin{aligned}
Q\phi&= \psi &&& \bar{Q}\bar{\phi}&= \bar{\psi}\\
Q\bar{\psi} &= *i\left( \sigma \wedge_\star  *\mathrm{d}\phi \right) &&& \bar{Q} \psi &= *i\left( \sigma \wedge_\star *\mathrm{d} \bar{\phi} \right),
\end{aligned}
\end{equation}
because $\sigma$ is uncharged with respect to our twists. We then have
\begin{equation}
\label{eq:sigmawedgehodgeexteriorstarversion}
\begin{aligned}
*(\sigma \wedge_\star  *\mathrm{d}\phi) = & *(\sigma \wedge *\mathrm{d}\phi) \\
= & \sigma_\mu \partial_\nu \phi *(\mathrm{d}x^\mu \wedge *\mathrm{d}x^\nu)\\
= & \sigma_\mu \partial_\nu \phi *(\mathrm{d}x^\mu \wedge_\star *\mathrm{d}x^\nu)\\
= & \sigma^\mu \partial_\mu \phi,
\end{aligned}
\end{equation}
where the second to last and last equality follow as in our discussion of the Laplacian in section \ref{subsec:Laplaciandiscussion}. This is the same as the undeformed expression \eqref{eq:sigmawedgehodgeexterior}, and the transformations \eqref{eq:freeWZvariationsindexfreewithstar} reduce to those of eqs. \eqref{eq:freeWZvariationsindexfree}. The last step is now to verify that the not-purely-algebraic cancellations -- those involving the kinetic terms -- of the undeformed model extend to the noncommutative setting.

Combining the just discussed undeformed nature of $\sigma \wedge_\star  *\mathrm{d}$ with the related fact that the Laplacian is undeformed, allows us to immediate conclude that the twisted supersymmetry variation of the free noncommutative WZ model vanishes. Concretely, this variation takes the form
\begin{equation}
\begin{aligned}
\delta S^\star_{WZ} =&
\int \mathrm{d}^2s\int \varepsilon \left(\mathrm{d} \bar{\phi} \wedge_\star * \mathrm{d} \psi\right) + \int\mathrm{d}^2s\mathrm{d}^2\bar{s}\mathrm{d}^2s'\int \varepsilon' \left(*(\sigma \wedge_\star *\mathrm{d} \bar{\phi})\star(\sigma \wedge_\star *\mathrm{d}\psi)\right),\\
\end{aligned}
\end{equation}
plus, again, a similar term proportional to $\bar{\varepsilon}$. Because $\sigma$ is star commutative, and our deformed Hodge star has the usual property \eqref{eq:Hodgestarflip}, we can again put this in the form of eqn. \eqref{eq:SWZundefvariation2}, with appropriate star products of course.  Then, as in the undeformed setting, using property \eqref{eq:sigmawedgehodgeexteriorstarversion} twice, the fact that $s_\alpha \left.s'\right.^\alpha$ and the volume form are not affected by star products, and the fact that the Laplacian is undeformed, the second term again cancels against the first.
%

If we add interaction terms to the WZ model, working with the usual auxiliary field there are no further modifications beyond including star products in the superpotential. If instead we integrate out the auxiliary field, the products of fields appearing in the usual supersymmetry transformations are correspondingly replaced by star products, as required to realize the supersymmetry algebra on shell, since the equations of motion now also contain star products. Algebraically, the presence of star products is of course natural, since we want (repeated) supersymmetry transformations to act via the twisted coproduct, in our twisted setting.

In summary, our discussion above illustrates how, in index-free notation, twisted supersymmetry transformations of star products of fields are formally identical to regular ones of regular products of fields, how the supersymmetry transformations pick up, or can be freely written via, star products, and finally, that our twisted differential structure is compatible with supersymmetry. This shows that twisted supersymmetry variations of noncommutative deformed actions are identical in form to those of the corresponding undeformed actions, and similarly vanish, establishing twisted supersymmetry for our noncommutative deformations of supersymmetric models.\footnote{In some cases we also need a star-Fierz identity, as discussed for SYM at the end of the the next section.}

\section{Noncommutative $\mathcal{N}=4$ SYM}

As a particular application of our construction, we would like to briefly discuss noncommutative versions of maximally supersymmetric Yang-Mills theory (SYM) with gauge group $\mathrm{U}(N)$, and comment on their possible AdS/CFT interpretation. This will also provide an example of theories with extended twisted superconformal symmetry.

\subsection{Action and component Lagrangian}

In component fields, the field content of SYM given by a gauge field $A_\mu$ with corresponding field strength $G_{\mu\nu}$, four fermionic fields and their antifields $\psi_{\alpha I}$ and $\tensor{\bar{\psi}}{_{\dot{\alpha}}^I}$, and six real scalar fields $\phi^m$, $m=1,\ldots,6$. All fields transform in the adjoint representation of the gauge group. The action of SYM is given by
\begin{align*}
S_{\tiny \mbox{SYM}}=
\int\mathrm{d}^4x~\tr&\left(-\frac{1}{4g_{\text{\tiny{YM}}}^2}G^{\mu\nu}G_{\mu\nu}-\frac{1}{2}D_\mu\phi^{m} D_\mu\phi_{m}-\frac{g_{\text{\tiny{YM}}}^2}{2}\com{\phi^{m}}{\phi^{n}}\com{\phi_{m}}{\phi_{n}}\right.\\
+&\left.\tensor{\bar{\psi}}{_{\dot{\alpha}}^I}\sigma^{\mu\dot{\alpha}\alpha}D_\mu\tensor{\psi}{_{\alpha I}}-\frac{ig_{\text{\tiny{YM}}}}{2}\sigma_m^{IJ}\tensor{\psi}{^{\alpha} _I}\com{\phi^{m}}{\psi_{\alpha J}} -i\frac{g_{\text{\tiny{YM}}}}{2}\sigma_{IJ}^m\tensor{\bar{\psi}}{_{\dot{\alpha}} ^I}\com{\phi_{m}}{\tensor{\bar{\psi}}{^{\dot{\alpha}J}}} \right),
\end{align*}
where we work in the normalization of \cite{Beisert:2004ry}, but in Lorentzian signature. Hence, the $SO(6)$ $\sigma$-matrices satisfy the conditions
\begin{equation}
\begin{aligned}
\sigma_m^{IJ}\sigma_{n,JK}+\sigma_n^{IJ}\sigma_{m,JK}&=\delta_{mn}\delta^I_K\\
\sigma_{m,IJ}&=\frac{1}{2}\epsilon_{IJKL}\sigma_m^{KL}\\
\sigma_m^{IJ}&=\sigma_m^{JI}.
\end{aligned}
\end{equation}

To deform this action, we first recast it in index-free notation as
\begin{align*}
S_{\tiny \mbox{SYM}}&=-\frac{1}{4g_{\text{\tiny{YM}}}^2}\tr\int G\wedge *G - \tr\int \frac{1}{2}\mathrm{D}\phi^{m} \wedge *\mathrm{D}\phi_{m}-\frac{g_{\text{\tiny{YM}}}^2}{4}\tr\int\mathrm{d}^4x~ \com{\phi^{m}}{\phi^{n}}\com{\phi_{m}}{\phi_{n}}\\
&+\tr\int\mathrm{d}^2s\mathrm{d}^2\bar{s}\int \bar{\psi}^I\sigma\wedge*\mathrm{D}\psi_I\\
&-\frac{ig_{\text{\tiny{YM}}}}{2}\tr\int\mathrm{d}^2s\int\mathrm{d}^4x~\sigma^{IJ}_m\psi_I\com{\phi^m}{\psi_J}-\frac{ig_{\text{\tiny{YM}}}}{2}\tr\int\mathrm{d}^2\bar{s}\int\mathrm{d}^4x~\sigma_{IJ}^m\bar{\psi}^I\com{\phi_m}{\bar{\psi}^J}.
\end{align*}
Written in terms of objects that manifestly transform properly under star gauge symmetry after the deformation, we can deform this action by simply replacing every point-wise product by a star product. This gives
\begin{align*}
S^\star_{\tiny \mbox{SYM}}&=-\frac{1}{2}\tr\int \mathrm{D}\phi^{m} \wedge_\star *\mathrm{D}\phi_{m}-\frac{1}{4g_{\text{\tiny{YM}}}^2}\tr\int G\wedge_\star *G-\frac{g_{\text{\tiny{YM}}}^2}{4}\tr\int\mathrm{d}^4x~ \scom{\phi^{m}}{\phi^{n}}\star\scom{\phi_{m}}{\phi_{n}}\\
&+\tr\int\mathrm{d}^2s\mathrm{d}^2\bar{s}\int \bar{\psi}^I\star\sigma\wedge_\star*\mathrm{D}\psi_I\\
&-\frac{ig_{\text{\tiny{YM}}}}{2}~\tr\int\mathrm{d}^2s\int\mathrm{d}^4x~\sigma_m^{IJ}\psi_I\star\scom{\phi^{m}}{\psi_J}-\frac{ig_{\text{\tiny{YM}}}}{2}~\tr\int\mathrm{d}^2\bar{s}\int\mathrm{d}^4x~\sigma_{IJ}^m\bar{\psi}^I\star\scom{\phi_{m}}{\bar{\psi}^J},
\end{align*}
where we recall that
\begin{align*}
\sigma=\sigma^\star_{\mu\alpha\dot{\alpha}} \star \left( s^\alpha \star \bar{s}^{\dot{\alpha}} \star \form{x^\mu}\right)=\sigma_{\mu\alpha\dot{\alpha}} s^\alpha \bar{s}^{\dot{\alpha}} \form{x^\mu}.
\end{align*}
In terms of component fields, the Lagrangian becomes
\begin{align*}
\mathcal{L}&=\tr\left(-\frac{1}{2}D_\mu\phi^m\star D^\mu\phi_m-\frac{1}{4g_{\text{\tiny{YM}}}^2} G_{\mu\nu} \star G^{\mu\nu}+\tensor{\bar{\psi}}{_{\dot{\alpha}}^I}\sigma^{\mu\dot{\alpha}\alpha} \star D_\mu\tensor{\psi}{_{\alpha I}}\right)\\
&-\frac{g_{\text{\tiny{YM}}}^2}{4}~\tr\left(\scom{\phi^m}{\phi^n} \star \scom{\phi_m}{\phi_n}\right)\\
&-\frac{i g_{\text{\tiny{YM}}}}{2}\sigma_m^{IJ}~\tr\left(\tensor{\bar{F}}{_\alpha^\gamma}\tensor{\psi}{^{\alpha}_I} \star \phi^{m} \star \tensor{\left(F_{op}\right)}{_\gamma^\beta}\psi_{\beta J}-\tensor{\left(\bar{F}_{op}\right)}{_\alpha^\beta}\tensor{\psi}{^{\alpha}_I} \star \tensor{F}{_\beta^\gamma}\psi_{\gamma J} \star \phi^{m}\right)\\
&-\frac{ig_{\text{\tiny{YM}}}}{2}\sigma^m_{IJ}~\tr\left( \tensor{F}{_{\dot{\beta}}^{\dot{\alpha}}} \tensor{\bar{\psi}}{_{\dot{\alpha}}^{I}} \star  \phi_{m} \star \tensor{\left(\bar{F}_{op}\right)}{_{\dot{\gamma}}^{\dot{\beta}}}\tensor{\bar{\psi}}{^{\dot{\gamma}J}} - \tensor{\left(F_{op}\right)}{_{\dot{\beta}}^{\dot{\alpha}}}\tensor{\bar{\psi}}{_{\dot{\alpha}}^{I}} \star \tensor{\bar{F}}{_{\dot{\gamma}}^{\dot{\beta}}}\tensor{\bar{\psi}}{^{\dot{\gamma}J}} \star \phi_{m} \right).
\end{align*}
Here we work in terms of component fields without the star superscript, as this gives the simpler expression, and because these are the fields with undeformed propagators, to which our planar equivalence theorem of section \ref{sec:filk1} applies.

This theory admits a planar limit where we take the rank $N$ of the gauge group to infinity, which plays a central role in the AdS/CFT correspondence. The notion of planarity induced by our noncommutative product, to which the planar equivalence theorem refers, is the same in this case, as we are dealing with single trace interactions.

\subsection{Twisted superconformal symmetry}

Our noncommutative versions of $\mathcal{N}=4$ SYM have twisted superconformal symmetry. The conformal algebra acts by the usual Lie derivatives on single fields, and becomes a symmetry of the action when acting via the twisted coproduct. The (extended) supersymmetry transformations are twisted themselves, and concretely given by\footnote{The star products on the $s^\alpha$ and $s'_\alpha$ terms in e.g. the $Q$ variation of $\psi$ do nothing. We included them to have uniform star products.}
\begin{equation}
\begin{aligned}
Q^K(\phi_{m}) &=\sigma_m^{IJ}\psi_J\\
\bar{Q}_K(\phi^{m}) &= \sigma^m_{IJ} \bar{\psi}^J\\
Q^J(\left.s'\right.^\alpha\psi_{\alpha I}) &= \frac{1}{2}\delta^J_I\int\mathrm{d}^2\bar{s}~ *\left(\sigma \wedge_\star \sigma' \wedge_\star *F\right) +\frac{ig}{2}~s^\alpha \star s'_\alpha \star  \scom{\phi^{m}}{\phi_{n}}\sigma_m^{JK}\sigma^n_{KI}\\
\bar{Q}_I(s^\alpha\psi_{\alpha J}) &= *\frac{1}{2}\sigma_m^{IJ}\left(\sigma \wedge_\star *D \phi^{m}\right)\\
Q^I(\bar{\psi}^J{\dot{\alpha}}\bar{s}^{\dot{\alpha}})&=*\frac{1}{2}\sigma_m^{IJ}\left(\sigma \wedge_\star *\mathrm{D}\phi^{m}\right)\\
\bar{Q}_J(\bar{\psi}^I_{\dot{\alpha}}\bar{s}'^{\dot{\alpha}}) &=\frac{1}{2} \delta_J^I \int \mathrm{d}^2s *\left(\sigma \wedge_\star \sigma' \wedge_\star *F\right) + \frac{ig}{2} \bar{s}^{\dot{\alpha}} \star \bar{s}'_{\dot{\alpha}} \star  \scom{\phi^{m}}{\phi_{n}}\sigma_m^{I K}\sigma^n_{KJ}\\
Q^I(A) &= ig~\int\mathrm{d}^2\bar{s}~ \sigma \star \bar{\psi}^I\\
\bar{Q}_I(A) &= ig~\int\mathrm{d}^2s~  \psi_I \star \sigma.
\end{aligned}
\end{equation}
These also act on star products of fields via the twisted coproduct (effectively undeformed in index-free notation). This gives a symmetry of the action due to relation \eqref{eq:CliffAlg} in combination with a star-Fierz identity, which for our class of twists is given by
\begin{equation}
*\left( \sigma \wedge_\star *\sigma' \right) = s^\alpha s'_\alpha \bar{s}_{\dot{\alpha}} \left.\bar{s}'\right.^{\dot{\alpha}}.
\end{equation}
This identity from star commutativity of $\sigma$, and the regular Fierz identity for the matrices $\sigma^\mu_{\alpha\dot{\alpha}}$. For completeness, the $\mathfrak{su}(4)$ $R$-symmetry of SYM is not affected by spacetime twists, and is manifestly present after our noncommutative deformations, in line with the algebraic picture, where a twist in the Poincar\'e algebra does not affect the Hopf algebra restricted to the $\mathfrak{su}(4)$ generators of the superconformal algebra $\mathfrak{psu}(2,2|4)$.

\subsection{AdS/CFT}

Our noncommutative deformations of SYM have a natural interpretation in terms of AdS/CFT, extending the well-known AdS/CFT dualities for the Groenewold-Moyal noncommutative deformation of SYM \cite{Maldacena:1999mh,Hashimoto:1999ut}, and the marginal real-$\beta$ Lunin-Maldacena deformation of SYM \cite{Lunin:2005jy,Frolov:2005dj}. Namely, on the string theory side there is a class of deformations of the AdS$_5\times$S$^5$ string specified by classical $r$ matrices, known as Yang-Baxter (YB) deformations \cite{Delduc:2013qra,Kawaguchi:2014qwa,vanTongeren:2015soa}, which in the homogeneous case have Drinfel'd twisted symmetry \cite{Vicedo:2015pna,vanTongeren:2015uha}.\footnote{YB models date back to \cite{Klimcik:2002zj}, see \cite{Hoare:2021dix} for a recent pedagogical review. Inhomogeneous YB deformations, as originally considered in \cite{Delduc:2013qra}, lead to a trigonometric quantum deformed symmetry algebra \cite{Delduc:2014kha,Delduc:2017brb}.} As originally conjectured in \cite{vanTongeren:2015uha}, our twist-noncommutative versions of SYM provide natural gauge theory duals to these homogeneous YB deformed strings, with each dual pair specified and determined by a classical $r$ matrix.

Our construction allows us to deform the spacetime symmetries of SYM, i.e. the isometries of AdS$_5$ on the string side, by considering noncommutative spacetimes.\footnote{Including the $R$ symmetry of SYM in the deformation leads to dipole deformations \cite{Dasgupta:2001zu,Guica:2017mtd}, the $\beta$ deformation \cite{Leigh:1995ep,Lunin:2005jy} mentioned earlier, and its three parameter generalization \cite{Frolov:2005dj}, all fitting with twisting the superconformal symmetry of SYM. We are not aware of existing AdS/CFT case studies of deformations involving supercharges, which should however again be determined by the corresponding twist. The basis such for ``noncommutativity in superspace'' is understood \cite{deBoer:2003dpn}, with nonzero RR forms in the brane geometry resulting in nonzero fermionic anticommutators.} Our current construction limits us to the Poincar\'e subalgebra of the conformal symmetry of SYM, and unimodular $r$ matrices. The unimodularity constraint is actually natural in the context of AdS/CFT, since homogeneous Yang-Baxter deformation of the AdS$_5\times$S$^5$ are one-loop \emph{Weyl invariant} if and only if the $r$ matrix is unimodular \cite{Borsato:2016ose,Hronek:2020skb}. The restriction to the Poincar\'e algebra on the other hand, is something we hope may be overcome by extending our construction. Nevertheless, from the point of view of brane constructions underlying AdS/CFT, Poincar\'e symmetry is a fundamental structure, as a stack of coincident D3 branes has Poincar\'e symmetry, while the larger conformal symmetry only emerges in the low energy limit.

For our Poincar\'e-based deformations we can immediately give deformed brane geometries that formally give rise to the conjectured dualities. For all abelian and almost-abelian cases this was previously discussed in \cite{vanTongeren:2016eeb}, by applying the TsT transformations corresponding to (almost-)abelian Yang-Baxter deformations \cite{Osten:2016dvf,Borsato:2016ose} to give the desired brane geometries. To complete this picture and include the single not-almost-abelian deformation of $r_{14}$, we can use the general equivalence between homogeneous Yang-Baxter deformations and non-abelian T duality transformations \cite{Hoare:2016wsk,Borsato:2016pas,Borsato:2017qsx}, and similarly apply the correspondingtransformation to the standard D3 brane background. By definition, in the low energy limit this \emph{formally} yields the desired Yang-Baxter deformations of AdS$_5\times$S$^5$ in the closed string sector, and the $B$ field corresponding to the desired noncommutativity in the open string sector. The main caveat here is that already in the Groenewold-Moyal case with time-like noncommutativity -- i.e. constant electric B fields in the brane construction -- this limit is more subtle and the ``CFT'' dual is actually a noncommutative open string theory \cite{Seiberg:2000ms,Gopakumar:2000na}. This is something we expect to apply more generally, and is important to analyze further in our broader setting. Nevertheless, also for noncommutative deformations involving time, we may hope that remnants of a duality to noncommutative field theory survive at the planar level. In any case, these subtleties do not apply to purely spacelike or null cases. A second concern is that, especially in cases without supersymmetry, the underlying brane configurations may be unstable. Nonetheless, especially in settings with integrability, it may still be possible to make meaningful comparisons between field and string theory, see e.g. \cite{Bajnok:2013wsa,Skrzypek:2022cgg}.

To be clear, nontrivial deformations remain even when excluding all non-supersymmetric or (potentially) timelike cases. A particularly interesting example of such a case is the light-cone Lorent deformation with $r = \alpha M_{+1}\wedge M_{+2}$, whose twist and deformed Levi-Civita symbol are given in eqs. \eqref{eq:lightconeLorentztwist} and \eqref{eq:lightconelorentzepsilon}. The corresponding Yang-Baxter deformation of AdS$_5\times$S$^5$ is given by
\begin{align}
ds^2 & = \frac{1}{z^2} \left(-
 \mathrm{d}x^+ \mathrm{d}x^- - \frac{\alpha^2 r^2 (x^- \mathrm{d}x^-)^2 + 2 \alpha^2 r (x^-)^3 \mathrm{d}x^- \mathrm{d}r   +  z^4 (\mathrm{d}r^2 + r^2 \mathrm{d}\theta^2)}{z^4 + \alpha^2 (x^-)^4} + \mathrm{d}z^2\right)\notag\\
B & = -2\alpha  \frac{x^- r}{z^4 + \alpha^2 (x^-)^4} (r \mathrm{d}x^- - x^-\mathrm{d}r)\wedge \mathrm{d}\theta \\
e^{2(\phi-\phi_0)} & = \frac{z^4}{z^4+\alpha^2 (x^-)^4},\notag
\end{align}
in the Poincar\'e patch, using light-cone $(x^\pm)$ and polar $(r,\theta)$ coordinates in the (03) and (12) planes respectively, further supported by RR forms.\footnote{This background is obtained by a TsT transformation of undeformed AdS$_5\times$S$^5$, in the shift-isometry coordinates associated to $M_{+1}$ and $M_{+2}$. This determines the RR forms as well. Alternatively, the full background can be extracted from the Yang-Baxter sigma model action following \cite{Borsato:2016ose}.} In addition to preserving 16 supercharges and having an apparently smooth decoupling limit at the level of the brane geometry, we see that this background has a bounded dilaton, presenting a solid base for our desired AdS/CFT interpretation. This continues to apply to the non-abelian deformation obtained by adding a further $p_+ \wedge (\beta p_1 +\gamma p_2)$ term to the light-cone Lorentz $r$ matrix, in which case it preserves 8 supercharges. Another example would be the null deformation associated to $r = M_{-1} \wedge p_-$ (16 supercharges) with non-abelian extension $p_1 \wedge p_2$ (8 supercharges). As mentioned in \cite{Meier:2023kzt}, the Lorentz deformation is spacelike inside the lightcone in the $(0,1)$ plane, but timelike outside it. The corresponding background preserves no supersymmetry, and has an apparently unbounded dilaton, meaning its AdS/CFT interpretation requires further investigation. On the other hand, algebraically speaking the Lorentz deformation, based on Cartan generators, should nicely fit the quantum integrable (Bethe Ansatz) description of the AdS$_5\times$S$^5$ superstring, while the other deformations just mentioned, involve nilpotent generators whose effect on e.g. the spectral problem is less clear. Interestingly, the Lorentz deformation is relate to the light-cone Lorentz deformation by a suitable infinite boost limit.\footnote{The $r$ matrix and twist of the Lorentz deformation become those of the light-cone Lorentz deformation upon an infinite boost in the $(03)$ plane, combined with a suitable rescaling of the deformation parameter. On the string side, geometrically the identification involves a rescaling of the coordinates and deformation parameter along the lines of \cite{Hoare:2016hwh}.}

\section{Conclusions}

We have constructed noncommutative star-gauge invariant actions to all orders in the noncommutativity, including fundamental and adjoint matter, for all Drinfel'd twists of the Poincar\'e algebra that have unimodular $r$ matrices. In terms of suitable index free objects, the noncommutative deformation amounts to simply replacing products with star products, as expected. Our construction relies on a twisted notion of Hodge duality, whose linearity leads to constraints on the $\mathcal{R}$ matrix, and hence the twist defining our noncommutative spaces, limiting us to Poincar\'e-based twists. Unimodularity on the other hand is the leading order constraint to have a cyclic star product under an integral, which in the context of AdS/CFT has an interesting connection to Weyl invariance of their proposed dual strings. The resulting twists satisfy a related all-order property guaranteeing this cyclicity in general. In component fields, our actions contain various nontrivial twist and $\mathcal{R}$-matrix factors, in addition to star products. We also proved a planar equivalence theorem for Feynman diagrams in our noncommutative field theories, and discussed how our theories have twisted Poincar\'e symmetry, or twisted conformal and super symmetry when applicable. We also explicitly discussed twist-noncommutative versions of four dimensional maximally supersymmetric Yang-Mills theory. While our motivations lie mainly in AdS/CFT, we hope that our construction is of broader interest, for instance by enabling the study of star-gauge theory on Lorentz deformed, as well as $\lambda$ and $\rho$-Minkowski space, and their nonabelian generalizations.\footnote{Our construction does not cover $\kappa$-Minkowski space \cite{Majid:1994cy}. This space was originally introduced as a noncommutative space admitting a covariant action of the $\kappa$-Poincar\'e group \cite{Lukierski:1991pn}, a proper quantum group, rather than a twist deformation. However, it can also arise in a twisted setting, see e.g. \cite{Borowiec:2008uj} and references therein. This requires an extension beyond the Poincar\'e algebra, however, and is therefore not covered by our current construction.}

There are many open questions both in terms of general noncommutative field theory, and specifically our motivations coming from integrability in AdS/CFT. Since we only constructed classical actions, the main questions surround the quantum behavior of these noncommutative field theories, in terms of e.g. general renormalizability, UV/IR mixing both in general and for maximally supersymmetric Yang-Mills theory in particular,\footnote{A priori, constant, linear, and quadratic noncommutativity can lead to different behavior, since e.g. their deformation parameters have different length dimensions. UV/IR mixing is known to be absent in SYM for the Groenewold-Moyal deformation \cite{Szabo:2001kg}. It has been studied for scalar fields and linear noncommutativity, for $\kappa$-Minkowski space in \cite{Grosse:2005iz} and $\lambda$-Minkowski space in \cite{DimitrijevicCiric:2018blz}.} the structure of the Seiberg-Witten map,\footnote{This has been understood order by order for arbitrary abelian twists in \cite{Aschieri:2011ng}. See also \cite{Blumenhagen:2018shf} and the recent \cite{Kupriyanov:2023zfh} for algebraic interpretations of the Seiberg-Witten map.} the survival of twisted symmetry at the quantum level, etcetera. Our planar equivalence theorem partially addresses such questions at the planar level, but also here further studies are warranted, in particular in relation to AdS/CFT and integrability.

Applied to maximally supersymmetric Yang-Mills theory, the noncommutative deformations we consider are conjectured to be dual to homogeneous Yang-Baxter deformations of the AdS$_5\times$S$^5$ superstring \cite{vanTongeren:2016eeb}, giving rise to a variety of integrable deformations of AdS/CFT. In this setting we would like to understand the twisting of the integrable structures arising in the computation of two and higher point correlation functions in planar SYM, and match them with the expected twists of the underlying integrable model \cite{Beisert:2005if,Arutyunov:2010gu,Ahn:2010ws,Kazakov:2015efa,vanTongeren:2021jhh}. The Lorentz deformation is the natural deformation to study here, as it involves the Cartan generators of the conformal symmetry of SYM, whose associated eigenvalues play a central role in quantum integrability in AdS/CFT.\footnote{While the other deformations are classically integrable on the string theory side, their quantum integrable description is currently unclear since the typical exact S matrix approach appears difficult \cite{vanTongeren:2021jhh}. The recently constructed classical spectral curve of these deformed strings \cite{Borsato:2021fuy} may provide a systematic path forward here.} We have defined what we believe to be a suitable spectral problem for two point correlation functions in Lorentz-deformed SYM, and are working to extract its twisted integrable structure via our planar equivalence theorem \cite{NCintegrabilitypaper}.

Independently of quantum questions, we would like to extend our construction to more general noncommutative deformations. We hope that our construction can be suitably modified for deformations going beyond the Poincar\'e algebra, at least provided we stay in the symmetry algebra of a given undeformed theory, such as deformations including the dilatation generator, for conformal theories. For SYM in particular, we ultimately hope to define deformations based the entire (super)conformal algebra, matching the possible homogeneous Yang-Baxter deformations of the AdS$_5\times$S$^5$ superstring. The space of these deformations would include dipole deformations \cite{Dasgupta:2001zu,Guica:2017mtd}, the real-$\beta$ deformation \cite{Leigh:1995ep} based on twisting the $R$-symmetry of SYM \cite{Lunin:2005jy}, but also new types of deformations involving supercharges. Independently of generalizing our construction, it is important to further clarify the link between unimodularity and graded cyclicity of integrals on the field theory side, both from an algebraic point of view, and in terms of the link to Weyl-invariance of the dual strings, ideally also at higher loops. Relatedly, it would be good to understand the role of twists containing supercharges, in our construction, since they affect unimodularity. For instance, unimodular jordanian deformations are possible when including supercharges \cite{vanTongeren:2019dlq,Borsato:2022ubq}, potentially opening the door to supersymmetric Yang-Mills theory on $\kappa$-Minkowski space, described by extending the bosonic jordanian twists of \cite{Borowiec:2008uj}. Coming back to integrability, our noncommutative versions of SYM are presumably also classically integrable in the spirit of \cite{Beisert:2017pnr,Beisert:2018zxs}, with the Yangian now twisted analogously to the one for $\beta$-deformed SYM \cite{Garus:2017bgl}, and it would be nice to work this out in detail. It would also be interesting to investigate whether our noncommutative deformations admit strong deformation limits with interesting properties, similarly to the fishnet limit of $\gamma_i$-deformed SYM \cite{Gurdogan:2015csr,Sieg:2016vap}.\footnote{For example, in the Groenewold-Moyal case, it is known that maximally noncommutative field theories are purely planar \cite{Szabo:2001kg}.} Beyond Yang-Mills theory, it would also be interesting to consider noncommutative deformations of Chern-Simons theory, in particular applied to AdS$_4$/CFT$_3$.

While we focused on theories with star-gauge symmetry as a natural structure coming from string theory, it would be also interesting to contrast them with the recent braided approach to noncommutative gauge theories of \cite{Ciric:2021rhi,Giotopoulos:2021ieg}, based on the underlying $L_\infty$ structure of noncommutative gauge theories  \cite{Blumenhagen:2018kwq,Giotopoulos:2021ieg}. Finally, more generally it would be nice to investigate noncommutative instantons and our deformed self-dual Yang-Mills equations, as well as for instance the (noncommutative) integrable models that would arise in their reductions.\footnote{This has been previously studied for the Groenewold-Moyal deformation, see e.g. \cite{Lechtenfeld:2004qh,Hamanaka:2006re,Hamanaka:2011jv} and references therein.}

\section*{Acknowledgements}

We would like to thank Riccardo Borsato, Ben Hoare, Anna Pacho\l{}, and Richard Szabo for discussions, and Richard Szabo for valuable comments on the draft. TM's research is funded by the Deutsche Forschungsgemeinschaft (DFG, German Research Foundation) - Projektnummer 417533893/GRK2575 ``Rethinking Quantum Field Theory''. The work of ST is supported by the German Research Foundation via the Emmy Noether program ``Exact Results in Extended Holography''. ST is supported by LT.

\appendix

\section{A non-factorized rank four twist}
\label{app:newtwist}

In this appendix we will find a non-almost-abelian twist for $r_{13}$ and, in particular, $r_{14}$. Before doing so, however, we will discuss their common algebraic structure, as well as their difference in the Poincar\'e algebra.

Both $r_{13}$ and $r_{14}$ are simple extensions of $r$ matrices appearing in  \cite{Borsato:2016ose}, in their discussion of unimodular $r$ matrices for the conformal algebra, $\mathfrak{so}(2,4)$. The rank four $r$ matrices discussed in \cite{Borsato:2016ose} are associated to four different four-dimensional algebras, three of which have a direct-sum structure and correspond to almost-abelian $r$ matrices. The remaining case includes our $r_{13}$ and $r_{14}$, and is denoted $\mathfrak{n}_4$ in \cite{Borsato:2016ose}. It refers to an algebra defined by
\begin{equation}
[\xi_1,\xi_2] = - \xi_4, \quad [\xi_4,\xi_2] = \xi_3,
\end{equation}
and $r$ matrix of the form
\begin{equation}
r_{\mathfrak{n}_4} = \xi_1 \wedge \xi_4 + \xi_2 \wedge \xi_3.
\end{equation}
Restricted to this algebra, the symmetry algebra of $r_{\mathfrak{n}_4}$ is spanned by $\xi_3$ and $\xi_4$. Adding the abelian $\alpha \,\xi_3 \wedge \xi_4$ to $r_{\mathfrak{n}_4}$ does not change the symmetry algebra, and cannot generate an $r$ matrix of rank two for any $\alpha$. Hence, this type of $r$ matrix is not almost abelian within $\mathfrak{n}_4$, and need not be in general. Whether $r_{\mathfrak{n}_4}$ can be factorized in almost abelian form in a larger algebra, depends on the embedding in this larger algebra. $r_{13}$ and $r_{14}$ correspond to different embeddings of $r_{\mathfrak{n}_4}$ in the Poincar\'e algebra.

Concretely, let us consider $r_{13}$, whose symmetries include $p_+$ and $p_1$, and add the (subordinate) abelian term $\alpha_3 p_+ \wedge p_1$ to form
\begin{equation}
r_{13}' = M_{+1}\wedge p_+ +  \alpha_2 p_+ \wedge p_2 + (M_{+2} + \alpha_1 p_- + \alpha_3 p_+) \wedge p_1
\end{equation}
With $\alpha_2 = 1$ and $\alpha_1 = - \alpha_3 = 1/2$, and adding zero in the form $p_1 \wedge p_1$, this becomes
\begin{equation}
r =  p_1 \wedge (M_{+2} + p_1 + p_3) + p_+ \wedge (M_{+1} + p_2)
\end{equation}
which is exactly of $\mathfrak{n}_4$ form, and coincides with $r_{17}$ in Table 1 of \cite{Borsato:2016ose}. This $r$ matrix can be factorized in almost abelian form \cite{Tolstoy:2008zz}, as indicated in Table \ref{tab:rmatrices}, essentially because its symmetry algebra includes suitable Lorentz generators.

The other case, our $r_{14}$, is similarly related to $r_{15}$ (and $r_{16}$) of \cite{Borsato:2016ose}, of the form
\begin{equation}
r =   p_1 \wedge p_3 + M_{+1} \wedge p_+.
\end{equation}
The symmetry algebra of this $r$ matrix (and our $r_{14}$ for generic parameter values) is spanned by $p_+$, $p_1$, and $p_2$. Adding an arbitrary abelian term constructed out of these does not change this, or make it possible to reduce the $r$ matrix to rank two, meaning this case is not almost abelian. This was also argued in \cite{Borsato:2016ose} at the level of the geometry of the corresponding Yang-Baxter deformations of the $\ads$ string sigma model, using the link between abelian Yang-Baxter deformations and TsT transformations \cite{Osten:2016dvf}. Relatedly, the author of \cite{Tolstoy:2008zz} was also not able to find a subordinate factorization, or twist, for this case.

While we are not able factorize $r_{14}$ to find a corresponding twist, we can avoid this by rewriting the factorized twist for $r_{13}$, and exploiting the shared algebraic structure of both $r$ matrices. Consider $r_{13}$
\begin{equation}
\begin{aligned}
r_{13}= & (M_{+1}-\alpha_2 p_2)\wedge p_+ + (M_{+2} + \alpha_1 p_-) \wedge p_1,\\
= & M_{+1}\wedge p_+ + (2 \alpha_1 p_1 + \alpha_2 p_+) \wedge p_2 + (M_{+2} + \alpha_1 (p_- + 2 p_2)) \wedge p_1.
\end{aligned}
\end{equation}
If we start from the second, almost abelian, form of $r_{13}$, we can construct a corresponding twist as
\begin{equation}
\mathcal{F}_{13} = e^{i \lambda M_{+1} \wedge p_+} e^{-i \lambda \alpha_2 p_2\wedge p_+} e^{-i \lambda 2 \alpha_1 p_2 \wedge p_1} e^{i \lambda (M_{+2} + \alpha_1 (p_- + 2 p_2))\wedge p_1},
\end{equation}
where we have split the middle abelian term in two, so that we can start to combine them with the other two terms. The first two terms immediately combine since everything commutes, while for the second we use the Baker-Campbell-Hausdorff formula to find
\begin{equation}
\begin{aligned}
\mathcal{F}_{13} = & e^{i \lambda (M_{+1}- \alpha_2 p_2) \wedge p_+} e^{i \lambda (M_{+2} + \alpha_1 p_-)\wedge p_1 + \frac{1}{2} (i\lambda)^2 [-2 \alpha_1 p_2 \wedge p_1,(M_{+2} + \alpha_1 (p_- + 2 p_2))\wedge p_1]},\\
 = & e^{i \lambda (M_{+1}- \alpha_2 p_2) \wedge p_+} e^{i \lambda (M_{+2} + \alpha_1 p_-)\wedge p_1 + i \lambda^2 \alpha_1 p_+ \odot (p_1)^2},
\end{aligned}
\end{equation}
where $\odot$ denotes the symmetric tensor product. In this form, the twist for $r_{13}$ involves only the generators in its support, and it satisfies the cocycle condition due to the associated algebra relations. In other words, in this form the twist applies to any $r$ matrix of the same algebraic type, regardless of a (lack of a) almost-abelian factorization in a larger algebra.

Concretely, consider $r_{14}$
\begin{equation}
r_{14} = p_+ \wedge M_{+1} + p_- \wedge \left( \alpha p_+ + \alpha_1 p_1\right) +\tilde{\alpha} p_+ \wedge p_2.
\end{equation}
Let us drop the subordinate abelian $p_+ \wedge p_2$ term for now, and consider
\begin{equation}
\begin{aligned}
\tilde{r}_{14} = &p_+ \wedge M_{+1} + p_- \wedge \left( \alpha p_+ + \alpha_1 p_1\right)\\
 = & M_{+1} \wedge p_+ + (\alpha_1 p_-) \wedge \left(-\frac{\alpha_2}{2\alpha^1} p_+ + p_1\right).
\end{aligned}
\end{equation}
where we relabeled $\alpha = - \alpha_2/2$ in the second equality. In this form, the generators appearing in $r_{13}$ and $r_{14}$ form identical algebras, in the order presented. We can then use the second form of $\mathcal{F}_{13}$ to give a twist for $\tilde{r}_{14}$ as
\begin{equation}
\tilde{\mathcal{F}}_{14} =  e^{i \lambda M_{+1} \wedge p_+} e^{i \lambda p_-\wedge (\alpha p_+ + \alpha_1 p_1) + i \lambda^2 \alpha_1^{-1} p_+ \odot \left(\alpha p_+ + \alpha_1 p_1\right)^2},
\end{equation}
where we reinstated $\alpha$. The second term is not well-defined in the limit $\alpha_1\rightarrow 0$, but in this case $\tilde{r}_{14}$ reduces to the abelian
\begin{equation}
r = p_+ \wedge (M_{+1} - \alpha p_-),
\end{equation}
with an obvious associated twist. The full twist for $r_{14}$ is now given by adding the subordinate abelian twist corresponding to $\hat{r}_{14} = \tilde{\alpha} p_+ \wedge p_2$, i.e.
\begin{equation}
\mathcal{F}_{14} = e^{i \lambda \hat{r}_{14}} \tilde{\mathcal{F}}_{14}.
\end{equation}
The new type of term appearing in this twist has different powers of (imaginary) vector fields and a symmetric tensor product. These combine to give a star product that continues to have the desired conjugation property.

\section{Twisted coproduct for supercharges}
\label{app:twistedsusycoproduct}
In this appendix we discuss the twisted coproduct of supercharges. For abelian twists the twisted coproduct immediately takes a nice form. Namely, since a twist acts on the undeformed coproduct by conjugation, for abelian twists this simply evaluates the relevant generators in the adjoint representation. Since our supercharges are suitably labeled, this adjoint transformation is just a Lorentz transformation of the corresponding indices, i.e. the adjoint transformation of $Q_\alpha \in \mathfrak{g}$ is the same as the corresponding fundamental ($\mathrm{sl}(2,\mathbb{C})$) transformation of $s_\alpha$. This means that in terms of the notation used in the main text, for Poincar\'e-based abelian twists, we have
\begin{equation}
\Delta_{\mathcal{F}}(Q_\alpha) = Q_\beta \otimes F^\beta{}_\alpha + (F_{op})^\beta{}_\alpha \otimes Q_\beta.
\end{equation}

To extend this to almost abelian twists, it is helpful to derive the above explicitly. We consider an arbitrary abelian twist
\begin{equation}
\mathcal{F} = e^{\lambda A \wedge B}.
\end{equation}
To evaluate the twisted coproduct
\begin{equation}
\Delta_{\mathcal{F}}(Q_\alpha) = \mathcal{F} (Q_\alpha \otimes 1 + 1 \otimes Q_\alpha) \mathcal{F}^{-1},
\end{equation}
we consider its first term, and split the twist in two exponents, to write
\begin{equation}
\begin{aligned}
\mathcal{F} (Q_\alpha \otimes 1) \mathcal{F}^{-1} & = \mbox{Ad}_{e^{\lambda A \otimes B}} \mbox{Ad}_{e^{-\lambda B \otimes A}} (Q_\alpha \otimes 1) \\
& =e^{\lambda \mbox{ad}_A \otimes B} e^{-\lambda \mbox{ad}_B \otimes A} (Q_\alpha \otimes 1)
\end{aligned}
\end{equation}
where we used $\mbox{Ad}_{e^X} = e^{\mbox{ad}_X}$, $\mbox{ad}_{A\otimes B}(C\otimes 1) = \mbox{ad}_A(C)\otimes B$, and $\mbox{ad}_A(B)=\mbox{ad}_B(A)=0$. Assuming $A$ and $B$ generate spacetime symmetries, $\mbox{ad}_A$ and $\mbox{ad}_B$ act as linear operators on the $Q_\alpha$, i.e. $\mbox{ad}_A(Q_\alpha) = A_\alpha{}^\beta Q_\beta$ and similarly for $B$. This action is of course the same as the corresponding Lie derivative on a spacetime spinor $s_\alpha$, i.e. $\mathcal{L}_A (s_\alpha) = A_\alpha{}^\beta s_\beta$. Since our twisted product involves $\bar{\mathcal{F}}$, and we defined
\begin{equation}
f \star s^\alpha=s^\beta \left(\bar{F}_{op}\right)_\beta{}^\alpha f,
\end{equation}
this means
\begin{equation}
\mathcal{F} (Q_\alpha \otimes 1) \mathcal{F}^{-1} = Q_\beta \otimes F^\beta{}_\alpha,
\end{equation}
where, strictly speaking, the Lie derivatives in $\bar{F}$ should be replaced by their corresponding abstract generators. However, we will soon evaluate them as Lie derivatives again anyway. Applying similar considerations to the second term, in total gives
\begin{equation}
\label{eq:twistedcoproductQ}
\Delta_{\mathcal{F}}(Q_\alpha) = Q_\beta \otimes F^\beta{}_\alpha + (F_{op})^\beta{}_\alpha \otimes Q_\beta.
\end{equation}

It seems natural to assume that this expression applies more generally, but we have not attempted to verify this for jordanian twists, which would go beyond our present considerations, and cannot explicitly verify it for twists of unknown explicit form. Equation \eqref{eq:twistedcoproductQ} does apply for all twists we consider -- in particular those being abelian or almost abelian -- since momenta act trivially on supercharges, and therefore only contribute to the generators (Lie derivatives) appearing in $F$ and $F_{op}$, but not to the matrix structure. Inspecting Table \ref{tab:rmatrices}, all Poincar\'e generators that can appear on the right hand side of eqn. \eqref{eq:twistedcoproductQ} therefore mutually commute. Concretely this means that for an almost abelian twist, taking it rank four for concreteness, we have
\begin{equation}
\begin{aligned}
\Delta_{\mathcal{F}}(Q_\alpha) & = \mbox{Ad}_{\tilde{\mathcal{F}}}\mbox{Ad}_{\hat{\mathcal{F}}}\Delta(Q_\alpha) \\
& =  \mbox{Ad}_{\tilde{\mathcal{F}}}(Q_\beta \otimes \hat{F}^\beta{}_\alpha + (\hat{F}_{op})^\beta{}_\alpha \otimes Q_\beta),
\end{aligned}
\end{equation}
where now the generators in $\tilde{\mathcal{F}}$ commute with $\hat{F}$ and $\hat{F}_{op}$. We can then repeat the derivation leading up to eqn. \eqref{eq:twistedcoproductQ} for the conjugation by $\tilde{\mathcal{F}}$, with $\hat{F}$ and $\hat{F}_{op}$ playing the role of $1$, to find eqn. \eqref{eq:twistedcoproductQ} also in our almost abelian cases. Finally, eqn. \eqref{eq:twistedcoproductQ} also applies to the twist for $r_{14}$, since any $p\otimes p$ terms do not contribute in the present context, leaving only its abelian $M_{+1} \wedge p_+$ term.

Next we want to apply this twisted coproduct to fields. In this case it is helpful to use
\begin{equation}
F^\beta{}_\alpha = \bar{F}_\alpha{}^\beta{}
\end{equation}
which is true for Poincar\'e based twists, to write
\begin{equation}
\label{eq:twistedcoproductQv2}
\Delta_{\mathcal{F}}(\epsilon^\alpha Q_\alpha) = \epsilon^\alpha Q_\beta \otimes \bar{F}_\alpha{}^\beta + F_\alpha{}^\beta \otimes \epsilon^\alpha Q_\beta,
\end{equation}
where we note that $F_{op} = \bar{F}$ also in our almost-abelian cases, since the contributions from the individual abelian terms commute. Then the variation of a star product of fields becomes
\begin{equation}
\delta_{\epsilon^\alpha Q_\alpha}(f \star g) = \epsilon^\alpha Q_\beta(f) \star \bar{F}_\alpha{}^\beta(g) + F_\alpha{}^\beta(f)\star \epsilon^\alpha Q_\beta(g).
\end{equation}
If, analogously to star partial derivatives, we now define $Q^\star_\alpha$ via
\begin{equation}
Q = s^\alpha Q_\alpha = s^\alpha \star Q^\star_\alpha
\end{equation}
we have
\begin{equation}
Q_\alpha(f) = \bar{F}_\alpha{}^\beta(Q^\star_\beta(f)).
\end{equation}
We can then rewrite the first term of $\delta_{\epsilon^\alpha Q_\alpha}(f \star g)$, dropping the inconsequential $\epsilon^\alpha$, as
\begin{equation}
\begin{aligned}
Q_\beta(f) \star \bar{F}_\alpha^\beta(g)  = \bar{F}_\beta{}^\gamma Q^\star_\gamma(f) \star \bar{F}_\alpha{}^\beta(g) = \bar{F}_\alpha{}^\gamma (Q^\star_\gamma(f) \star g),
\end{aligned}
\end{equation}
by the analogue of eqn. \eqref{eq:R2indexdistributionoverstarproduct} with $R$ in the fundamental representation replaced by $\bar{F}$ in the spinor one.\footnote{Replacing $\bar{R}$ by $R$ in eqn. \eqref{eq:R2indexdistributionoverstarproduct} changes the index contraction from inner to outer. Then, for abelian twists $R$ and $\bar{F}$ are the same object with deformation parameters related by a factor of two, so the same relation holds for $\bar{F}$. For almost-abelian twists, it continues to hold by the commutativity of the individual abelian contributions to $\bar{F}$.}
If we now multiply this expression by $s^\alpha$, we get
\begin{equation}
s^\alpha \bar{F}_\alpha{}^\gamma (Q^\star_\gamma(f) \star g)  = s^\alpha \star (Q^\star_\gamma(f) \star g) = (s^\alpha \star Q^\star_\gamma(f)) \star g= Q(f) \star g
\end{equation}
by associativity of the star product, and definition of $Q$ and $Q^\star_\alpha$. By writing $Q = \tilde{Q}^\star_\alpha \star s^\alpha $, or keeping track of an $R$ matrix, similar considerations allow us to rewrite the second term in the coproduct, and in total we find
\begin{equation}
\delta_Q(f\star g) \equiv \mu_\mathcal{F}(\Delta_{\mathcal{F}}(s^\alpha Q_\alpha)(f,g)) = Q(f) \star g + f \star Q(g),
\end{equation}
so that phrased in terms of $Q$, the twisted coproduct effectively behaves untwisted, as used to demonstrate twisted supersymmetry in the main text.




\begin{thebibliography}{10}
\ifx\href\asklfhas\newcommand{\href}[2]{#2}\fi
\ifx\arxivref\asklfhas\newcommand{\arxivref}[2]{\href{http://arxiv.org/abs/#1}{#2}}\fi
\ifx\doiref\asklfhas\newcommand{\doiref}[2]{\href{http://dx.doi.org/#1}{#2}}\fi
\raggedright
\small
\parskip 0pt

\bibitem{Doplicher:1994tu}
S.~Doplicher, K.~Fredenhagen and J.~E.~Roberts,
\textit{``{The Quantum structure of space-time at the Planck scale and quantum
  fields}''},
\textsf{\doiref{10.1007/BF02104515}{Commun.~Math.~Phys.~172,~187~(1995)}},
\texttt{\arxivref{hep-th/0303037}{hep-th/0303037}}.

\bibitem{Arzano:2021scz}
M.~Arzano and J.~Kowalski-Glikman,
\textit{``{Deformations of Spacetime Symmetries}: {Gravity, Group-Valued
  Momenta, and Non-Commutative Fields}''}.

\bibitem{Addazi:2021xuf}
A.~Addazi et~al.,
\textit{``{Quantum gravity phenomenology at the dawn of the multi-messenger
  era\textemdash{}A review}''},
\textsf{\doiref{10.1016/j.ppnp.2022.103948}{Prog.~Part.~Nucl.~Phys.~125,~103948~(2022)}},
\texttt{\arxivref{2111.05659}{arxiv:2111.05659}}.

\bibitem{Seiberg:1999vs}
N.~Seiberg and E.~Witten,
\textit{``{String theory and noncommutative geometry}''},
\textsf{\doiref{10.1088/1126-6708/1999/09/032}{JHEP~9909,~032~(1999)}},
\texttt{\arxivref{hep-th/9908142}{hep-th/9908142}}.

\bibitem{Douglas:2001ba}
M.~R.~Douglas and N.~A.~Nekrasov,
\textit{``{Noncommutative field theory}''},
\textsf{\doiref{10.1103/RevModPhys.73.977}{Rev.~Mod.~Phys.~73,~977~(2001)}},
\texttt{\arxivref{hep-th/0106048}{hep-th/0106048}}.

\bibitem{Szabo:2001kg}
R.~J.~Szabo,
\textit{``{Quantum field theory on noncommutative spaces}''},
\textsf{\doiref{10.1016/S0370-1573(03)00059-0}{Phys.~Rept.~378,~207~(2003)}},
\texttt{\arxivref{hep-th/0109162}{hep-th/0109162}}.

\bibitem{Maldacena:1997re}
J.~M.~Maldacena,
\textit{``{The Large N limit of superconformal field theories and
  supergravity}''},
\textsf{\doiref{10.1023/A:1026654312961}{Adv.~Theor.~Math.~Phys.~2,~231~(1998)}},
\texttt{\arxivref{hep-th/9711200}{hep-th/9711200}}.

\bibitem{Maldacena:1999mh}
J.~M.~Maldacena and J.~G.~Russo,
\textit{``{Large N limit of noncommutative gauge theories}''},
\textsf{\doiref{10.1088/1126-6708/1999/09/025}{JHEP~9909,~025~(1999)}},
\texttt{\arxivref{hep-th/9908134}{hep-th/9908134}}.

\bibitem{Hashimoto:1999ut}
A.~Hashimoto and N.~Itzhaki,
\textit{``{Noncommutative Yang-Mills and the AdS / CFT correspondence}''},
\textsf{\doiref{10.1016/S0370-2693(99)01037-0}{Phys.~Lett.~B~465,~142~(1999)}},
\texttt{\arxivref{hep-th/9907166}{hep-th/9907166}}.

\bibitem{Snyder:1946qz}
H.~S.~Snyder,
\textit{``{Quantized space-time}''},
\textsf{\doiref{10.1103/PhysRev.71.38}{Phys.~Rev.~71,~38~(1947)}}.

\bibitem{Madore:2000en}
J.~Madore, S.~Schraml, P.~Schupp and J.~Wess,
\textit{``{Gauge theory on noncommutative spaces}''},
\textsf{\doiref{10.1007/s100520050012}{Eur.~Phys.~J.~C~16,~161~(2000)}},
\texttt{\arxivref{hep-th/0001203}{hep-th/0001203}}.

\bibitem{Kontsevich:1997vb}
M.~Kontsevich,
\textit{``{Deformation quantization of Poisson manifolds. 1.}''},
\textsf{\doiref{10.1023/B:MATH.0000027508.00421.bf}{Lett.~Math.~Phys.~66,~157~(2003)}},
\texttt{\arxivref{q-alg/9709040}{q-alg/9709040}}.

\bibitem{drinfeld_YBESolutions_1983}
V.~Drinfel'd,
\textit{``On constant quasi-classical solutions of the Yang-Baxter quantum
  equation''},
\textsf{Sov.~Math.~Dokl.~28~(1983)~667~66,~V.~Drinfel'd}.

\bibitem{Aschieri:2009zz}
P.~Aschieri, M.~Dimitrijevic, P.~Kulish, F.~Lizzi and J.~Wess,
\textit{``{Noncommutative spacetimes: Symmetries in noncommutative geometry and
  field theory}''}.

\bibitem{Chaichian:2004za}
M.~Chaichian, P.~P.~Kulish, K.~Nishijima and A.~Tureanu,
\textit{``{On a Lorentz-invariant interpretation of noncommutative space-time
  and its implications on noncommutative QFT}''},
\textsf{\doiref{10.1016/j.physletb.2004.10.045}{Phys.~Lett.~B~604,~98~(2004)}},
\texttt{\arxivref{hep-th/0408069}{hep-th/0408069}}.

\bibitem{Wess:2003da}
J.~Wess,
\textit{``{Deformed coordinate spaces: Derivatives}''},
\texttt{\arxivref{hep-th/0408080}{hep-th/0408080}},
in: \textit{``{1st Balkan Workshop on Mathematical, Theoretical and
  Phenomenological Challenges Beyond the Standard Model}: {Perspectives of
  Balkans Collaboration}''},
122--128p.

\bibitem{Chaichian:2004yh}
M.~Chaichian, P.~Presnajder and A.~Tureanu,
\textit{``{New concept of relativistic invariance in NC space-time: Twisted
  Poincare symmetry and its implications}''},
\textsf{\doiref{10.1103/PhysRevLett.94.151602}{Phys.~Rev.~Lett.~94,~151602~(2005)}},
\texttt{\arxivref{hep-th/0409096}{hep-th/0409096}}.

\bibitem{Vicedo:2015pna}
B.~Vicedo,
\textit{``{Deformed integrable \ensuremath{\sigma}-models, classical R-matrices
  and classical exchange algebra on Drinfel\textquoteright{}d doubles}''},
\textsf{\doiref{10.1088/1751-8113/48/35/355203}{J.~Phys.~A~48,~355203~(2015)}},
\texttt{\arxivref{1504.06303}{arxiv:1504.06303}}.

\bibitem{vanTongeren:2015uha}
S.~J.~van~Tongeren,
\textit{``{Yang\textendash{}Baxter deformations, AdS/CFT, and
  twist-noncommutative gauge theory}''},
\textsf{\doiref{10.1016/j.nuclphysb.2016.01.012}{Nucl.~Phys.~B~904,~148~(2016)}},
\texttt{\arxivref{1506.01023}{arxiv:1506.01023}}.

\bibitem{Borsato:2021fuy}
R.~Borsato, S.~Driezen and J.~L.~Miramontes,
\textit{``{Homogeneous Yang-Baxter deformations as undeformed yet twisted
  models}''},
\textsf{\doiref{10.1007/JHEP04(2022)053}{JHEP~2204,~053~(2022)}},
\texttt{\arxivref{2112.12025}{arxiv:2112.12025}}.

\bibitem{vanTongeren:2016eeb}
S.~J.~van~Tongeren,
\textit{``{Almost abelian twists and AdS/CFT}''},
\textsf{\doiref{10.1016/j.physletb.2016.12.002}{Phys.~Lett.~B~765,~344~(2017)}},
\texttt{\arxivref{1610.05677}{arxiv:1610.05677}}.

\bibitem{Dimitrijevic:2011jg}
M.~Dimitrijevic and L.~Jonke,
\textit{``{A Twisted look on kappa-Minkowski: U(1) gauge theory}''},
\textsf{\doiref{10.1007/JHEP12(2011)080}{JHEP~1112,~080~(2011)}},
\texttt{\arxivref{1107.3475}{arxiv:1107.3475}}.

\bibitem{Dimitrijevic:2014dxa}
M.~Dimitrijevic, L.~Jonke and A.~Pachol,
\textit{``{Gauge Theory on Twisted $\kappa$-Minkowski: Old Problems and
  Possible Solutions}''},
\textsf{\doiref{10.3842/SIGMA.2014.063}{SIGMA~10,~063~(2014)}},
\texttt{\arxivref{1403.1857}{arxiv:1403.1857}}.

\bibitem{Hersent:2022gry}
K.~Hersent, P.~Mathieu and J.-C.~Wallet,
\textit{``{Gauge theories on quantum spaces}''},
\textsf{\doiref{10.1016/j.physrep.2023.03.002}{Phys.~Rept.~1014,~1~(2023)}},
\texttt{\arxivref{2210.11890}{arxiv:2210.11890}}.

\bibitem{Mathieu:2020ccc}
P.~Mathieu and J.-C.~Wallet,
\textit{``{Gauge theories on \ensuremath{\kappa}-Minkowski spaces: twist and
  modular operators}''},
\textsf{\doiref{10.1007/JHEP05(2020)112}{JHEP~2005,~112~(2020)}},
\texttt{\arxivref{2002.02309}{arxiv:2002.02309}}.

\bibitem{Ciric:2017rnf}
M.~D.~\'Ciri\'c, N.~Konjik and A.~Samsarov,
\textit{``{Noncommutative scalar quasinormal modes of the
  Reissner\textendash{}Nordstr\"om black hole}''},
\textsf{\doiref{10.1088/1361-6382/aad201}{Class.~Quant.~Grav.~35,~175005~(2018)}},
\texttt{\arxivref{1708.04066}{arxiv:1708.04066}}.

\bibitem{Meier:2023kzt}
T.~Meier and S.~J.~van~Tongeren,
\textit{``{Quadratic twist-noncommutative gauge theory}''},
\texttt{\arxivref{2301.08757}{arxiv:2301.08757}}.

\bibitem{Filk:1996dm}
T.~Filk,
\textit{``{Divergencies in a field theory on quantum space}''},
\textsf{\doiref{10.1016/0370-2693(96)00024-X}{Phys.~Lett.~B~376,~53~(1996)}}.

\bibitem{Beisert:2010jr}
N.~Beisert et~al.,
\textit{``{Review of AdS/CFT Integrability: An Overview}''},
\textsf{\doiref{10.1007/s11005-011-0529-2}{Lett.~Math.~Phys.~99,~3~(2012)}},
\texttt{\arxivref{1012.3982}{arxiv:1012.3982}}.

\bibitem{NCintegrabilitypaper}
T.~Meier and S.~J.~van~Tongeren.

\bibitem{Tolstoy:2008zz}
V.~N.~Tolstoy,
\textit{``{Twisted quantum deformations of Lorentz and Poincare algebras}''},
\textsf{Bulg.~J.~Phys.~35,~441~(2008)},
\texttt{\arxivref{0712.3962}{arxiv:0712.3962}}.

\bibitem{Chari:1994pz}
V.~Chari and A.~Pressley,
\textit{``{A guide to quantum groups}''},
Cambridge university press (1995).

\bibitem{Giaquinto:1994jx}
A.~Giaquinto and J.~J.~Zhang,
\textit{``{Bialgebra actions, twists, and universal deformation formulas}''},
\textsf{\doiref{10.1016/S0022-4049(97)00041-8}{J.Pure~Appl.Algebra~128,~133~(1998)}},
\texttt{\arxivref{hep-th/9411140}{hep-th/9411140}}.

\bibitem{Gutt:1983}
S.~Gutt,
\textit{``An explicit*-product on the cotangent bundle of a Lie group''},
\textsf{Letters~in~Mathematical~Physics~7,~249~(1983)}.

\bibitem{Gracia-Bondia:2001ynb}
J.~M.~Gracia-Bondia, F.~Lizzi, G.~Marmo and P.~Vitale,
\textit{``{Infinitely many star products to play with}''},
\textsf{\doiref{10.1088/1126-6708/2002/04/026}{JHEP~0204,~026~(2002)}},
\texttt{\arxivref{hep-th/0112092}{hep-th/0112092}}.

\bibitem{DimitrijevicCiric:2018blz}
M.~Dimitrijevic~Ciric, N.~Konjik, M.~A.~Kurkov, F.~Lizzi and P.~Vitale,
\textit{``{Noncommutative field theory from angular twist}''},
\textsf{\doiref{10.1103/PhysRevD.98.085011}{Phys.~Rev.~D~98,~085011~(2018)}},
\texttt{\arxivref{1806.06678}{arxiv:1806.06678}}.

\bibitem{lizzi:2021dud}
F.~Lizzi and P.~Vitale,
\textit{``{Time Discretization From Noncommutativity}''},
\textsf{\doiref{10.1016/j.physletb.2021.136372}{Phys.~Lett.~B~818,~136372~(2021)}},
\texttt{\arxivref{2101.06633}{arxiv:2101.06633}}.

\bibitem{Gubitosi:2021itz}
G.~Gubitosi, F.~Lizzi, J.~J.~Relancio and P.~Vitale,
\textit{``{Double quantization}''},
\textsf{\doiref{10.1103/PhysRevD.105.126013}{Phys.~Rev.~D~105,~126013~(2022)}},
\texttt{\arxivref{2112.11401}{arxiv:2112.11401}}.

\bibitem{Fabiano:2023uhg}
G.~Fabiano, G.~Gubitosi, F.~Lizzi, L.~Scala and P.~Vitale,
\textit{``{Bicrossproduct vs. twist quantum symmetries in noncommutative
  geometries: the case of $\varrho$-Minkowski}''},
\texttt{\arxivref{2305.00526}{arxiv:2305.00526}}.

\bibitem{Lukierski:2005fc}
J.~Lukierski and M.~Woronowicz,
\textit{``{New Lie-algebraic and quadratic deformations of Minkowski space from
  twisted Poincare symmetries}''},
\textsf{\doiref{10.1016/j.physletb.2005.11.052}{Phys.~Lett.~B~633,~116~(2006)}},
\texttt{\arxivref{hep-th/0508083}{hep-th/0508083}}.

\bibitem{Borsato:2016ose}
R.~Borsato and L.~Wulff,
\textit{``{Target space supergeometry of $\eta$ and $\lambda$-deformed
  strings}''},
\textsf{\doiref{10.1007/JHEP10(2016)045}{JHEP~1610,~045~(2016)}},
\texttt{\arxivref{1608.03570}{arxiv:1608.03570}}.

\bibitem{Aschieri:2005zs}
P.~Aschieri, M.~Dimitrijevic, F.~Meyer and J.~Wess,
\textit{``{Noncommutative geometry and gravity}''},
\textsf{\doiref{10.1088/0264-9381/23/6/005}{Class.~Quant.~Grav.~23,~1883~(2006)}},
\texttt{\arxivref{hep-th/0510059}{hep-th/0510059}}.

\bibitem{Aschieri:2009ky}
P.~Aschieri and L.~Castellani,
\textit{``{Noncommutative D=4 gravity coupled to fermions}''},
\textsf{\doiref{10.1088/1126-6708/2009/06/086}{JHEP~0906,~086~(2009)}},
\texttt{\arxivref{0902.3817}{arxiv:0902.3817}}.

\bibitem{Meyer:1994wi}
U.~Meyer,
\textit{``{Wave equations on q Minkowski space}''},
\textsf{\doiref{10.1007/BF02101524}{Commun.~Math.~Phys.~174,~457~(1995)}},
\texttt{\arxivref{hep-th/9404054}{hep-th/9404054}}.

\bibitem{Majid:1994mh}
S.~Majid,
\textit{``{q epsilon tensor for quantum and braided spaces}''},
\textsf{\doiref{10.1063/1.531098}{J.~Math.~Phys.~36,~1991~(1995)}},
\texttt{\arxivref{hep-th/9406157}{hep-th/9406157}}.

\bibitem{Schenkel:2011biz}
A.~Schenkel,
\textit{``{Noncommutative Gravity and Quantum Field Theory on Noncommutative
  Curved Spacetimes}''},
\texttt{\arxivref{1210.1115}{arxiv:1210.1115}}.

\bibitem{STOLIN1999285}
A.~Stolin,
\textit{``Rational solutions of the classical Yang-Baxter equation and quasi
  Frobenius Lie algebras''},
\textsf{\doiref{https://doi.org/10.1016/S0022-4049(97)00217-X}{Journal~of~Pure~and~Applied~Algebra~137,~285~(1999)}},
\href{https://www.sciencedirect.com/science/article/pii/S002240499700217X}{\texttt{https://www.sciencedirect.com/science/article/pii/S002240499700217X}}.

\bibitem{Hoare:2021dix}
B.~Hoare,
\textit{``{Integrable deformations of sigma models}''},
\textsf{\doiref{10.1088/1751-8121/ac4a1e}{J.~Phys.~A~55,~093001~(2022)}},
\texttt{\arxivref{2109.14284}{arxiv:2109.14284}}.

\bibitem{zakrzewski:1997}
S.~Zakrzewski,
\textit{``Poisson {Structures} on the {Poincaré} {Group}''},
\textsf{\doiref{10.1007/s002200050091}{Communications~in~Mathematical~Physics~185,~285~(1997)}},
\href{https://doi.org/10.1007/s002200050091}{\texttt{https://doi.org/10.1007/s002200050091}}.

\bibitem{Jurco:2001rq}
B.~Jurco, L.~Moller, S.~Schraml, P.~Schupp and J.~Wess,
\textit{``{Construction of nonAbelian gauge theories on noncommutative
  spaces}''},
\textsf{\doiref{10.1007/s100520100731}{Eur.~Phys.~J.~C~21,~383~(2001)}},
\texttt{\arxivref{hep-th/0104153}{hep-th/0104153}}.

\bibitem{Ciric:2021rhi}
M.~D.~\'Ciri\'c, G.~Giotopoulos, V.~Radovanovi\'c and R.~J.~Szabo,
\textit{``{Braided $L_{\infty}$-Algebras, Braided Field Theory and
  Noncommutative Gravity}''},
\texttt{\arxivref{2103.08939}{arxiv:2103.08939}}.

\bibitem{Giotopoulos:2021ieg}
G.~Giotopoulos and R.~J.~Szabo,
\textit{``{Braided symmetries in noncommutative field theory}''},
\textsf{\doiref{10.1088/1751-8121/ac5dad}{J.~Phys.~A~55,~353001~(2022)}},
\texttt{\arxivref{2112.00541}{arxiv:2112.00541}}.

\bibitem{SpinorLie}
Y.~CHOQUET-BRUHAT and C.~DEWITT-MORETTE,
\textit{``Analysis, Manifolds and Physics''},
North-Holland (2000),
Amsterdam,
433-524p.

\bibitem{Beisert:2004ry}
N.~Beisert,
\textit{``{The Dilatation operator of N=4 super Yang-Mills theory and
  integrability}''},
\textsf{\doiref{10.1016/j.physrep.2004.09.007}{Phys.~Rept.~405,~1~(2004)}},
\texttt{\arxivref{hep-th/0407277}{hep-th/0407277}}.

\bibitem{Lunin:2005jy}
O.~Lunin and J.~M.~Maldacena,
\textit{``{Deforming field theories with U(1) x U(1) global symmetry and their
  gravity duals}''},
\textsf{\doiref{10.1088/1126-6708/2005/05/033}{JHEP~0505,~033~(2005)}},
\texttt{\arxivref{hep-th/0502086}{hep-th/0502086}}.

\bibitem{Frolov:2005dj}
S.~Frolov,
\textit{``{Lax pair for strings in Lunin-Maldacena background}''},
\textsf{\doiref{10.1088/1126-6708/2005/05/069}{JHEP~0505,~069~(2005)}},
\texttt{\arxivref{hep-th/0503201}{hep-th/0503201}}.

\bibitem{Delduc:2013qra}
F.~Delduc, M.~Magro and B.~Vicedo,
\textit{``{An integrable deformation of the $AdS_5 \times S^5$ superstring
  action}''},
\textsf{\doiref{10.1103/PhysRevLett.112.051601}{Phys.~Rev.~Lett.~112,~051601~(2014)}},
\texttt{\arxivref{1309.5850}{arxiv:1309.5850}}.

\bibitem{Kawaguchi:2014qwa}
I.~Kawaguchi, T.~Matsumoto and K.~Yoshida,
\textit{``{Jordanian deformations of the $AdS_5 x S^5$ superstring}''},
\textsf{\doiref{10.1007/JHEP04(2014)153}{JHEP~1404,~153~(2014)}},
\texttt{\arxivref{1401.4855}{arxiv:1401.4855}}.

\bibitem{vanTongeren:2015soa}
S.~J.~van~Tongeren,
\textit{``{On classical Yang-Baxter based deformations of the AdS$_{5}$
  \texttimes{} S$^{5}$ superstring}''},
\textsf{\doiref{10.1007/JHEP06(2015)048}{JHEP~1506,~048~(2015)}},
\texttt{\arxivref{1504.05516}{arxiv:1504.05516}}.

\bibitem{Klimcik:2002zj}
C.~Klimcik,
\textit{``{Yang-Baxter sigma models and dS/AdS T duality}''},
\textsf{\doiref{10.1088/1126-6708/2002/12/051}{JHEP~0212,~051~(2002)}},
\texttt{\arxivref{hep-th/0210095}{hep-th/0210095}}.

\bibitem{Delduc:2014kha}
F.~Delduc, M.~Magro and B.~Vicedo,
\textit{``{Derivation of the action and symmetries of the $q$-deformed $AdS_{5}
  \times S^{5}$ superstring}''},
\textsf{\doiref{10.1007/JHEP10(2014)132}{JHEP~1410,~132~(2014)}},
\texttt{\arxivref{1406.6286}{arxiv:1406.6286}}.

\bibitem{Delduc:2017brb}
F.~Delduc, T.~Kameyama, M.~Magro and B.~Vicedo,
\textit{``{Affine $q$-deformed symmetry and the classical Yang-Baxter
  $\sigma$-model}''},
\textsf{\doiref{10.1007/JHEP03(2017)126}{JHEP~1703,~126~(2017)}},
\texttt{\arxivref{1701.03691}{arxiv:1701.03691}}.

\bibitem{Dasgupta:2001zu}
K.~Dasgupta and M.~M.~Sheikh-Jabbari,
\textit{``{Noncommutative dipole field theories}''},
\textsf{\doiref{10.1088/1126-6708/2002/02/002}{JHEP~0202,~002~(2002)}},
\texttt{\arxivref{hep-th/0112064}{hep-th/0112064}}.

\bibitem{Guica:2017mtd}
M.~Guica, F.~Levkovich-Maslyuk and K.~Zarembo,
\textit{``{Integrability in dipole-deformed $\boldsymbol{\mathcal{N}=4}$ super
  Yang\textendash{}Mills}''},
\textsf{\doiref{10.1088/1751-8121/aa8491}{J.~Phys.~A~50,~39~(2017)}},
\texttt{\arxivref{1706.07957}{arxiv:1706.07957}}.

\bibitem{Leigh:1995ep}
R.~G.~Leigh and M.~J.~Strassler,
\textit{``{Exactly marginal operators and duality in four-dimensional N=1
  supersymmetric gauge theory}''},
\textsf{\doiref{10.1016/0550-3213(95)00261-P}{Nucl.~Phys.~B~447,~95~(1995)}},
\texttt{\arxivref{hep-th/9503121}{hep-th/9503121}}.

\bibitem{deBoer:2003dpn}
J.~de~Boer, P.~A.~Grassi and P.~van~Nieuwenhuizen,
\textit{``{Noncommutative superspace from string theory}''},
\textsf{\doiref{10.1016/j.physletb.2003.08.071}{Phys.~Lett.~B~574,~98~(2003)}},
\texttt{\arxivref{hep-th/0302078}{hep-th/0302078}}.

\bibitem{Hronek:2020skb}
S.~Hronek and L.~Wulff,
\textit{``{Relaxing unimodularity for Yang-Baxter deformed strings}''},
\textsf{\doiref{10.1007/JHEP10(2020)065}{JHEP~2010,~065~(2020)}},
\texttt{\arxivref{2007.15663}{arxiv:2007.15663}}.

\bibitem{Osten:2016dvf}
D.~Osten and S.~J.~van~Tongeren,
\textit{``{Abelian Yang\textendash{}Baxter deformations and TsT
  transformations}''},
\textsf{\doiref{10.1016/j.nuclphysb.2016.12.007}{Nucl.~Phys.~B~915,~184~(2017)}},
\texttt{\arxivref{1608.08504}{arxiv:1608.08504}}.

\bibitem{Hoare:2016wsk}
B.~Hoare and A.~A.~Tseytlin,
\textit{``{Homogeneous Yang-Baxter deformations as non-abelian duals of the
  $\mathrm{AdS}_5 \sigma$-model}''},
\textsf{\doiref{10.1088/1751-8113/49/49/494001}{J.~Phys.~A~49,~494001~(2016)}},
\texttt{\arxivref{1609.02550}{arxiv:1609.02550}}.

\bibitem{Borsato:2016pas}
R.~Borsato and L.~Wulff,
\textit{``{Integrable Deformations of $T$-Dual $\sigma$ Models}''},
\textsf{\doiref{10.1103/PhysRevLett.117.251602}{Phys.~Rev.~Lett.~117,~251602~(2016)}},
\texttt{\arxivref{1609.09834}{arxiv:1609.09834}}.

\bibitem{Borsato:2017qsx}
R.~Borsato and L.~Wulff,
\textit{``{On non-abelian T-duality and deformations of supercoset string
  sigma-models}''},
\textsf{\doiref{10.1007/JHEP10(2017)024}{JHEP~1710,~024~(2017)}},
\texttt{\arxivref{1706.10169}{arxiv:1706.10169}}.

\bibitem{Seiberg:2000ms}
N.~Seiberg, L.~Susskind and N.~Toumbas,
\textit{``{Strings in background electric field, space / time noncommutativity
  and a new noncritical string theory}''},
\textsf{\doiref{10.1088/1126-6708/2000/06/021}{JHEP~0006,~021~(2000)}},
\texttt{\arxivref{hep-th/0005040}{hep-th/0005040}}.

\bibitem{Gopakumar:2000na}
R.~Gopakumar, J.~M.~Maldacena, S.~Minwalla and A.~Strominger,
\textit{``{S duality and noncommutative gauge theory}''},
\textsf{\doiref{10.1088/1126-6708/2000/06/036}{JHEP~0006,~036~(2000)}},
\texttt{\arxivref{hep-th/0005048}{hep-th/0005048}}.

\bibitem{Bajnok:2013wsa}
Z.~Bajnok, N.~Drukker, A.~Heged\"us, R.~I.~Nepomechie, L.~Palla, C.~Sieg and
  R.~Suzuki,
\textit{``{The spectrum of tachyons in AdS/CFT}''},
\textsf{\doiref{10.1007/JHEP03(2014)055}{JHEP~1403,~055~(2014)}},
\texttt{\arxivref{1312.3900}{arxiv:1312.3900}}.

\bibitem{Skrzypek:2022cgg}
T.~Skrzypek,
\textit{``{Integrability treatment of AdS/CFT orbifolds}''},
\texttt{\arxivref{2211.03806}{arxiv:2211.03806}}.

\bibitem{Hoare:2016hwh}
B.~Hoare and S.~J.~van~Tongeren,
\textit{``{On jordanian deformations of AdS$_5$ and supergravity}''},
\textsf{\doiref{10.1088/1751-8113/49/43/434006}{J.~Phys.~A~49,~434006~(2016)}},
\texttt{\arxivref{1605.03554}{arxiv:1605.03554}}.

\bibitem{Majid:1994cy}
S.~Majid and H.~Ruegg,
\textit{``{Bicrossproduct structure of kappa Poincare group and noncommutative
  geometry}''},
\textsf{\doiref{10.1016/0370-2693(94)90699-8}{Phys.~Lett.~B~334,~348~(1994)}},
\texttt{\arxivref{hep-th/9405107}{hep-th/9405107}}.

\bibitem{Lukierski:1991pn}
J.~Lukierski, H.~Ruegg, A.~Nowicki and V.~N.~Tolstoi,
\textit{``{Q deformation of Poincare algebra}''},
\textsf{\doiref{10.1016/0370-2693(91)90358-W}{Phys.~Lett.~B~264,~331~(1991)}}.

\bibitem{Borowiec:2008uj}
A.~Borowiec and A.~Pachol,
\textit{``{kappa-Minkowski spacetime as the result of Jordanian twist
  deformation}''},
\textsf{\doiref{10.1103/PhysRevD.79.045012}{Phys.~Rev.~D~79,~045012~(2009)}},
\texttt{\arxivref{0812.0576}{arxiv:0812.0576}}.

\bibitem{Grosse:2005iz}
H.~Grosse and M.~Wohlgenannt,
\textit{``{On kappa-deformation and UV/IR mixing}''},
\textsf{\doiref{10.1016/j.nuclphysb.2006.05.004}{Nucl.~Phys.~B~748,~473~(2006)}},
\texttt{\arxivref{hep-th/0507030}{hep-th/0507030}}.

\bibitem{Aschieri:2011ng}
P.~Aschieri and L.~Castellani,
\textit{``{Noncommutative gravity coupled to fermions: second order expansion
  via Seiberg-Witten map}''},
\textsf{\doiref{10.1007/JHEP07(2012)184}{JHEP~1207,~184~(2012)}},
\texttt{\arxivref{1111.4822}{arxiv:1111.4822}}.

\bibitem{Blumenhagen:2018shf}
R.~Blumenhagen, M.~Brinkmann, V.~Kupriyanov and M.~Traube,
\textit{``{On the Uniqueness of L$_\infty$ bootstrap: Quasi-isomorphisms are
  Seiberg-Witten Maps}''},
\textsf{\doiref{10.1063/1.5048352}{J.~Math.~Phys.~59,~123505~(2018)}},
\texttt{\arxivref{1806.10314}{arxiv:1806.10314}}.

\bibitem{Kupriyanov:2023zfh}
V.~Kupriyanov and A.~Sharapov,
\textit{``{What is the Seiberg-Witten map exactly?}''},
\texttt{\arxivref{2302.07175}{arxiv:2302.07175}}.

\bibitem{Beisert:2005if}
N.~Beisert and R.~Roiban,
\textit{``{Beauty and the twist: The Bethe ansatz for twisted N=4 SYM}''},
\textsf{\doiref{10.1088/1126-6708/2005/08/039}{JHEP~0508,~039~(2005)}},
\texttt{\arxivref{hep-th/0505187}{hep-th/0505187}}.

\bibitem{Arutyunov:2010gu}
G.~Arutyunov, M.~de~Leeuw and S.~J.~van~Tongeren,
\textit{``{Twisting the Mirror TBA}''},
\textsf{\doiref{10.1007/JHEP02(2011)025}{JHEP~1102,~025~(2011)}},
\texttt{\arxivref{1009.4118}{arxiv:1009.4118}}.

\bibitem{Ahn:2010ws}
C.~Ahn, Z.~Bajnok, D.~Bombardelli and R.~I.~Nepomechie,
\textit{``{Twisted Bethe equations from a twisted S-matrix}''},
\textsf{\doiref{10.1007/JHEP02(2011)027}{JHEP~1102,~027~(2011)}},
\texttt{\arxivref{1010.3229}{arxiv:1010.3229}}.

\bibitem{Kazakov:2015efa}
V.~Kazakov, S.~Leurent and D.~Volin,
\textit{``{T-system on T-hook: Grassmannian Solution and Twisted Quantum
  Spectral Curve}''},
\textsf{\doiref{10.1007/JHEP12(2016)044}{JHEP~1612,~044~(2016)}},
\texttt{\arxivref{1510.02100}{arxiv:1510.02100}}.

\bibitem{vanTongeren:2021jhh}
S.~J.~van~Tongeren and Y.~Zimmermann,
\textit{``{Do Drinfeld twists of $AdS_5 \times S^5$ survive light-cone
  quantization?}''},
\textsf{\doiref{10.21468/SciPostPhysCore.5.2.028}{SciPost~Phys.~Core~5,~028~(2022)}},
\texttt{\arxivref{2112.10279}{arxiv:2112.10279}}.

\bibitem{vanTongeren:2019dlq}
S.~J.~van~Tongeren,
\textit{``{Unimodular jordanian deformations of integrable superstrings}''},
\textsf{\doiref{10.21468/SciPostPhys.7.1.011}{SciPost~Phys.~7,~011~(2019)}},
\texttt{\arxivref{1904.08892}{arxiv:1904.08892}}.

\bibitem{Borsato:2022ubq}
R.~Borsato and S.~Driezen,
\textit{``{All Jordanian deformations of the $AdS_5 \times S^5$
  superstring}''},
\texttt{\arxivref{2212.11269}{arxiv:2212.11269}}.

\bibitem{Beisert:2017pnr}
N.~Beisert, A.~Garus and M.~Rosso,
\textit{``{Yangian Symmetry and Integrability of Planar N=4 Supersymmetric
  Yang-Mills Theory}''},
\textsf{\doiref{10.1103/PhysRevLett.118.141603}{Phys.~Rev.~Lett.~118,~141603~(2017)}},
\texttt{\arxivref{1701.09162}{arxiv:1701.09162}}.

\bibitem{Beisert:2018zxs}
N.~Beisert, A.~Garus and M.~Rosso,
\textit{``{Yangian Symmetry for the Action of Planar $\mathcal N=$ 4 Super
  Yang-Mills and $\mathcal N=$ 6 Super Chern-Simons Theories}''},
\textsf{\doiref{10.1103/PhysRevD.98.046006}{Phys.~Rev.~D~98,~046006~(2018)}},
\texttt{\arxivref{1803.06310}{arxiv:1803.06310}}.

\bibitem{Garus:2017bgl}
A.~Garus,
\textit{``{Untwisting the symmetries of $\beta$-deformed Super-Yang--Mills}''},
\textsf{\doiref{10.1007/JHEP10(2017)007}{JHEP~1710,~007~(2017)}},
\texttt{\arxivref{1707.04128}{arxiv:1707.04128}}.

\bibitem{Gurdogan:2015csr}
O.~G\"urdo\u{g}an and V.~Kazakov,
\textit{``{New Integrable 4D Quantum Field Theories from Strongly Deformed
  Planar $\mathcal N = $ 4 Supersymmetric Yang-Mills Theory}''},
\textsf{\doiref{10.1103/PhysRevLett.117.201602}{Phys.~Rev.~Lett.~117,~201602~(2016)}},
\texttt{\arxivref{1512.06704}{arxiv:1512.06704}},
[Addendum: Phys.Rev.Lett. 117, 259903 (2016)].

\bibitem{Sieg:2016vap}
C.~Sieg and M.~Wilhelm,
\textit{``{On a CFT limit of planar $\gamma_i$-deformed $\mathcal{N}=4$ SYM
  theory}''},
\textsf{\doiref{10.1016/j.physletb.2016.03.004}{Phys.~Lett.~B~756,~118~(2016)}},
\texttt{\arxivref{1602.05817}{arxiv:1602.05817}}.

\bibitem{Blumenhagen:2018kwq}
R.~Blumenhagen, I.~Brunner, V.~Kupriyanov and D.~L\"ust,
\textit{``{Bootstrapping non-commutative gauge theories from L$_\infty$
  algebras}''},
\textsf{\doiref{10.1007/JHEP05(2018)097}{JHEP~1805,~097~(2018)}},
\texttt{\arxivref{1803.00732}{arxiv:1803.00732}}.

\bibitem{Lechtenfeld:2004qh}
O.~Lechtenfeld, L.~Mazzanti, S.~Penati, A.~D.~Popov and L.~Tamassia,
\textit{``{Integrable noncommutative Sine-Gordon model}''},
\textsf{\doiref{10.1016/j.nuclphysb.2004.10.050}{Nucl.~Phys.~B~705,~477~(2005)}},
\texttt{\arxivref{hep-th/0406065}{hep-th/0406065}}.

\bibitem{Hamanaka:2006re}
M.~Hamanaka,
\textit{``{Noncommutative Ward's conjecture and integrable systems}''},
\textsf{\doiref{10.1016/j.nuclphysb.2006.02.014}{Nucl.~Phys.~B~741,~368~(2006)}},
\texttt{\arxivref{hep-th/0601209}{hep-th/0601209}}.

\bibitem{Hamanaka:2011jv}
M.~Hamanaka,
\textit{``{Noncommutative Solitons and Quasideterminants}''},
\textsf{\doiref{10.1088/0031-8949/89/03/038006}{Phys.~Scripta~89,~038006~(2014)}},
\texttt{\arxivref{1101.0005}{arxiv:1101.0005}}.

\end{thebibliography}



\end{document}